\documentclass[twocolumn, trackchanges]{aastex7}

\accepted{December 3, 2025}

\usepackage{natbib}
\usepackage{gensymb}
\usepackage{amsmath, amssymb}
\usepackage{bigints}
\usepackage{indentfirst}
\usepackage{threeparttable,booktabs}
\usepackage{afterpage}
\usepackage{txfonts}
\usepackage[T1]{fontenc}
\usepackage{multirow}
\usepackage{footnote}
\usepackage{graphicx}
\usepackage{epstopdf}
\usepackage{savesym}
\usepackage{hyperref}
\usepackage{longtable}
\usepackage{breakurl}
\usepackage{xcolor}
\usepackage{ulem}
\usepackage{soul}
\usepackage{float}
\usepackage{mathrsfs}
\usepackage{breqn} 
\usepackage{bm}
\usepackage{epstopdf}
\usepackage{rotating}

\def\HI{H~\textsc{i} }
\def\HII{H~\textsc{ii} }
\def\CII{[C~\textsc{ii}] }

\graphicspath{{./}{figures/}}

\begin{document}

\title{A High-resolution Study of the Cold Neutral Medium in and around 30 Doradus} 

\correspondingauthor{Min-Young Lee}
\email{mlee@kasi.re.kr}

\author[orcid=0000-0002-7374-7864,gname=Gyueun, sname='Park']{Gyueun Park} 
\affiliation{Korea Astronomy \& Space Science Institute, 776 Daedeok-daero, Yuseong-gu, Daejeon 34055, Republic of Korea}
\affiliation{Department of Astronomy and Space Science, University of Science and Technology, 217 Gajeong-ro, Yuseong-gu, 
             Daejeon 34113, Republic of Korea}
\email{gpark@kasi.re.kr}

\author[orcid=0000-0002-9888-0784]{Min-Young Lee}
\affiliation{Korea Astronomy \& Space Science Institute, 776 Daedeok-daero, Yuseong-gu, Daejeon 34055, Republic of Korea}
\affiliation{Department of Astronomy and Space Science, University of Science and Technology, 217 Gajeong-ro, Yuseong-gu, 
             Daejeon 34113, Republic of Korea}
\email{mlee@kasi.re.kr}

\author[orcid=0000-0002-6300-7459]{John M. Dickey}
\affiliation{School of Natural Sciences, Private Bag 37, University of Tasmania, Hobart, TAS, 7001, Australia}
\email{}

\author[orcid=0000-0001-9504-7386]{Nick M. Pingel}
\affiliation{Department of Astronomy, University of Wisconsin-Madison, 475 N. Charter St., Madison, WI 53703, USA}
\email{}

\author[orcid=0000-0002-4899-4169]{James Dempsey} 
\affiliation{Research School of Astronomy and Astrophysics, The Australian National University, Canberra, ACT 2611, Australia} 
\email{}

\author[orcid=0000-0002-9214-8613]{Helga D\'enes}
\affiliation{School of Physical Sciences and Nanotechnology, Yachay Tech University, Hacienda San Jos\'e S/N, 100119, Urcuqu\'{\i}, Ecuador}
\affiliation{{College of Sciences and Engineering, Universidad San Francisco de Quito, Quito, Ecuador}}
\email{}

\author[orcid=0000-0002-1495-760X]{Steven Gibson} 
\affiliation{Department of Physics and Astronomy, Western Kentucky University, Bowling Green, KY 42101, USA} 
\email{}

\author[orcid=0000-0001-7105-0994]{Katie Jameson} 
\affiliation{Caltech Owens Valley Radio Observatory, Pasadena, CA 91125, USA} 
\email{}

\author[orcid=0000-0002-6637-9987]{Ian Kemp}
\affiliation{International Centre for Radio Astronomy Research (ICRAR), Curtin University, Bentley, WA 6102, Australia}
\affiliation{{CSIRO Space and Astronomy, 26 Dick Perry Avenue, Kensington, 6151, WA, Australia}}
\email{}

\author[orcid=0000-0003-2896-3725]{Chang-Goo Kim}
\affiliation{Department of Astrophysical Sciences, Princeton University, 4 Ivy Lane, Princeton, NJ 08544, USA}
\email{}

\author[orcid=0000-0002-4814-958X]{Denis Leahy}
\affiliation{Department of Physics \& Astronomy, University of Calgary, Calgary, AB T2N 1N4, Canada}
\email{}

\author[orcid=0000-0002-3810-1806]{Bumhyun Lee}
\affiliation{Department of Astronomy, Yonsei University, 50 Yonsei-ro, Seodaemun-gu, Seoul 03722, Republic of Korea}
\email{}

\author[orcid=0000-0001-6846-5347]{Callum Lynn}
\affiliation{Research School of Astronomy and Astrophysics, The Australian National University, Canberra, ACT 2611, Australia}
\email{}

\author[orcid=0000-0003-0742-2006]{Yik Ki Ma}
\affiliation{Research School of Astronomy and Astrophysics, The Australian National University, Canberra, ACT 2611, Australia}
\affiliation{{Max-Planck-Institut f\"ur Radioastronomie, Auf dem H\"ugel 69, 53121 Bonn, Germany}}
\email{}

\author[orcid=0000-0002-5501-232X]{Antoine Marchal}
\affiliation{Research School of Astronomy and Astrophysics, The Australian National University, Canberra, ACT 2611, Australia} 
\email{}

\author[orcid=0000-0003-2730-957X]{Naomi M. McClure-Griffiths}
\affiliation{Research School of Astronomy and Astrophysics, The Australian National University, Canberra, ACT 2611, Australia}
\email{}

\author[orcid=0000-0001-5621-1577]{Eric Muller}
\affiliation{Research School of Astronomy and Astrophysics, The Australian National University, Canberra, ACT 2611, Australia}
\email{}

\author[orcid=0000-0002-2712-4156]{Hiep Nguyen}
\affiliation{Research School of Astronomy and Astrophysics, The Australian National University, Canberra, ACT 2611, Australia}
\email{}

\author[orcid=0000-0002-3418-7817]{Sne\v zana Stanimirovi\'c}
\affiliation{Department of Astronomy, University of Wisconsin-Madison, 475 N. Charter St., Madison, WI 53703, USA} 
\email{}

\author[orcid=0000-0002-1272-3017]{Jacco Th. Van Loon}
\affiliation{Lennard-Jones Laboratories, Keele University, ST5 5BG, UK}
\email{}

\collaboration{all}{The GASKAP-\HI collaboration}

\begin{abstract}
	\noindent
	With the aim of evaluating the roles of the cold neutral medium (CNM) in the cloud-scale baryon cycle, 
	we perform a high-resolution study of the CNM in and around the extreme star-forming region 30 Doradus (30 Dor). 
	For our study, we use Galactic Australian Square Kilometre Array Pathfinder \HI Survey data 
	and produce \HI emission and absorption cubes on 7~pc scales. 
	To examine the CNM structures {toward} 30 Dor, we decompose the \HI absorption cube into 862 Gaussian components 
	and find that these components are distributed at four velocity ranges (B1, B2, B3, and B4, respectively):  
	200--230~km~s$^{-1}$, 230--260~km~s$^{-1}$, 260--277~km~s$^{-1}$, and 277--300~km~s$^{-1}$. 
	We derive line-of-sight average spin temperatures and opacity-corrected total \HI column densities 
	and show that the B1--B4 structures have systematically different properties, indicating that they are physically distinct. 
	As for the nature of the observed CNM structures, we find that {B2 is associated with the main dense structure  
	where ionized, atomic, and molecular gases are concentrated. 
	B3 and B4 trace inflows whose combined mass flux rate of 0.14~$M_{\odot}$~yr$^{-1}$ is comparable to the current star formation rate, 
	while B1 probes outflows with a much lower mass flux rate of 0.007~$M_{\odot}$~yr$^{-1}$. 
	Interestingly, the \HI column densities in B1--B4 are nearly uniform with a factor of two spatial variations,} 
	implying the presence of \HI shielding layers for H$_{2}$ formation. 
\end{abstract}


\section{Introduction} \label{s:intro}

Molecular clouds (MCs) are the coldest and densest component of the interstellar medium (ISM) and the primary stellar nurseries \citep[e.g.,][]{kennicutt12}.
Recent theoretical and observational studies have demonstrated that a range of processes on the scales of individual MCs control the evolution of galaxies
\citep[e.g.,][]{krumholz05, kruijssen14, chevance2020, kim2022}. 
These processes encompass the accretion of raw materials, the transition from atomic (H~\textsc{i}) to molecular hydrogen (H$_{2}$), 
star formation, and the destruction of MCs via stellar feedback, forming the complex cloud-scale baryon cycle. 

\HI is a fundamental element in the cloud-scale baryon cycle and consists of the cold and warm neutral medium (CNM and WNM) 
with typical densities and kinetic temperatures of 
($n$, $T_{\textrm{k}}$)~$\sim$~(7--70~cm$^{-3}$, 60--260~K) and (0.2--0.9~cm$^{-3}$, 5000--8300~K) for solar neighborhood conditions 
\citep[e.g.,][]{mckee1977, wolfire2003, bialy2019}. 
The segregation into the CNM and WNM is not very strict, as observations have shown that 
20--30\% of \HI exist in an intermediate regime of the thermally unstable medium 
\citep[UNM; e.g.,][]{heiles2003, murray2018}. 
Among the three phases of H~\textsc{i}, the CNM is expected to play an important role in the H~\textsc{i}-to-H$_{2}$ transition and star formation, 
thanks to its high density and low temperature. 
This expectation has been corroborated by recent observations that found molecular gas preferentially forms 
in environments with low spin temperatures $T_{\textrm{s}}$~$\lesssim$~100~K and 
high CNM-to-total \HI column density ratios $f$(CNM)~$\sim$~0.6 
\citep{rybarczyk2022, park2023}. 

In spite of its importance in the cloud-scale baryon cycle, the CNM in and around individual MCs remains largely unexplored. 
This is because the CNM has been mainly studied via \HI absorption measurements toward background continuum sources 
that are distributed over a large area of sky.
To address the limitation of previous observations and shed light on the roles of the CNM in the cloud-scale baryon cycle, 
we focused on 30 Doradus (30 Dor hereafter) in the Large Magellanic Cloud (LMC), 
whose extensive continuum emission enabled us to spatially map \HI absorption across the surface of the giant \HII region.

Our target 30 Dor is the most extreme star-forming region in the Local Group of galaxies. 
It hosts more than 1000 OB-type and Wolf-Rayet stars \citep[e.g.,][]{doran2013} 
and produces $\sim$500 times more ionizing photons than the Orion Nebula \citep[e.g.,][]{pellegrini2010}. 
At the center of 30 Dor lies the super star cluster R136 with an exceedingly high stellar mass density of $\gtrsim 10^{7}$~$M_{\odot}$~pc$^{-3}$ \citep{selman2013}. 
This extraordinary star cluster harbors the most massive stars known ($\gtrsim$~150~$M_{\odot}$; \citealt{crowther2010})
and constitutes almost half of the stellar mass in 30 Dor \citep{doran2013}.
Strong stellar winds and supernova explosions (SNe) created vast cavities and bubbles 
that fill 30 Dor with bright H$\alpha$ and X-ray emission \citep[e.g.,][]{chu1994, townsley2006a}. 
Overall 30 Dor has been well examined at multi-wavelengths for star formation activities, as well as for gas and dust contents
\citep[e.g.,][]{townsley2006a, pellegrini2010, doran2013, indebetouw2013, roman-duval2014, lee2019, wong2022, nayak2023}, 
thanks to its proximity to the Milky Way (50~kpc; \citealt{pietrzynski2013}) 
and unique half-solar metallicity condition \citep{russell1992}. 
Our study fills a critical gap in the previous studies at 1.4~GHz, 
providing the first comprehensive view on the properties of the CNM in and around 30 Dor. 

\begin{figure}
    \centering
    \includegraphics[scale=0.3]{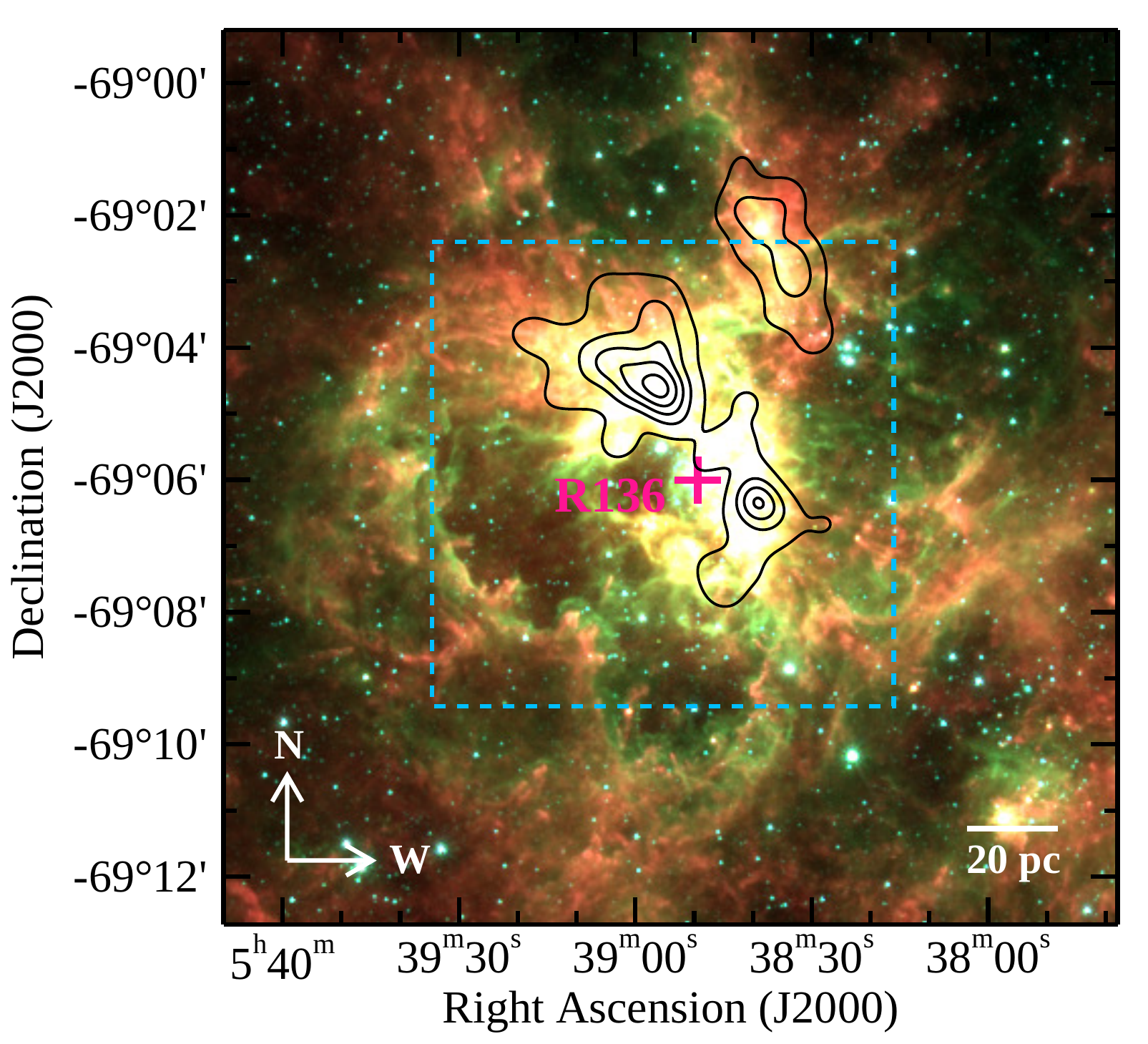}
	\caption{\label{f:rgb_image} Three-color composite image of 30 Dor 
	         (\textit{Spitzer} 8, 4.5, and 3.6~$\micron$ emission in red, green, and blue, respectively; \citealt{meixner2006}). 
		 The CO integrated intensity image from \citet{wong2022} is overlaid as the black contours 
		 with levels ranging from 10\% to 90\% of the maximum value of 44.1~K~km~s$^{-1}$ with increments of 20\%. 
		 The blue box outlines the coverage of the GASKAP-\HI data, while the red cross marks the location of the central star cluster R136.}
\end{figure}

The Galactic Australian Square Kilometre Array Pathfinder \HI Survey 
\citep[GASKAP-H~\textsc{i};][]{dickey2013} is the key to the success of our study. 
This survey aims to reveal the structure, kinematics, and thermodynamics of \HI in the Milky Way, Magellanic Clouds, and their surroundings  
with an unprecedented combination of sensitivity and resolution using the Australian Square Kilometre Array Pathfinder (ASKAP) telescope. 
The most innovative aspect of the ASKAP telescope would be the phased array feed (PAF) 
that is mounted on the focal planes of 36 12m telescopes \citep{hotan2021}. 
Up to 36 dual-polarization beams can be formed for each PAF-mounted telescope, 
resulting in a high survey speed with a wide field-of-view (FoV) of $\sim$30~deg$^{2}$ at 1.4~GHz.
The baselines range from 22~m to 6~km (630 in total), but most of them fall in two groups: 
one with lengths of 400--1200~m and the other with lengths of 2--3~km. 
The first peak of the baseline distribution gives an excellent brightness sensitivity for detecting weak \HI emission at beam sizes of 30--60$''$, 
while the second peak enables us to obtain sensitive \HI absorption spectra on 10$''$ scales.
The capabilities of the GASKAP-H~\textsc{i} survey were confirmed via Phase I and II pilot surveys \citep{mcclure-griffiths2018, pingel2022}, 
and the full survey operation began in late 2024. 
Our study is based on the Phase II pilot observations of the LMC, 
whose coverage for 30 Dor is outlined in Figure~\ref{f:rgb_image}. 

This paper is organized as follows. 
In Section \ref{s:data}, we describe the GASKAP-H~\textsc{i} observations and data reduction processes 
and present [C~\textsc{ii}] 158 $\mu$m and $^{12}$CO(2--1) (CO hereafter) data. 
In Sections \ref{s:results} and \ref{s:analyses}, we examine the structure of the CNM in and around 30 Dor based on Gaussian decomposition of \HI absorption spectra 
and derive the physical properties of the CNM such as line-of-sight (LOS) average spin temperatures and opacity-corrected total \HI column densities. 
In addition, we probe the multi-phase structures of 30 Dor based on a comparison between the CNM, [C~\textsc{ii}], and CO spatial distributions 
and investigate how different the CNM in and around 30 Dor is compared to that in the Milky Way and the less star-forming part of the LMC. 
Finally, we discuss and summarize our results in Sections \ref{s:discussion} and \ref{s:summary}.

\section{Data} \label{s:data}

\subsection{\HI} \label{s:hi}

\subsubsection{GASKAP-\HI Observations} \label{s:hi_obs}

The GASKAP-\HI observations of the LMC were carried out as part of the Phase II pilot survey 
and were divided into five zones with Scheduling Blocks (SBs)\footnote{The data and associated diagnostic reports 
are accessible from the CSIRO ASKAP Science Data Archive (CASDA; \url{https://data.csiro.au/collections/domain/casdaObservation/search/}).} 
of 33047, 38373, 38791, 38814, and 38845. 
Out of the total 108 observations for the SB33047 field (36 beams $\times$ 3 interleaves), 
five measurement sets were chosen for imaging, 
whose specifications are listed in Table~\ref{t:MS_info}.
We refer the readers to Section~2 of \citet{pingel2022} for details on the GASKAP-\HI observations.
 
The raw data were calibrated using ASKAPsoft \citep{guzman2019}. 
This custom pipeline calibrates the bandpass, flags bad data, and performs self-calibration (amplitude and phase). 
We refer the readers to Section~11 of \citet{hotan2021} and Section~3.1 of \citet{pingel2022} for details on the calibration process.
In the subsequent sections, we describe how we processed the calibrated data 
to produce \HI absorption (``high-resolution'') and emission (``low-resolution'') cubes, as well as continuum maps.

\startlongtable
\begin{deluxetable*}{c c c c c c c c c c}
    \centering
    \tablecaption{\label{t:MS_info} Specifications of the GASKAP-\HI Observations (SB33047)}
    \tablewidth{0pt}
    \setlength{\tabcolsep}{8pt}
    \tabletypesize{\small}
    \tablehead{
    \colhead{Interleave} & \colhead{Beam} & \colhead{$t_{\textrm{on}}$} & \colhead{$\nu_{0}$} & \colhead{$\Delta \nu$} & \colhead{BW} & \colhead{$\alpha_{\textrm{J2000}}$} & \colhead{$\delta_{\textrm{J2000}}$} & \colhead{Obs. Start} & \colhead{Obs. End} \\
    \colhead{ } & \colhead{ } & \colhead{(hrs)} & \colhead{(MHz)} & \colhead{(kHz)} & \colhead{(kHz)} & \colhead{(hh:mm:ss)} & \colhead{(dd:mm:ss)} & \colhead{(UTC; hrs)} & \colhead{(UTC; hrs)} \\
    \colhead{(1)} & \colhead{(2)} & \colhead{(3)} & \colhead{(4)} & \colhead{(5)} & \colhead{(6)} & \colhead{(7)} & \colhead{(8)} & \colhead{(9)} & \colhead{(10)}} 
    \startdata
    A & 21 & 3.54 & 1419.7228 & 1.157 & 2444.4 & 05:40:46.8 & $-$69:28:08 & 2021.10.27, 16.28 & 2021.10.28, 2.61 \\
    A & 27 & 3.54 & 1419.7228 & 1.157 & 2444.4 & 05:35:42.3 & $-$68:41:29 & 2021.10.27, 16.28 & 2021.10.28, 2.61 \\
    B & 20 & 3.57 & 1419.7228 & 1.157 & 2444.4 & 05:35:40.0 & $-$69:12:40 & 2021.10.27, 16.54 & 2021.10.28, 2.86 \\
    C & 20 & 3.61 & 1419.7228 & 1.157 & 2444.4 & 05:35:37.7 & $-$69:43:50 & 2021.10.27, 16.79 & 2021.10.28, 3.12 \\
    C & 27 & 3.61 & 1419.7228 & 1.157 & 2444.4 & 05:40:41.8 & $-$68:56:58 & 2021.10.27, 16.79 & 2021.10.28, 3.12 \\
    \enddata
    \tablecomments{(1) Interleave type; (2) Beam number; (3) On-source integration time; 
                             (4) Central frequency; (5) Spectral resolution (corresponding to a velocity resolution of 0.24~km~s$^{-1}$); 
                             (6) Total bandwidth; (7, 8) Pointing center in the equatorial coordinate system (right ascension and declination in J2000); 
                             (9, 10) Observation date and time range.}
\end{deluxetable*}

\subsubsection{High-Resolution Imaging} \label{s:imaging}

\begin{deluxetable}{l c c}[t]
\centering
\tablecaption{\label{t:mapping_parms} Imaging parameters}
    \tablewidth{0pt}
    \setlength{\tabcolsep}{8pt}
    \tabletypesize{\small}
    \tablehead{
    \colhead{Parameter} & \multicolumn{2}{c}{Resolution} \\
    \colhead{} & \colhead{High (13$''$)} & \colhead{Low (30$''$)}
    }
    \startdata
    (1) imsize (pixels) & 3600 $\times$ 3600 & 1024 $\times$ 1024 \\
    (2) cell ($''$) & 1 & 7 \\
    (3) reffreq (GHz) & 1.42040571183 & 1.42040561183 \\
    (4) width (km~s$^{-1}$) & 0.977192 & 0.244212 \\
    (5) restfreq (GHz) & 1.420405752000 & 1.420405752000 \\
    (6) gridder & mosaic &  IDG{\tablenotemark{\footnotesize{a}}} \\
    (7) deconvolver & multiscale & multiscale \\
    (8) scales & 0, 12, 24, 36 & auto{\tablenotemark{\footnotesize{b}}} \\
    (9) weighting & Briggs & Briggs \\
    (10) robust & 0.5 & 0 \\
    (11) niter & 1E+08 & 1E+04 \\
    (12) threshold (mJy) & 6.3 & 15 \\
    (13) uvrange & all baselines & all baselines \\
    \enddata
    \tablecomments{(1) Image size; (2) Pixel size; (3) Reference frequency; (4) Channel width; (5) Rest frequency; (6) Gridding option; 
                             (7) Minor cycle algorithm; (8) Scale sizes for the multiscale algorithm; (9) Weighting scheme; 
                             (10) Robustness parameter for the Briggs weighting; (11) Maximum number of iterations; (12) Stopping threshold; 
                             (13) \textit{uv} range for data selection.}
    \tablenotetext{a}{\small Image Domain Gridder; \citet{van2018}.}
    \tablenotetext{b}{\small In this mode, WSClean automatically sets the scales and uses as many as necessary. 
                                The first two scales include the delta scale $(0)$ and the synthesized beam in terms of pixels. 
                                Subsequent scales are added by continually multiplying the synthesized beam scale by a factor of two until the scale size becomes larger than the image size.}
\end{deluxetable}

We employed the \textit{tclean} task in Common Astronomy Software Application (CASA; \citealt{CASA2022}) 
to image a $1{\degree} \times 1{\degree}$ cube centered on 
$(\alpha, \delta)_{\textrm{J2000}}$ = (05$^{\textrm{h}}$38$^{\textrm{m}}$36.6$^{\textrm{s}}$, $-$69$\degree$05$'$56$''$). 
This cube is close to the size of the ASKAP primary beam at 1.4~GHz and has a velocity resolution of $\sim$1~km~s$^{-1}$
(four times the native spectral resolution)\footnote{In this paper, all velocities are quoted in the kinematic local standard of rest (LSRK) frame.}. 
For deconvolution, we used the multiscale CLEAN algorithm \citep{cornwell2008}. 
Specifically, we generated the cube with a pixel size of 1$''$, scale sizes in increments of 12 from 0 to 36 
(0$''$--36$''$ for the diameter of the clean component; 0--2.8 times the synthesized beam), and a maximum iteration count of 1E+08.
In addition, we adopted a stopping threshold of 6.3~mJy based on three times the median standard deviation of the off-line channels 
(150--200~km~s$^{-1}$ and 320--380~km~s$^{-1}$) in the dirty image.  
As the gridding option, we selected mosaic since our visibilities include many different pointing centers based on the PAFs \citep{hotan2021}.
Finally, we weighted the gridded data using the Briggs scheme with a robustness of 0.5 
and produced the cube with a synthesized beam of 13.3$''$~$\times$~12.3$''$.
The adopted \textit{tclean} parameters are summarized in Table~\ref{t:mapping_parms}.

\begin{figure}
    \centering
    \includegraphics[scale=0.25]{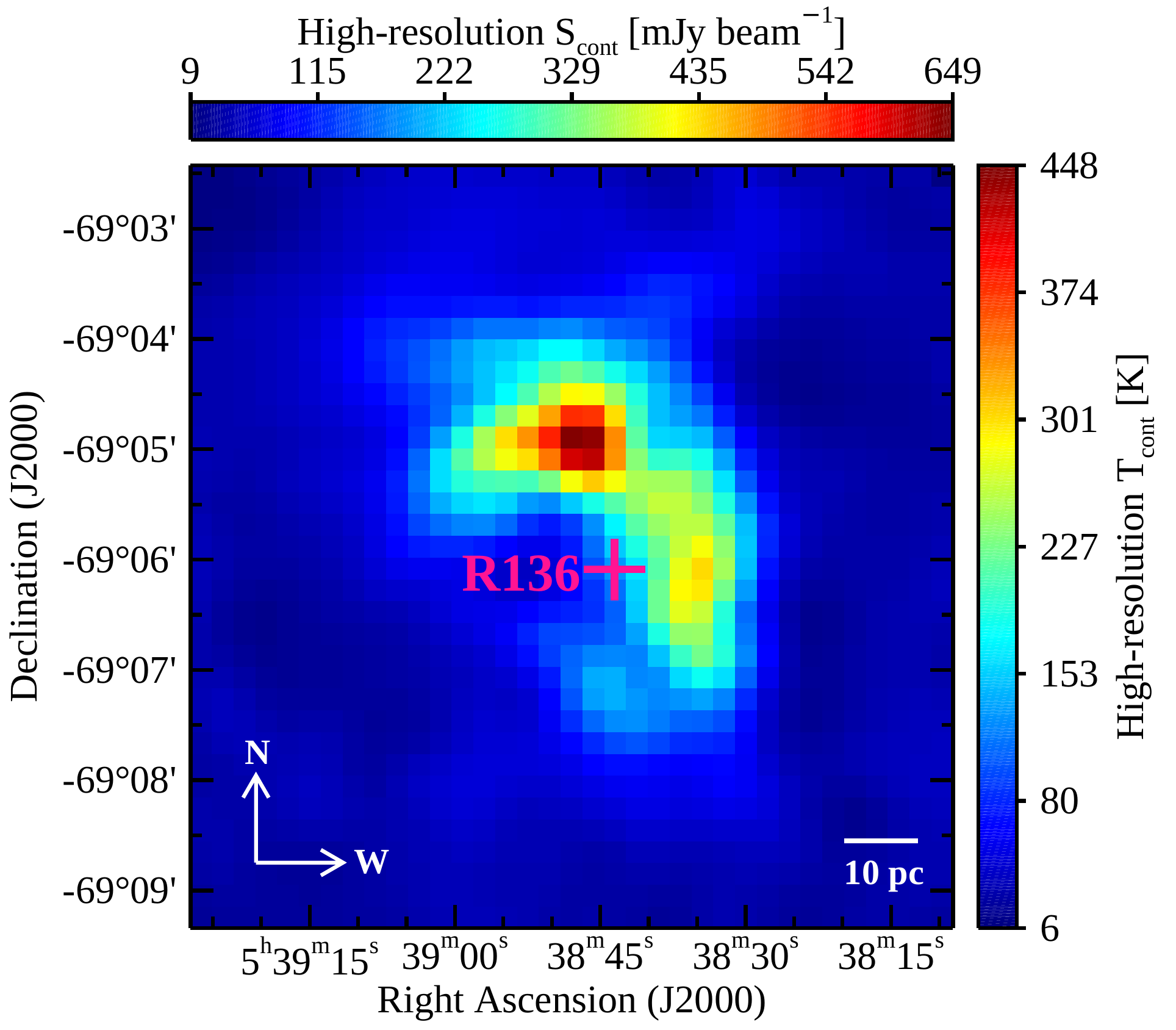} 
    \caption{\label{f:cont_pair} {1.4 GHz continuum image at a resolution of 30$''$ with a pixel size of 12$''$. 
	     The location of R136 is indicated as the red cross.}}
\end{figure}

To produce a high-resolution cube from which absorption spectra are extracted, 
one would need to resolve away the \HI emission, while keeping as much of the continuum as possible. 
These are two conflicting goals, since excluding short baselines to resolve away the \HI emission results in a decrease in the continuum, 
which in turn increases the noise in the optical depth ($\bm{\tau}$). 
To evaluate the impact of baseline selection, we produced three cubes with different short-spacing cut-offs: 
1.055~km (5~k$\lambda$), 316~m (1.5~k$\lambda$), 0 (all baselines).
As the beam size changes from 8.8$''$ to 12.1$''$ to 13.3$''$,
the peak continuum brightness temperature increases from 342~K to 429~K to 623~K, 
while the \HI emission begins to leak into the cubes.
Based on these results, we decided to include all baselines to extract as many absorption spectra as possible
and corrected for the emission leakage later based on the average spectrum of the pixels surrounding 30 Dor (Section~\ref{s:leakage}). 

For our subsequent analyses, we smoothed the processed cube to 30$''$ scales to match the low-resolution cube (Section~\ref{s:ems}), 
rebinned it with a pixel size of 12$''$ for Nyquist sampling, 
and address the final product as the ``high-resolution cube'' throughout this paper.
Since we did not subtract the continuum in the \textit{uv}-domain,
the high-resolution cube contains \HI emission and absorption along with the extended continuum.  
To produce the continuum image, we estimated the continuum level $T_{\textrm{cont}}$ from the off-line channels of 150--200~km~s$^{-1}$ and 320--380~km~s$^{-1}$
and cut out a 7$'$~$\times$~7$'$ region centered on 
$(\alpha, \delta)_{\textrm{J2000}}$ = (05$^{\textrm{h}}$38$^{\textrm{m}}$47.9$^{\textrm{s}}$, $-$69$\degree$05$'$56$''$) 
that includes all pixels where the continuum is bright enough to measure accurate \HI absorption 
($>$~83~K or 120~mJy~beam$^{-1}$; Figure~\ref{f:cont_pair}). 
No primary beam correction was applied, 
as the 7$'$~$\times$~7$'$ region is sufficiently small to have a correction value of unity. 


\subsubsection{Low-Resolution Imaging} \label{s:ems}

An emission cube must be sensitive to structures on all angular scales, 
since a knowledge of the \HI emission that arises from the absorbing gas is necessary to compute the spin temperature $T_{\textrm{s}}$. 
However, like all radio interferometers, ASKAP suffers from missing short-spacings that inherently filter out structures on angular scales 
larger than the maximum recoverable scale ($\sim$32$'$), including the total power at $u=v=0$. 

We created an emission cube based on the imaging pipeline in \citet{pingel2022} that utilizes WSClean, 
the command-line imaging package for radio data \citep{offringa2014}. 
To efficiently process the ASKAP data, we imaged each spectral plane in computing jobs distributed across 
the resources available through the Center for High Throughput Computing at the University of Wisconsin-Madison
and produced a combined cube by stitching together the deconvolved spectral planes. 
We used WSClean to ensure that the large-scale structures spanning across multiple ASKAP beams are accurately reconstructed through a joint-deconvolution process.
In addition to the imaging parameters summrized in Table~\ref{t:mapping_parms}, 
we configured WSClean with \textit{--mgain~0.7} to ensure that 70\% of the residual flux is cleaned per major cycle 
and \textit{--taper--gaussian~14arcsec} to convolve the cleaned image to a final resolution of 30$''$. 
This resolution provides the best compromise between the surface brightness sensitivity to large-scale diffuse structures and the final root-mean-square (rms) noise. 
After deconvolution, we added the missing short-spacings by feathering the single-dish data from the Galactic All Sky Survey (GASS; \citealt{mcclure-griffiths2009, kalberla2015}) 
using the \textit{feather} task in CASA. 
For this purpose, we set the \textit{sdfactor} parameter to 1.0, 
dictating that the flux scale of the single-dish data is calibrated so that 
the only difference in the flux between the Parkes and ASKAP data arises from their different beam areas. 
We refer the readers to Section 3.2 of \citet{pingel2022} for details on the feathering process 
and note that the continuum was not subtracted from the resultant low-resolution cube.

\subsubsection{Extended Continuum Emission} \label{s:extended_continuum}

The low-resolution cube was produced from the GASKAP and Parkes \HI observations and hence contains all spatial frequencies. 
In contrast, the low-resolution continuum image is missing large-scale flux 
from the diffuse halo around 30 Dor with a diameter of $\sim$8$'$--20$'$, 
since the single-dish continuum data were not included. 
In our continuum image, the diffuse halo is resolved away at radii greater than $\sim$3.5$'$, 
probably due to a clean bowl around the bright core of 30 Dor.
This suggests that the continuum brightness of 30 Dor is underestimated even in our low-resolution image 
and should be supplemented by an offset of $\sim$36~mJy~beam$^{-1}$ to match the 1.4~GHz flux density of 31.3~$\pm$~1.6~Jy 
measured by \citet{haynes1986} with the Parkes telescope.
The offset corresponds to an extra continuum brightness of 25~K for our 30$''$ beam (small compared to the peak of 452~$\pm$~5 K) 
and was added to the low-resolution cube for subsequent analyses.

\subsubsection{Leakage Spectrum in the High-Resolution Cube} \label{s:leakage}

\begin{figure*}
    \centering
    \includegraphics[scale=0.25]{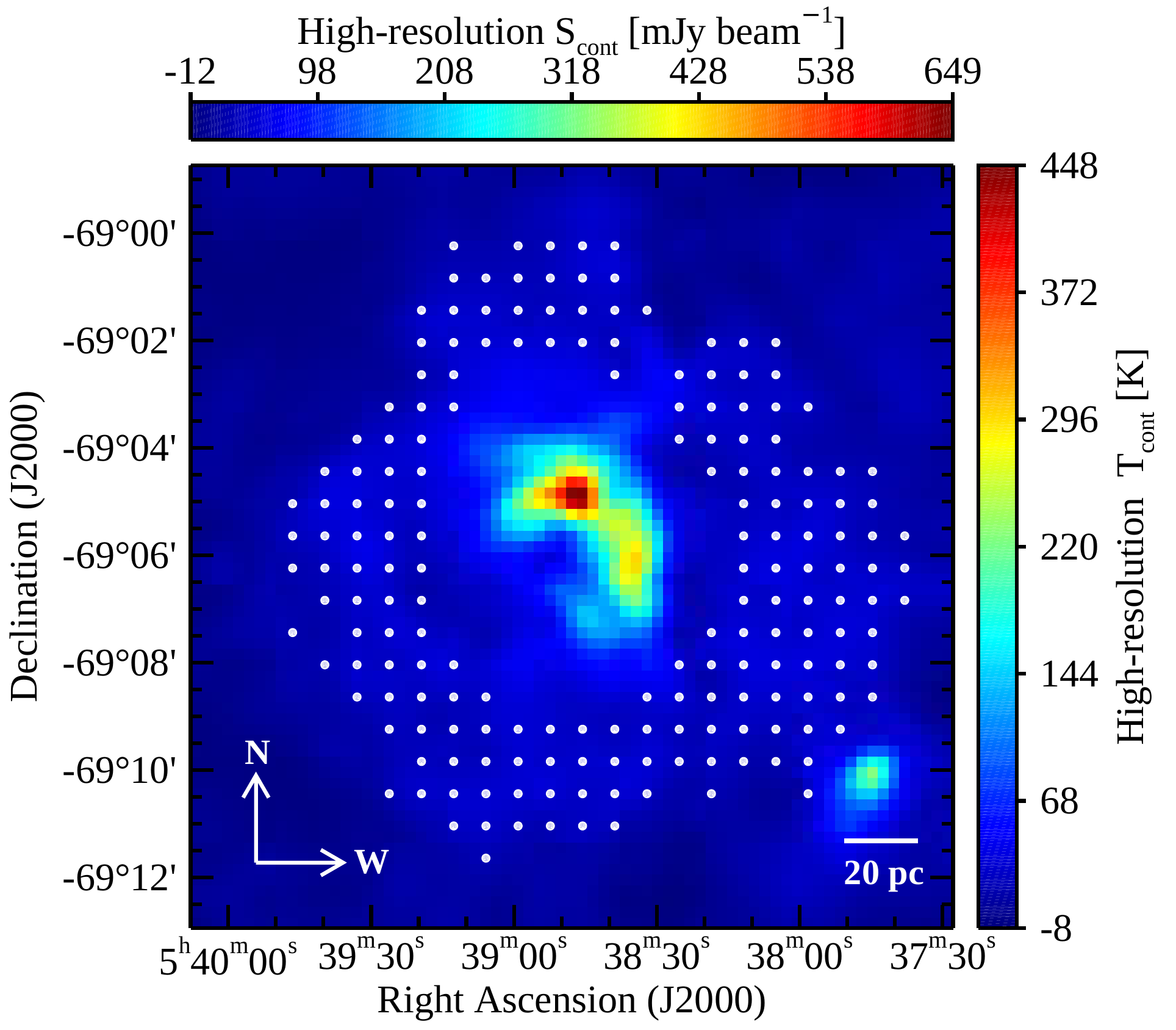} \hspace{0.5cm}
    \includegraphics[scale=0.25]{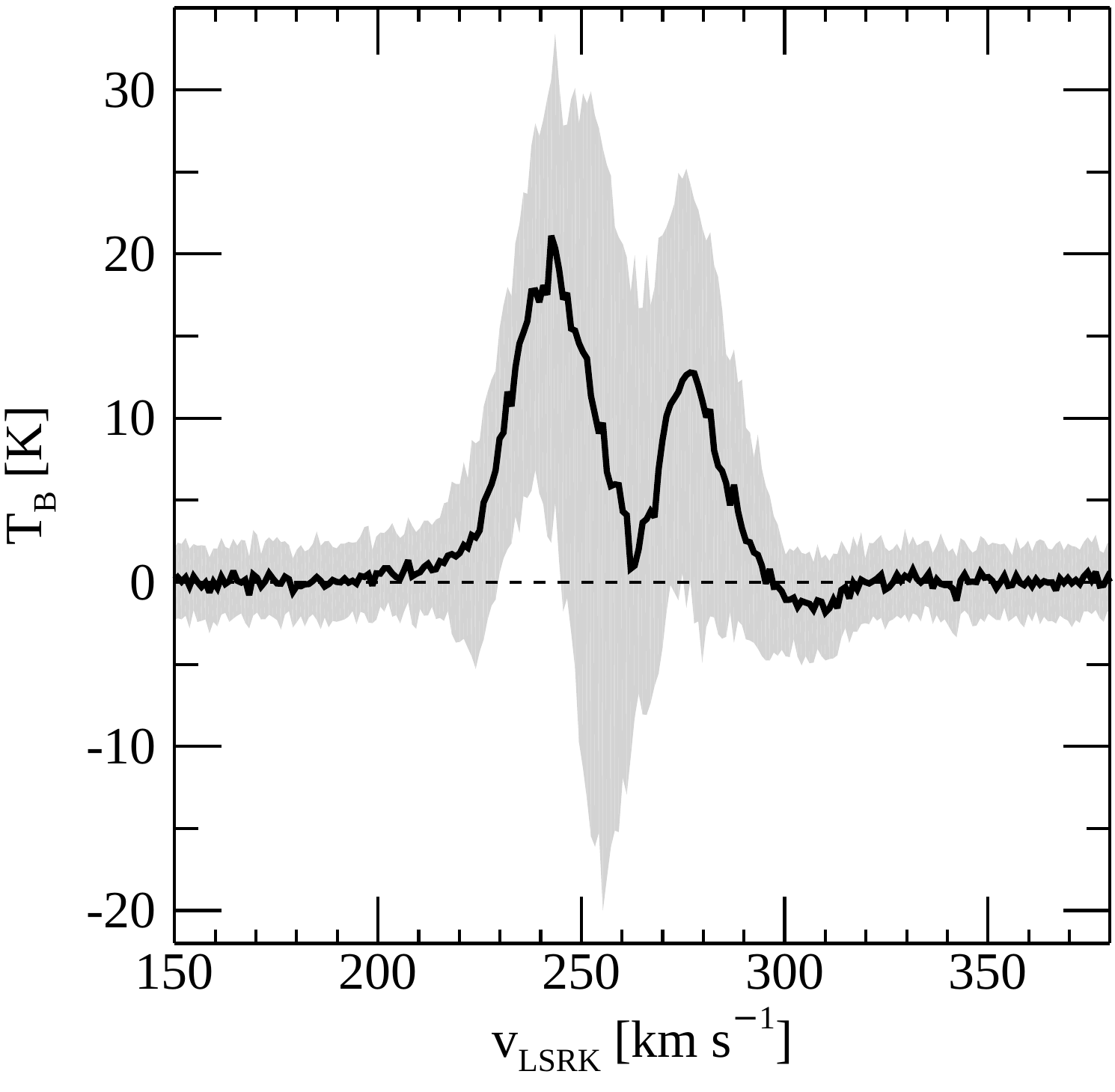}
    \caption{\label{f:off_source1} Derivation of the leakage spectrum.  
             (Left) 184 off-source points that were used to derive the leakage spectrum in the high-resolution cube are shown as the white dots. 
	     {The colorscale continuum image is the same as the one in Figure \ref{f:cont_pair}, but with a slightly larger coverage.} 
             (Right) Median of the spectra at the off-source points is shown in black, 
              while the 1$\sigma$ error computed from the distribution of the brightness temperatures in each velocity channel is indicated as the gray envelope. 
              This error dominates the precision of our absorption and emission-only spectra.}
\end{figure*}

The high-resolution cube filters out most of the \HI emission, but not all (Section~\ref{s:imaging}). 
To estimate the \HI emission that remains in the direction of 30 Dor, 
we considered a set of 184 points with low continuum brightness (15~K to 40~K) in an annulus around the source center. 
Figure~\ref{f:off_source1} shows the locations of these off-source points,  
as well as the median, {16th, and 84th} percentiles of the distribution of the brightness temperatures 
in each velocity channel of the off-source spectra (residual continuum at each point subtracted from baseline fitting). 
Here the median off-source emission represents our estimate of the emission leakage into the high-resolution cube, 
while the 1$\sigma$ envelope indicates the noise introduced by emission fluctuations 
that dominates other sources of error such as the radiometer noise 
(e.g., see the level of baselines at velocities below 200~km~s$^{-1}$ and above 350~km~s$^{-1}$). 

\subsubsection{Absorption Spectra and Associated Uncertainties} \label{s:optical_depth_spec} 

We subtracted the leakage spectrum from all pixels in the high-resolution cube 
and divided the product by the continuum to calculate the absorption spectra $e^{-\tau(\varv)}$
toward the pixels with continuum brightness higher than 83~K.  
We did not consider the pixels with lower continuum brightness, 
since their absorption spectra are not sensitive enough to derive the emission-only spectra $T_{\textrm{em-only}}(\varv)$ (Section~\ref{s:em_only}). 
Toward the pixels above our threshold, we also estimated the 1$\sigma$ uncertainties in the measured $t = 1 - e^{-\tau(\varv)}$ as follows:  

\begin{equation} \label{eq:sigma_t} 
	\sigma_{\textrm{t}}(\varv) = \frac{1}{T_{\textrm{cont}}}\sqrt{\sigma_{\textrm{n}}^{2} + \sigma_{\textrm{l}}(\varv)^{2}}
\end{equation} 

\noindent where $T_{\textrm{cont}}$ is the continuum brightness temperature, 
$\sigma_{\textrm{n}}$ is the radiometer noise measured off-line, 
and $\sigma_{\textrm{l}}(\varv)$ is the error in the leakage spectrum.

\begin{figure*}
    \centering
    \includegraphics[scale=0.33]{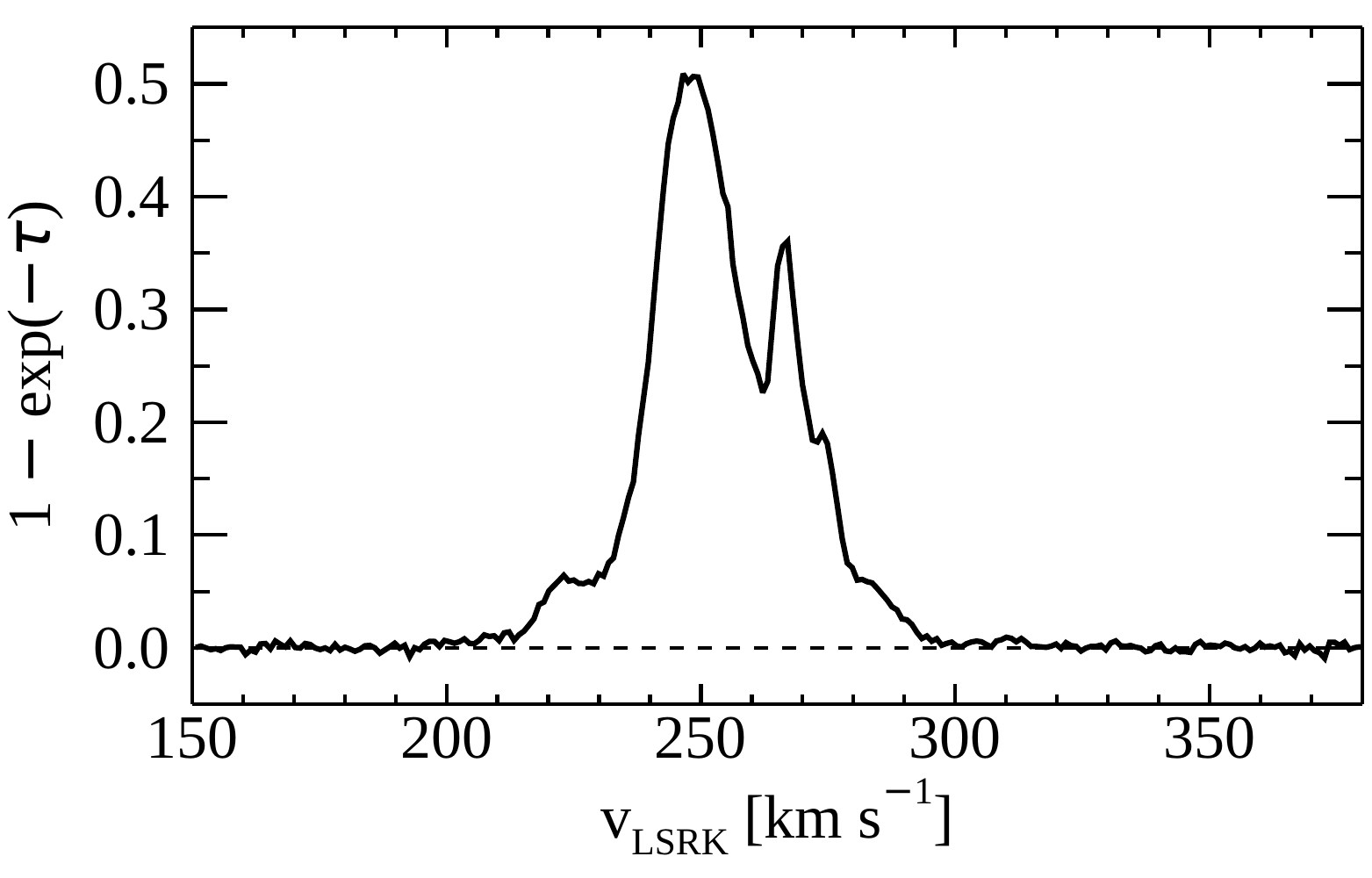} \hspace{0.2cm}
    \includegraphics[scale=0.33]{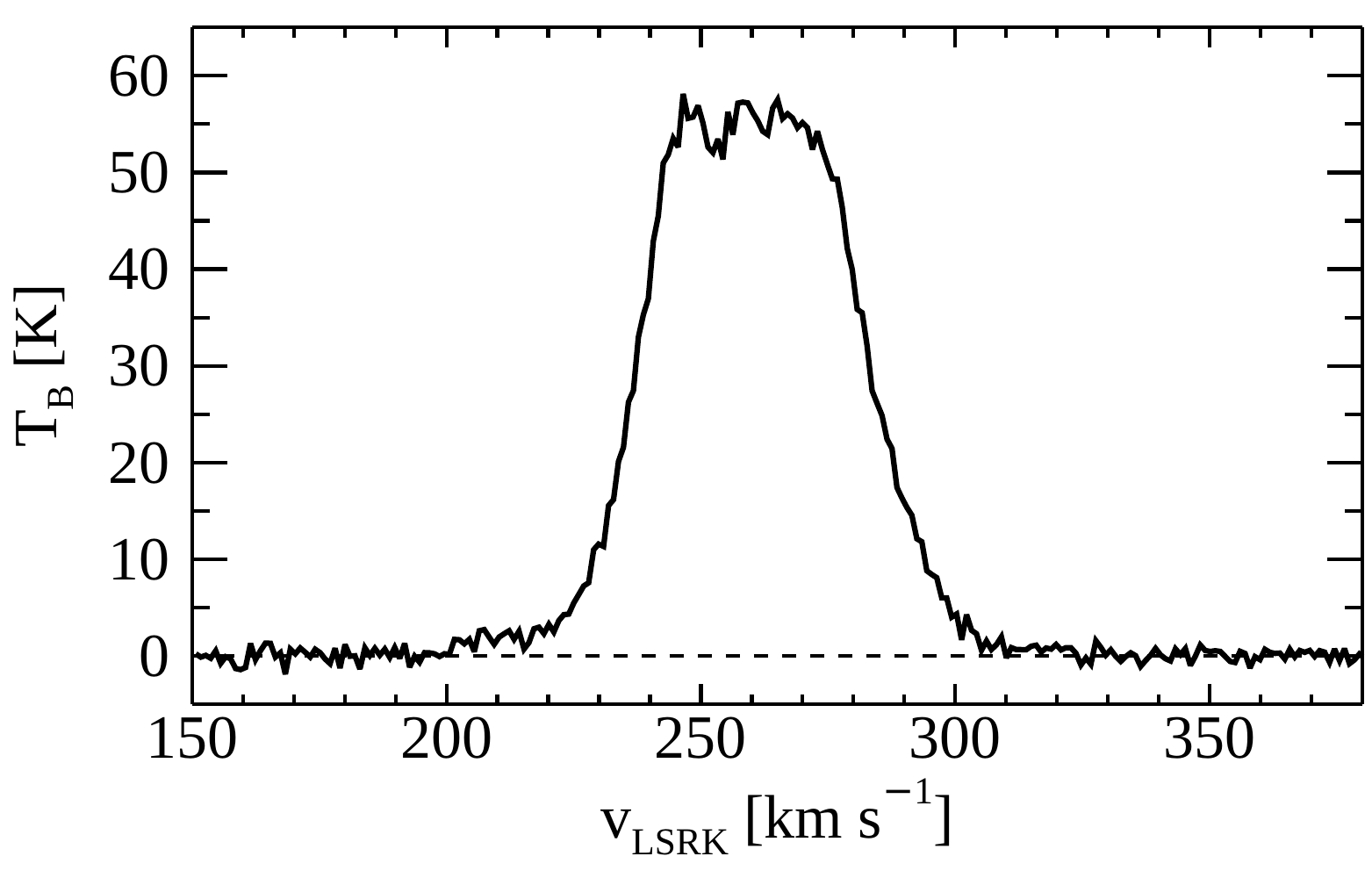}
    \caption{\label{f:avg_spec} (Left) 1/$\sigma_{\textrm{emt}}^{2}$-weighted mean $1 - e^{-\tau(\varv)}$ spectrum. 
             (Right) 1/$\sigma_{\textrm{em-only}}^{2}$-weighted mean emission-only spectrum.
	     To produce these two spectra, the same unmasked 225 pixels in the absorption and emission-only cubes were used.}
\end{figure*}

\begin{figure*}
	\centering
	\includegraphics[scale=0.25]{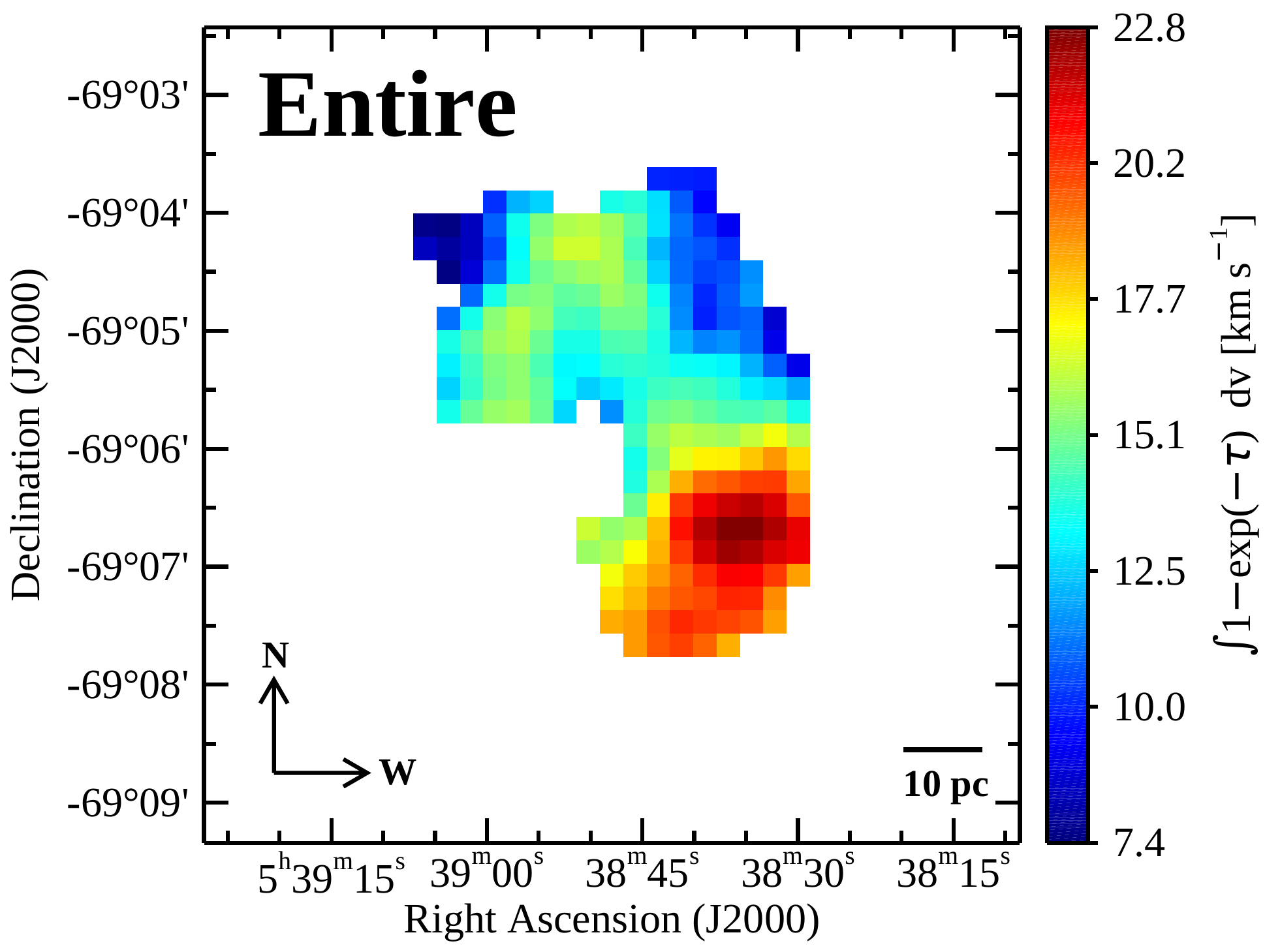} \hspace{0.3cm}
	\includegraphics[scale=0.25]{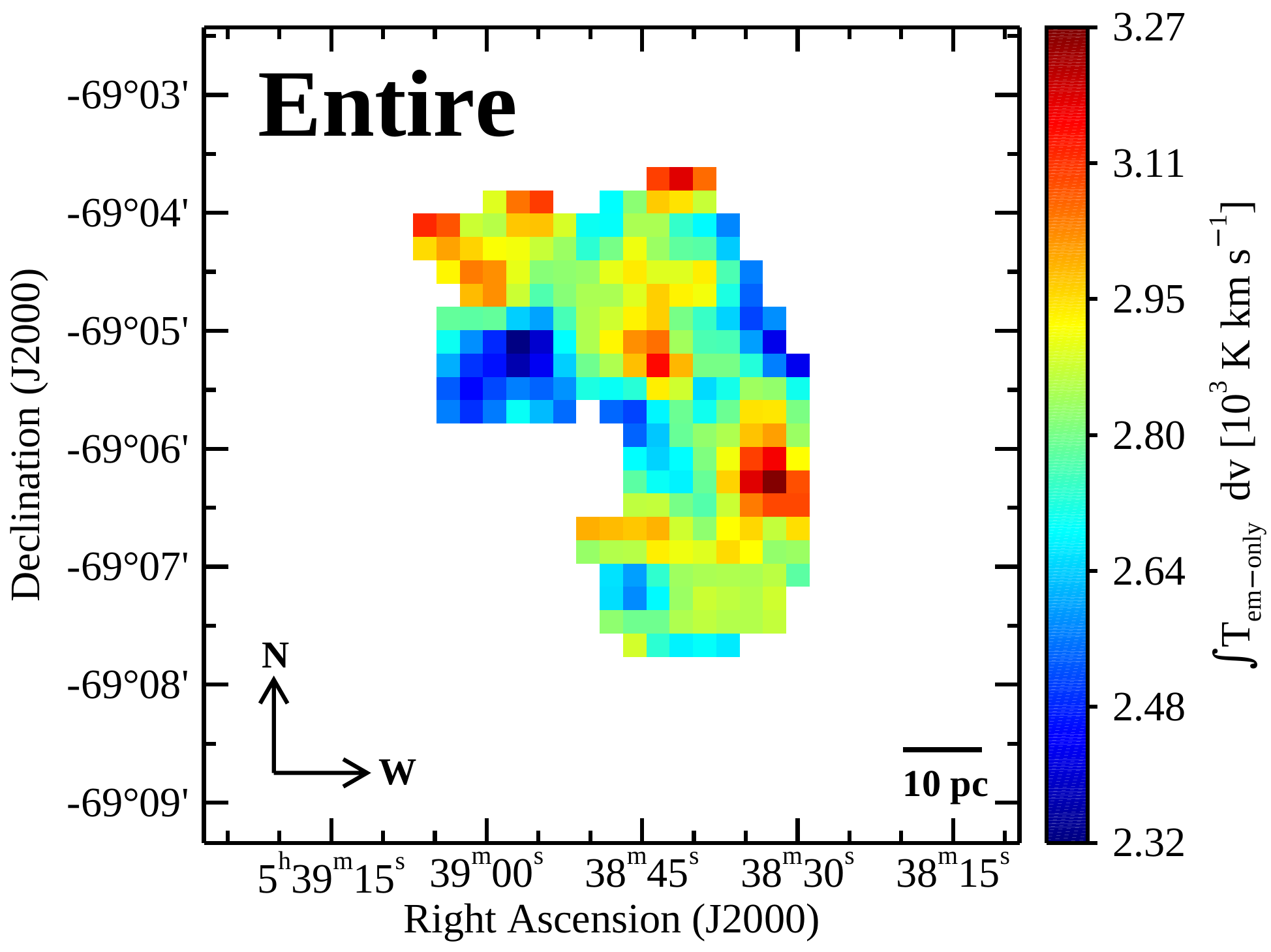} 
	\caption{\label{f:int_abs_em} Spatial distributions of the integrated $1 - e^{-\tau(\varv)}$ (left) and $T_{\rm{em-only}}$ (right). 
	The entire velocity range of 200--300~km~s$^{-1}$ over which \HI absorption and emission are clearly detected was used to produce the maps.}
\end{figure*}

Finally, we masked pixels that meet the following criteria and used 225 unmasked pixels for our subsequent analyses:   
(1) pixels with spurious negative features in $1 - e^{-\tau(\varv)}$ that likely result from emission fluctuations;   
(2) pixels where the maximum of $1 - e^{-\tau(\varv)}$ exceeds 0.982 ($\tau(\varv) > 4$); and 
(3) pixels with no detectable absorption (peak in $1 - e^{-\tau(\varv)}$~<~3$\sigma_{\textrm{emt}}$; 
$\sigma_{\textrm{emt}}$ = rms noise in $1 - e^{-\tau(\varv)}$). 
For the unmasked 225 pixels, the $1/\sigma_{\textrm{emt}}^{2}$-weighted mean $1 - e^{-\tau(\varv)}$ spectrum
and the spatial distribution of the integrated $1 - e^{-\tau(\varv)}$ are shown in Figures~\ref{f:avg_spec} and \ref{f:int_abs_em}. 

\subsubsection{Emission-only Spectra and Associated Uncertainties} \label{s:em_only} 

The \HI emission spectrum measured against a background continuum source with $T_{\textrm{cont}}$ can be written as 

\begin{equation} \label{eq:em_abs}
	T_{\textrm{B}}(\varv) = T_{\textrm{s}}(\varv) (1 - e^{-\tau(\varv)}) + T_{\textrm{cont}} e^{-\tau(\varv)}.
\end{equation}

\noindent If we assume that the optical depth in each pixel of the high-resolution cube 
is responsible for absorbing the continuum in the corresponding pixel in the low-resolution cube, 
we can separate the emission and absorption and derive an emission-only cube on 30$''$ scales as follows:

\begin{equation} \label{eq:em_only}
	T_{\textrm{em-only}}(\varv) = T_{\textrm{B}}(\varv) - T_{\textrm{cont}} e^{-\tau(\varv)} 
\end{equation}

\noindent where $T_{\textrm{cont}}$ and $e^{-\tau(\varv)}$ are measured in the low-resolution and high-resolution cubes on 30$''$ scales, respectively. 

\begin{figure}
	\includegraphics[scale=0.3]{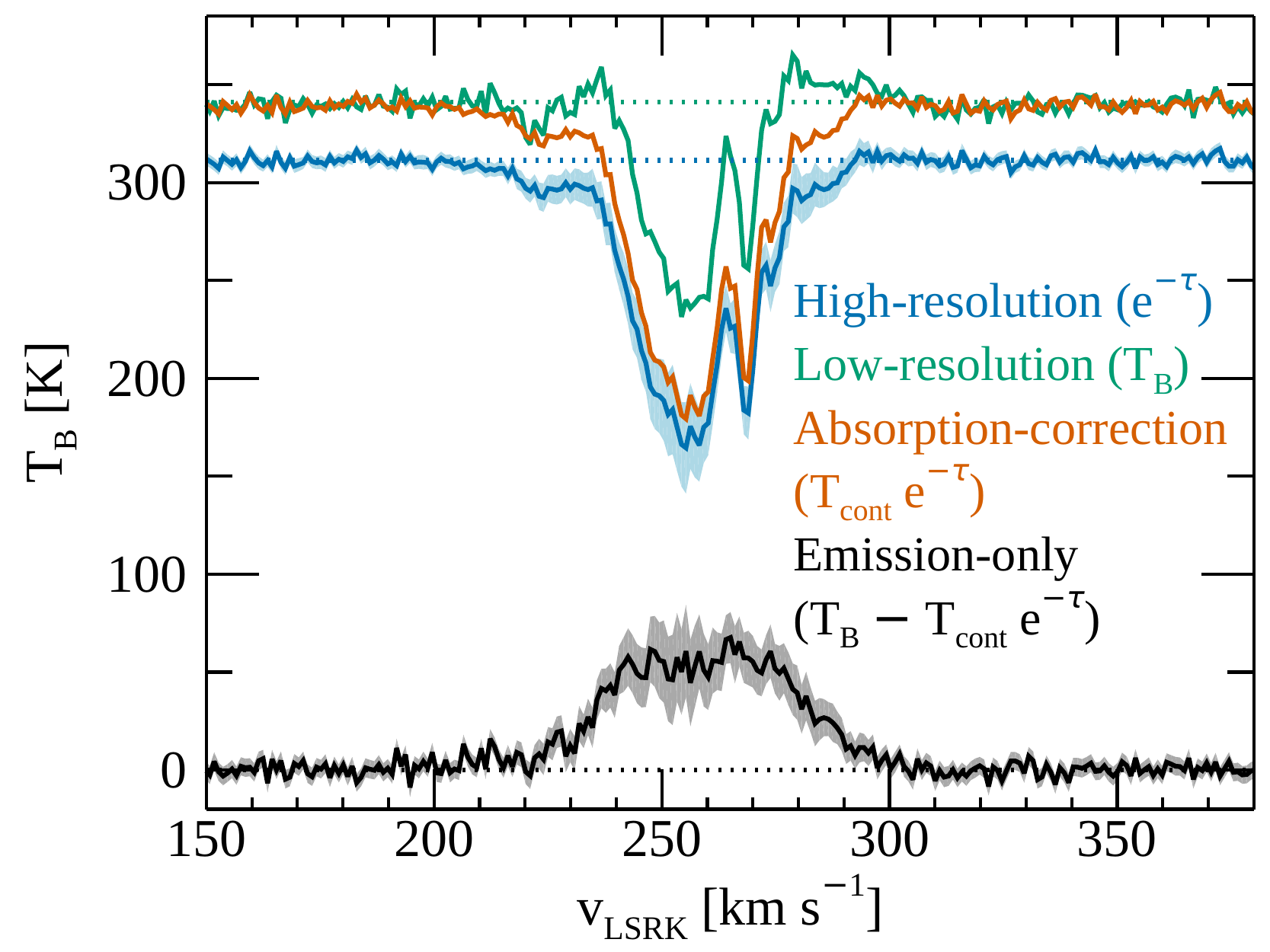}
	\caption{\label{f:em_only_deriv} Derivation of the emission-only spectrum for the pixel at 
	$(\alpha, \delta)_{\textrm{J2000}}$ = (05$^{\textrm{h}}$38$^{\textrm{m}}$45.7$^{\textrm{s}}$, $-$69$\degree$05$'$20$''$). 
	The blue curve shows the spectrum from the high-resolution cube along with the 1$\sigma$ envelope 
	that arises from both the radiometer noises and the uncertainties in the leakage spectrum. 
	Dividing the blue curve by the continuum ($\sim$310 K) results in the absorption spectrum $e^{-\tau(\varv)}$.
	The green curve is the corresponding spectrum measured in the low-resolution cube.
	As described in Section~\ref{s:em_only}, the tan curve is obtained by multiplying the absorption spectrum by the continuum of the green curve ($\sim$340 K),
	and subtracting the tan curve from the green curve leads to the black curve, which contains emission only.
	The error envelopes of the blue and black curves originate mostly from the uncertainties in the leakage spectrum.}
\end{figure}

Figure~\ref{f:em_only_deriv} {illustrates} our derivation of the emission-only spectra. 
The blue curve shows the absorption spectrum measured in the high-resolution cube along with the error envelope from the leakage spectrum (Section \ref{s:leakage}). 
The corresponding spectrum measured in the low-resolution cube is shown in green, 
which blends emission with absorption of the continuum of $\sim$340~K. 
Taking this continuum times $e^{-\tau(\varv)}$ gives the tan curve, 
and subtracting this from the green curve gives the black curve at the bottom (emission-only spectrum).

In addition, we calculated the 1$\sigma$ uncertainties $\sigma_{\textrm{o}}(\varv)$ in the emission-only spectra as follows: 

\begin{equation} \label{eq:sigma_o} 
	\sigma_{\textrm{o}}(\varv) = \sqrt{\sigma_{\textrm{B}}^2 + \sigma_{\textrm{n}}^2 + \sigma_{\textrm{l}}(\varv)^{2}}
\end{equation}

\noindent where $\sigma_{\textrm{B}}$ is the rms noise in $T_{\textrm{B}}(\varv)$ as defined in Equation (\ref{eq:em_abs}). 
For this calculation, we assumed that $T_{\textrm{cont}}$ is measured with very low noise,
which is reasonable considering that we averaged roughly 80 channels on either side of the emission line. 
The accurarcy of $T_{\textrm{cont}}$ is mainly limited by the lack of short-spacing information for the continuum (Section~\ref{s:extended_continuum}), 
but this does not introduce errors in the absorption and emission-only spectra.

For our subsequent analyses, we selected 225 pixels with clear detection 
(peak in $T_{\textrm{em-only}}(\varv)$~$>$~3$\sigma_{\textrm{em-only}}$; 
$\sigma_{\textrm{em-only}}$ = rms noise in $T_{\textrm{em-only}}(\varv)$)
and show the $1/\sigma_{\textrm{em-only}}^{2}$-weighted mean spectrum and  
spatial distribution of the integrated $T_{\textrm{em-only}}(\varv)$ of these pixels in Figures~\ref{f:avg_spec} and \ref{f:int_abs_em}. 
We note that the selected 225 pixels in the absorption and emission-only cubes
\footnote{The final \HI absorption and emission-only cubes, as well as their 1$\sigma$ uncertainty cubes (resolution = 30$''$ and pixel size = 12$''$), are available online.}
are the same. 

\subsection{\CII} \label{s:cii}

To compare with \HI in 30 Dor, we used the \CII 158~$\micron$ observations from \citet{okada2019} 
that were obtained using SOFIA with angular and velocity resolutions of 16$''$ and 1~km~s$^{-1}$. 
These data were smoothed and regridded to have a resolution of 30$''$ and a pixel size of 12$''$ for our analyses. 

\subsection{$^{12}$CO(2--1)} \label{s:co21}

We utilized the CO data from \citet{wong2022} that were obtained with ALMA. 
The 12m, 7m, and total power arrays were used for the CO observations, 
and the final angular and velocity resolutions were 1.8$''$ and 0.3~km~s$^{-1}$, respectively. 
To match our \HI data, we smoothed and regridded the CO cube to have a resolution of 30$''$ and a pixel size of 12$''$. 

\section{Results} \label{s:results}

In this section, we estimate various properties of the CNM (e.g., velocity, optical depth, spin temperature, etc.) 
and examine their global distributions.

\begin{deluxetable*}{c c c c c c c c}
\centering
\tablecaption{\label{t:gaussian_individual} Derived Gaussian Parameters for the Total 225 Pixels}
    \tablewidth{0pt}
    \setlength{\tabcolsep}{15pt}
    \tabletypesize{\small}
    \tablehead{
    \colhead{X} & \colhead{Y} & \colhead{$t_{\textrm{peak}}$} & \colhead{$\sigma_{t_{\textrm{peak}}}$} &     
    \colhead{$\varv_{0}$} & \colhead{$\sigma_{\varv_{0}}$} & 
    \colhead{FWHM} & \colhead{$\sigma_{\textrm{FWHM}}$} \\ 
    \colhead{} & \colhead{} & \colhead{} & \colhead{} &
    \colhead{(km~s$^{-1}$)} & \colhead{(km~s$^{-1}$)} & 
    \colhead{(km~s$^{-1}$)} & \colhead{(km~s$^{-1}$)} \\
    \colhead{(1)} & \colhead{(2)} & \colhead{(3)} & \colhead{(4)} &
    \colhead{(5)} & \colhead{(6)} &
    \colhead{(7)} & \colhead{(8)}}
    \startdata
     9 & 25 & 0.23 & 0.01 & 253.50 & 0.59 & 33.83 & 1.56 \\
     9 & 25 & 0.05 & 0.01 & 297.53 & 2.04 & 19.64 & 4.95 \\
     9 & 26 & 0.24 & 0.01 & 258.40 & 0.60 & 33.86 & 1.52 \\ 
    \enddata
    \tablecomments{\small (1, 2) X and Y pixel positions in the high-resolution absorption cube; 
	                        (3, 4) Peak in $1 - e^{-\tau}$ and its 1$\sigma$ uncertainty;
			        (5, 6) Central velocity and its 1$\sigma$ uncertainty;
				(7, 8) FWHM and its 1$\sigma$ uncertainty. \\
				(This table is available in its entirety in a machine-readable form in the online journal. 
				 A portion is shown here for guidance regarding its form and content)}
\end{deluxetable*}

\begin{deluxetable}{c c c}
\centering
\tablecaption{\label{t:gaussian_global} Global CNM Properties from Gaussian Fitting}
    \tablewidth{0pt}
    \setlength{\tabcolsep}{15pt}
    \tabletypesize{\small}
    \tablehead{
    \colhead{Property} & \colhead{Range} & \colhead{Median} \\
    \colhead{(1)} & \colhead{(2)} & \colhead{(3)}}
    \startdata
     $t_{\textrm{peak}}$ & 0.02--0.97 & 0.24 \\
     $\varv_{0}$ (km~s$^{-1}$) & 203.9--297.5 & 255.9 \\
     FWHM (km~s$^{-1}$) & 2.4--47.3 & 11.4 \\ 
    \enddata
    \tablecomments{\small (1) CNM properties including the peak in $1 - e^{-\tau}$, central velocity, and FWHM; 
			         (2, 3) Range and median value.}
\end{deluxetable}

\begin{figure*}
    \centering
    \includegraphics[scale=0.35]{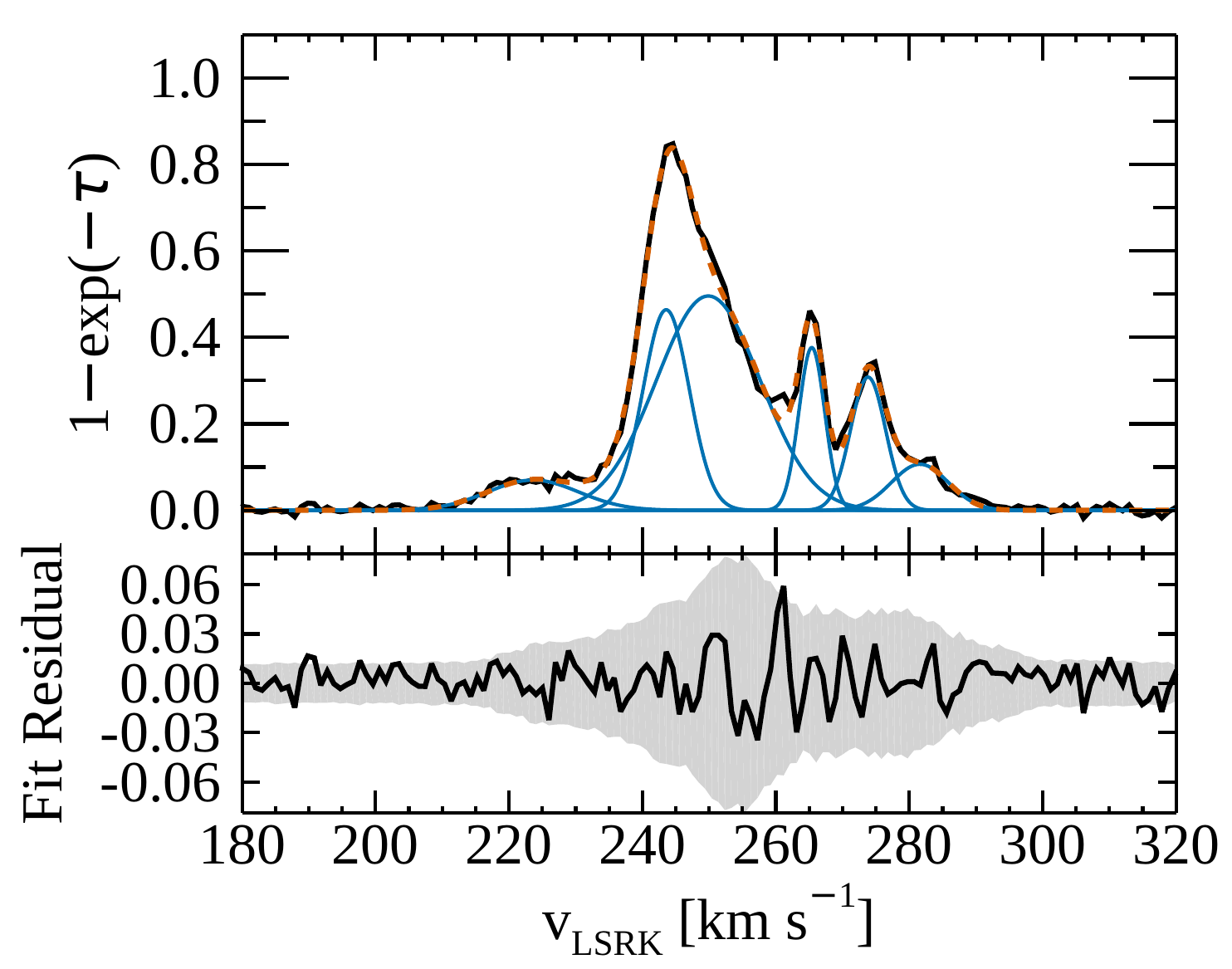} \hspace{0.2cm}
    \includegraphics[scale=0.35]{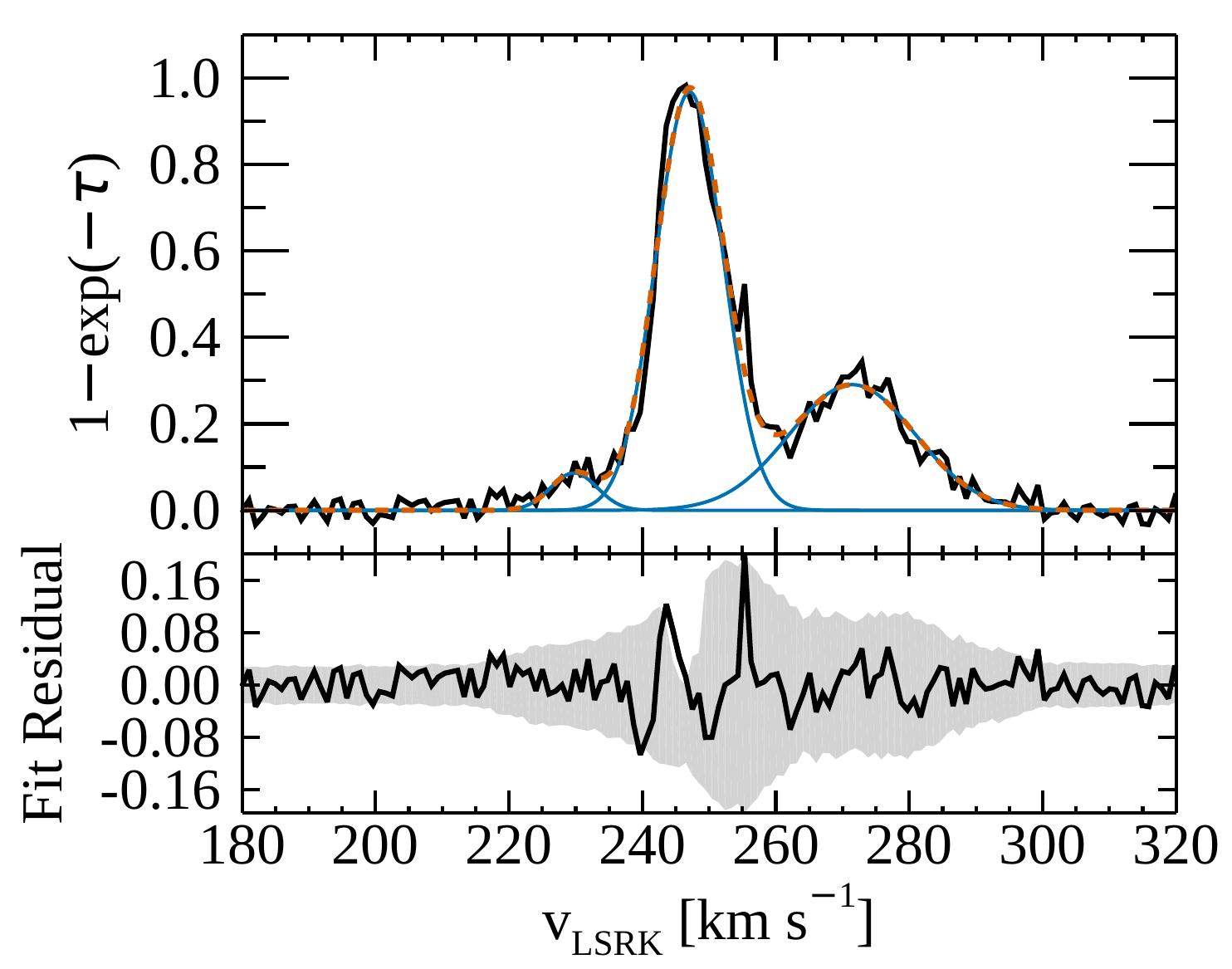}
    \caption{\label{f:gfit_example} (Left) Absorption spectrum from the pixel at 
             $(\alpha, \delta)_{\textrm{J2000}}$ = (05$^{\textrm{h}}$38$^{\textrm{m}}$34.4$^{\textrm{s}}$, $-$69$\degree$06$'$20$''$). 
             This pixel was selected to demonstrate our Gaussian fitting with the maximum number of components (six). 
             In the top panel, the observed absorption spectrum is shown in black, 
             while the individual components and the total fit are overlaid in blue and tan, respectively.   
	     The residuals from fitting are presented in the bottom panel along with the $\sigma_{\textrm{t}}(\varv)$ envelope in gray. 
	     (Right) Same as the left panel, but for the pixel with the highest value of the fitted $1-e^{-\tau}$ (0.97) at  
	     $(\alpha, \delta)_{\textrm{J2000}}$ = (05$^{\textrm{h}}$38$^{\textrm{m}}$38.9$^{\textrm{s}}$, $-$69$\degree$07$'$32$''$).}
\end{figure*} 

\begin{figure*}
    \centering
    \includegraphics[scale=0.45]{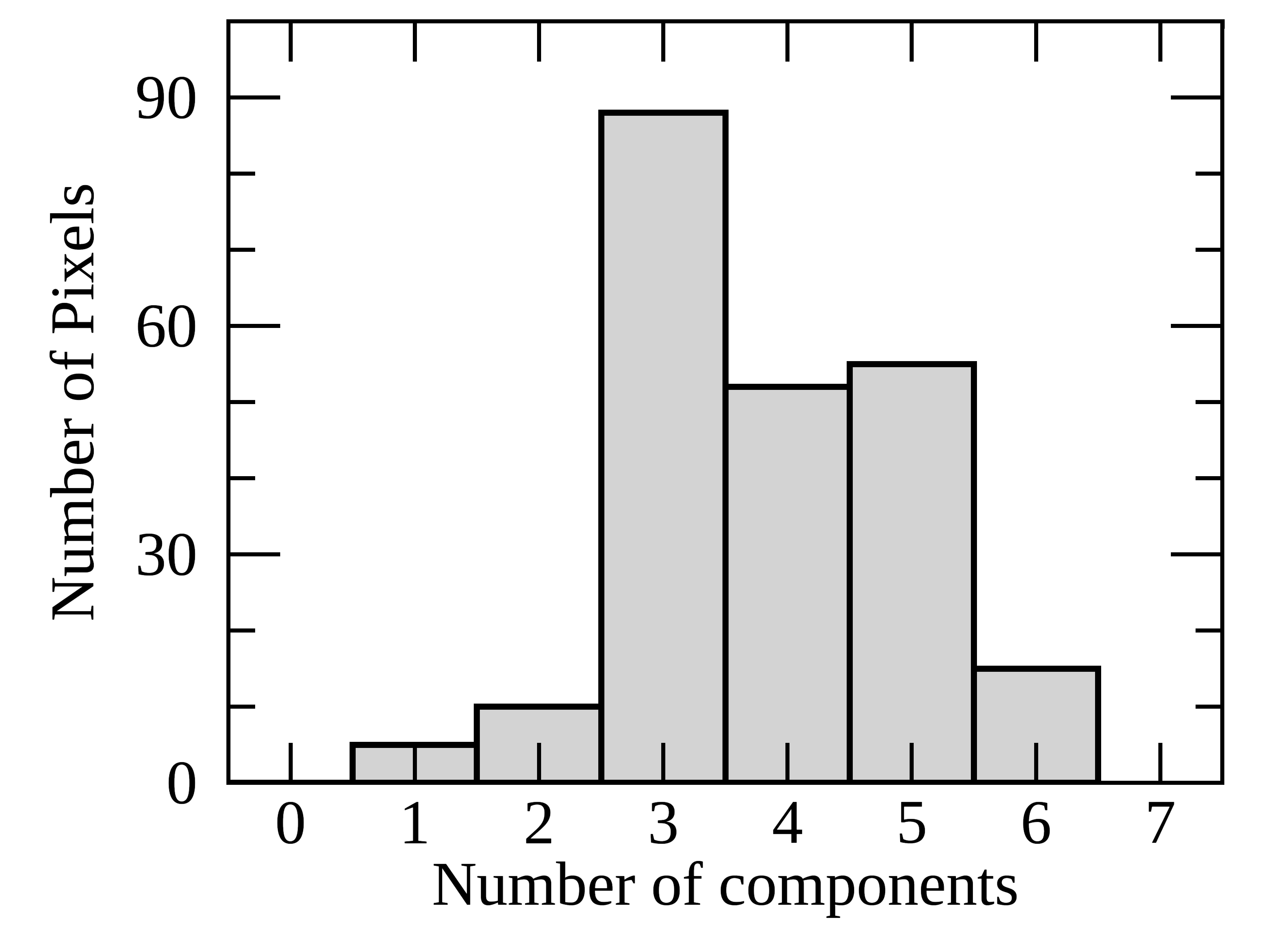} \hspace{0.2cm}
    \includegraphics[scale=0.45]{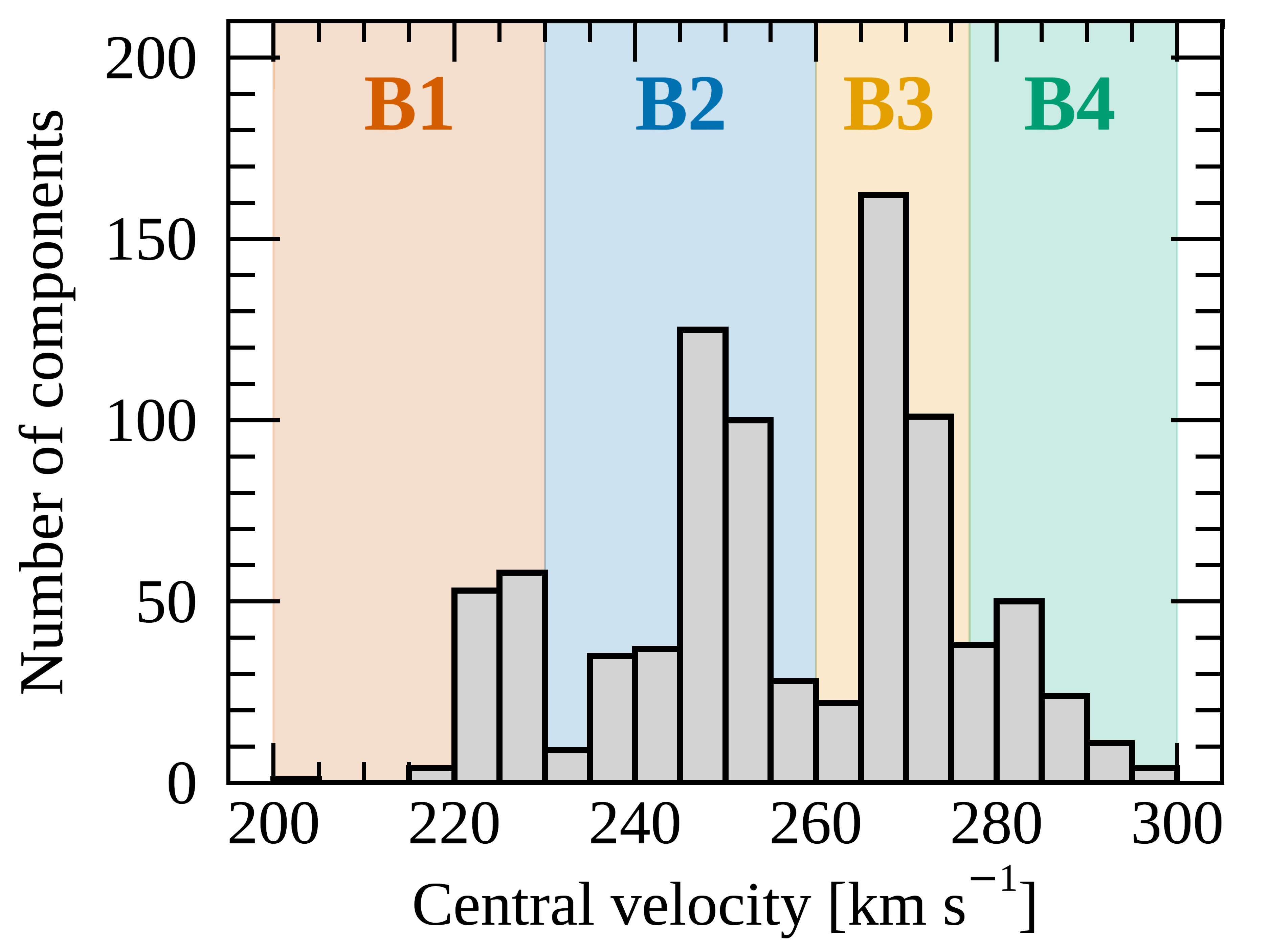}
    \includegraphics[scale=0.45]{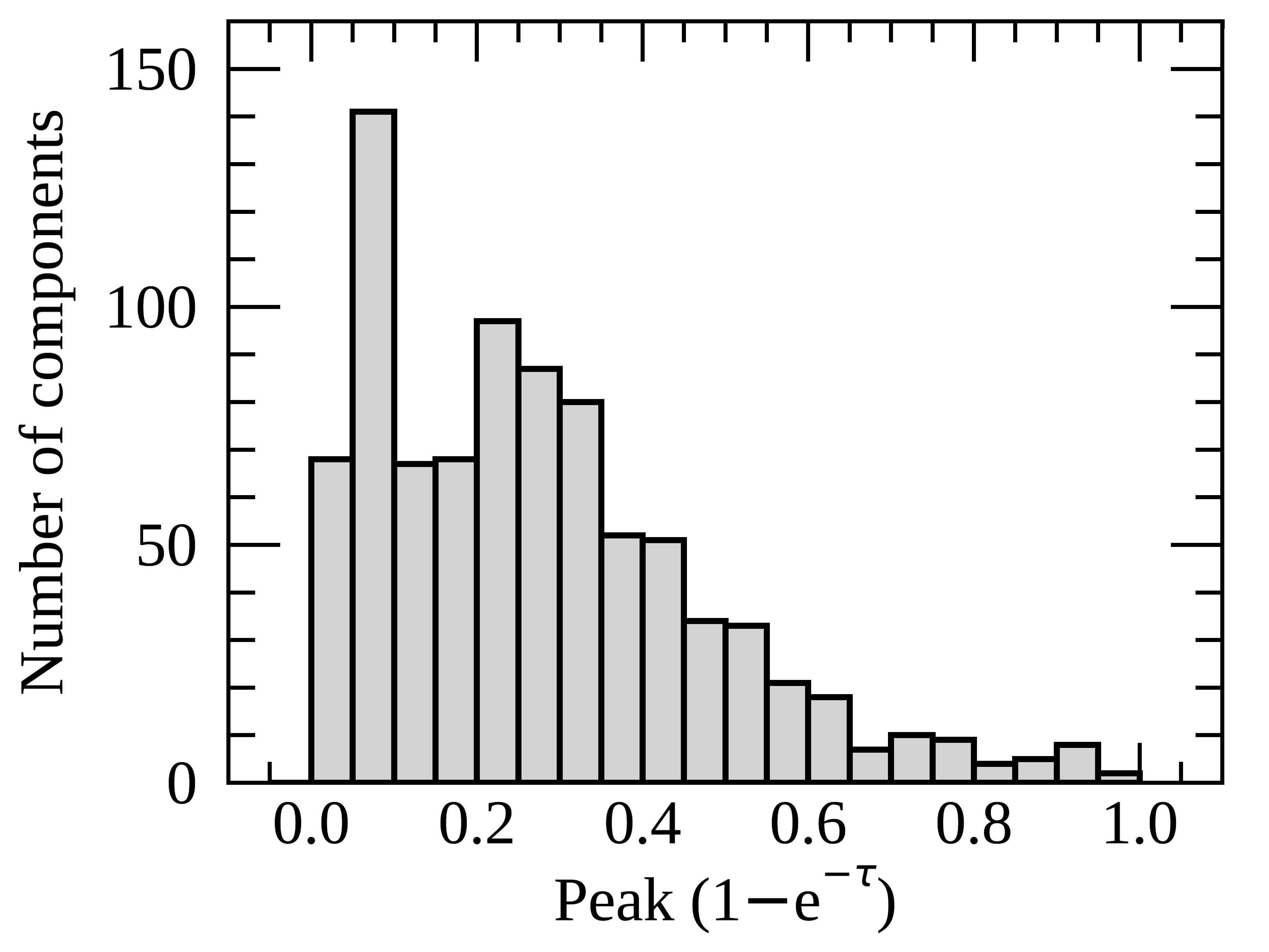} \hspace{0.2cm}
    \includegraphics[scale=0.45]{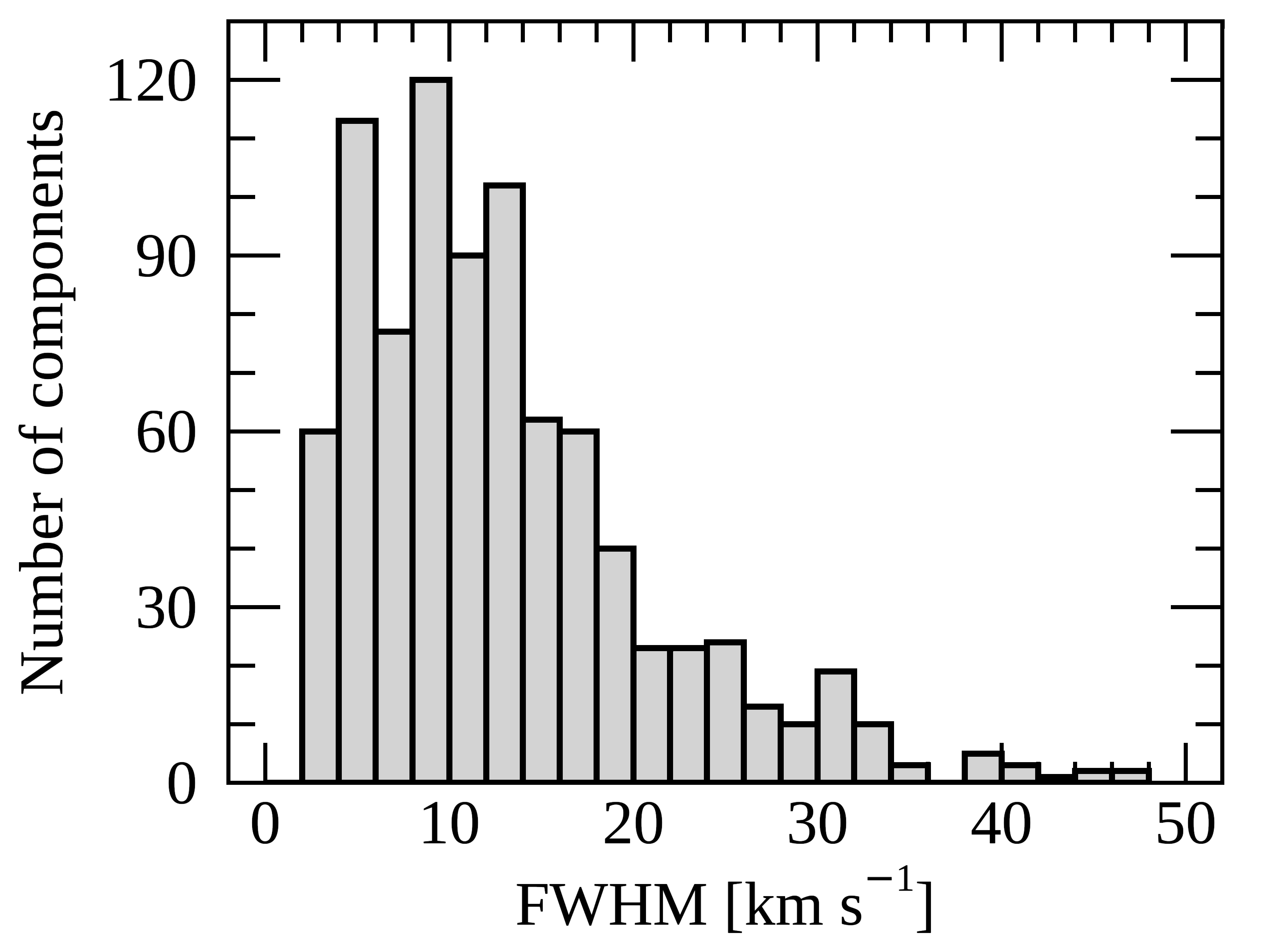}
    \caption{\label{f:hist_gauss} Histograms of the Gaussian decomposition results. 
             (Top) Number of the fitted Gaussian components and central velocity. 
             In the central velocity panel, the four velocity bands are shown in different colors 
             (B1, B2, B3, and B4 in tan, blue, yellow, and green, respectively). 
             (Bottom) Peak $1 - e^{-\tau}$ and FWHM.}
\end{figure*}

\begin{figure} 
    \includegraphics[scale=0.45]{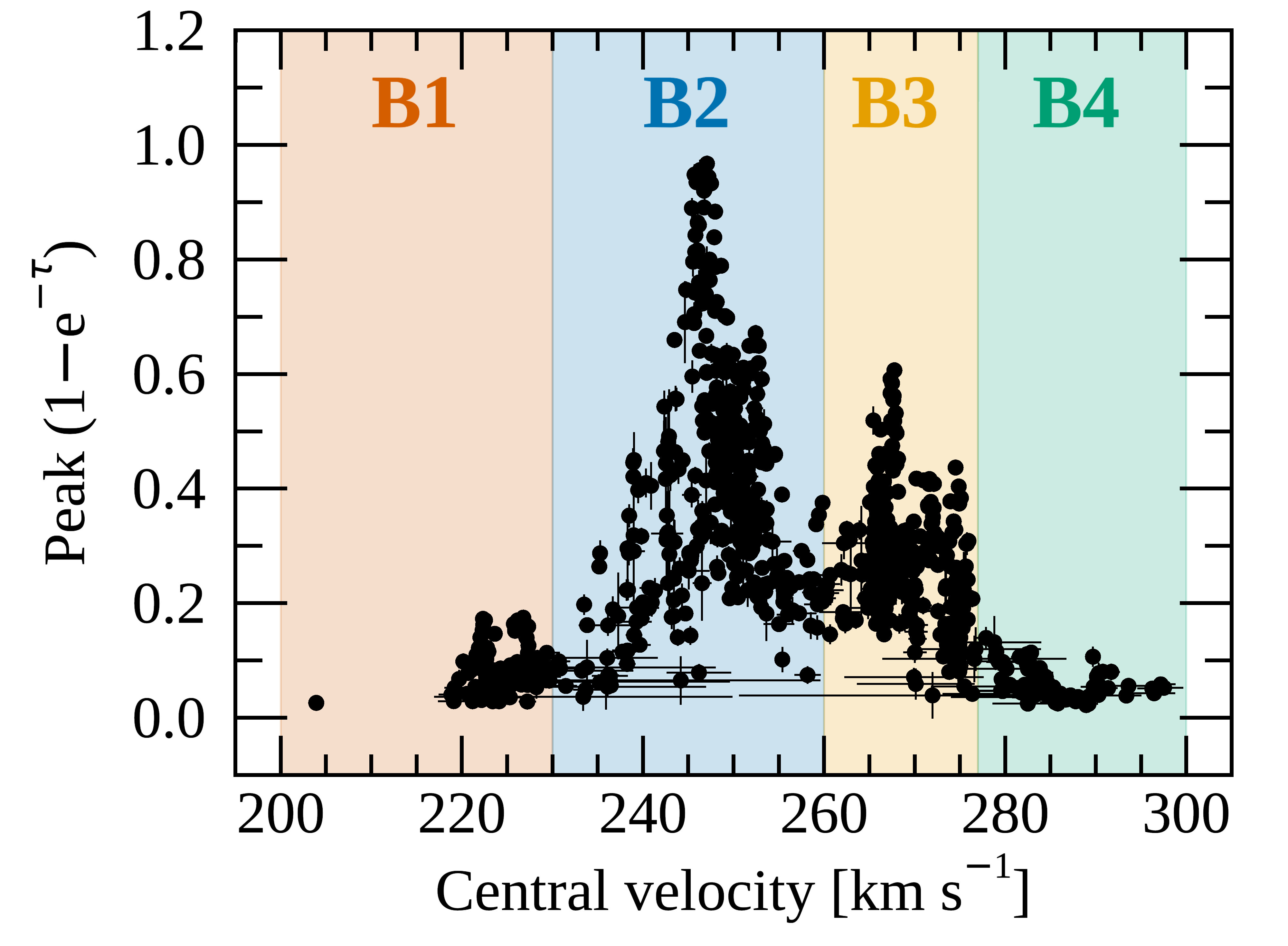}
    \caption{\label{f:tau_vs_vlsrk_per_bands} Peak $1 - e^{-\tau}$ as a function of the central velocity for the 862 individual Gaussian components.
             The four velocity bands (B1 to B4) are indicated in different colors.}
\end{figure}

\subsection{Gaussian Decomposition of the \HI Absorption Spectra} \label{s:gfit}

The average absorption spectrum of 30 Dor clearly shows distinct features in velocity (Figure~\ref{f:avg_spec}). 
To analyze these features separately, we performed Gaussian decomposition of the high-resolution $1 - e^{-\tau(\varv)}$ cube. 
For each unmasked pixel, we started with one Gaussian and added additional components 
until the residuals from fitting were within $\pm \sigma_{\textrm{t}}(\varv)$. 
We then derived the peaks, central velocities, and full widths at half maximum (FWHMs) of the fitted Gaussian components.
Two example spectra are shown in Figure~\ref{f:gfit_example} to demonstrate our Gaussian decomposition, 
and the results for the total 225 pixels are summarized in Tables~\ref{t:gaussian_individual} and \ref{t:gaussian_global} and 
Figure~\ref{f:hist_gauss}. 

We found one to six Gaussian components toward the observed LOSs, resulting in 862 in total. 
These components have the peak values of 0.02--0.97 in $1 - e^{-\tau}$ (median of 0.24) and FWHMs of 2.4--47.3 km~s$^{-1}$ (median of 11.4~km~s$^{-1}$). 
Both parameters show approximately normal distributions with long tails beyond the peak value of $\sim$0.5 and FWHM of $\sim$20~km~s$^{-1}$.
As for the central velocities, the fitted components are distributed at a wide range of 204--298~km~s$^{-1}$, 
likely indicating the complex CNM kinematics in and around 30 Dor.
To better understand the velocity distribution of the CNM components, 
we additionally examined how the peak values in $1 - e^{-\tau}$ change with the central velocities  
and found that there are largely four distinct groups with two strong peaks at 
$\sim$250~km~s$^{-1}$ and $\sim$270~km~s$^{-1}$ (Figure~\ref{f:tau_vs_vlsrk_per_bands}).
Based on this result, we divided the entire velocity range into four bands, 
i.e., B1 for 200--230~km~s$^{-1}$, B2 for 230--260~km~s$^{-1}$, B3 for 260--277~km~s$^{-1}$, and B4 for 277--300~km~s$^{-1}$, 
and used them for our subsequent analyses.

Since \HI absorption is measured against the main \HII complex at {B2} velocities (Section~\ref{s:multi-phase_structure} for details), 
B1, {B3}, and B4 are CNM structures in front of {B2}.  
In particular, we interpret {B1 as outflows and B3 and B4 as inflows}, 
considering that {B1 is blueshifted (moving toward us) and B3 and B4 are redshifted (moving away from us) with respect to B2}. 
We will revisit the nature of the CNM structures at B2 and B3 velocities in Section~\ref{s:multi-phase_structure}
and discuss the implication of the outflows and inflows for the evolution of 30 Dor in Section~\ref{s:outflow_inflow}.

\subsection{Derivation of Physical Properties} \label{s:ts_derivation}

\subsubsection{LOS Average Spin Temperature} \label{s:tspin_calculation}

\begin{figure*}
    \centering
    \includegraphics[scale=0.28]{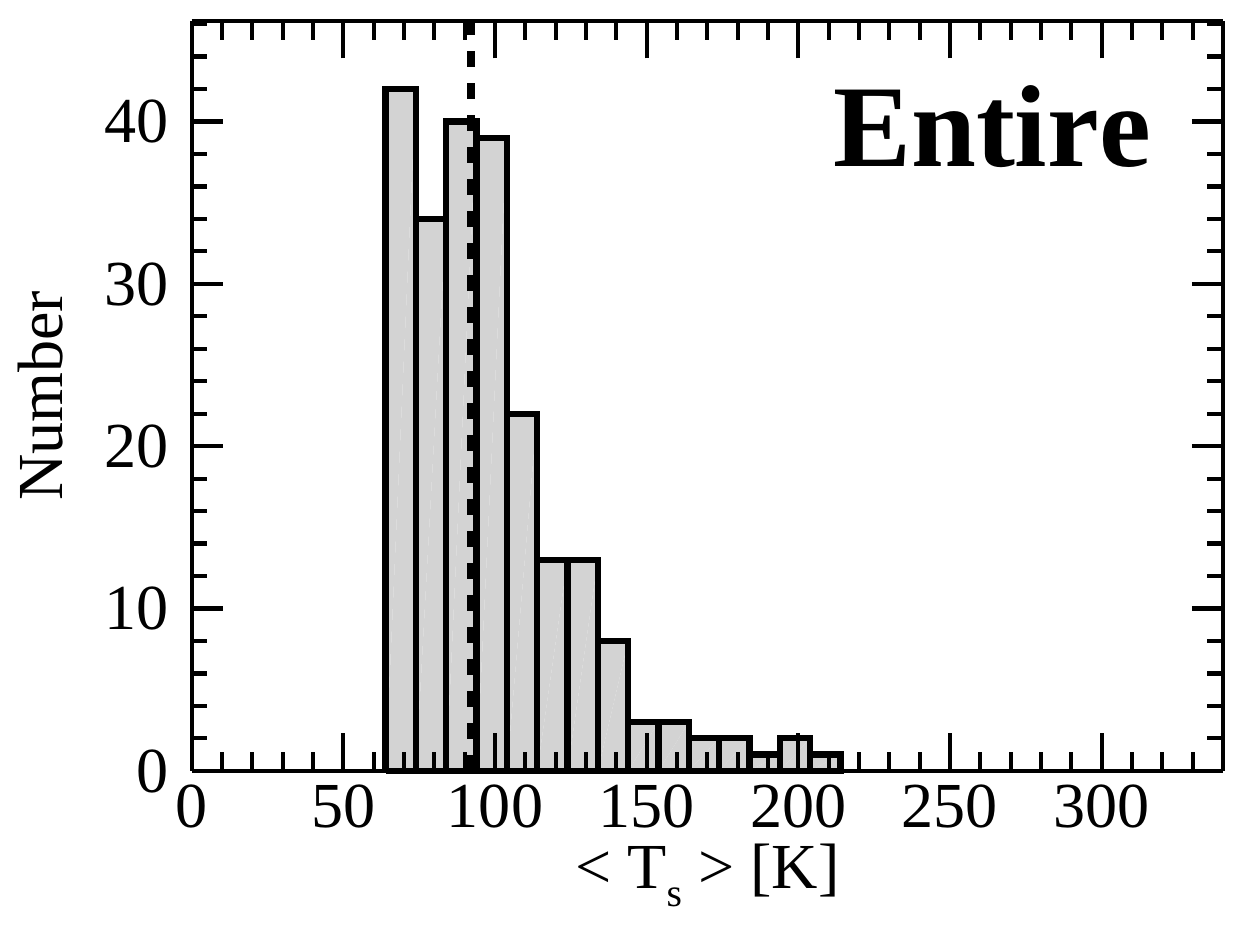} 
    \includegraphics[scale=0.28]{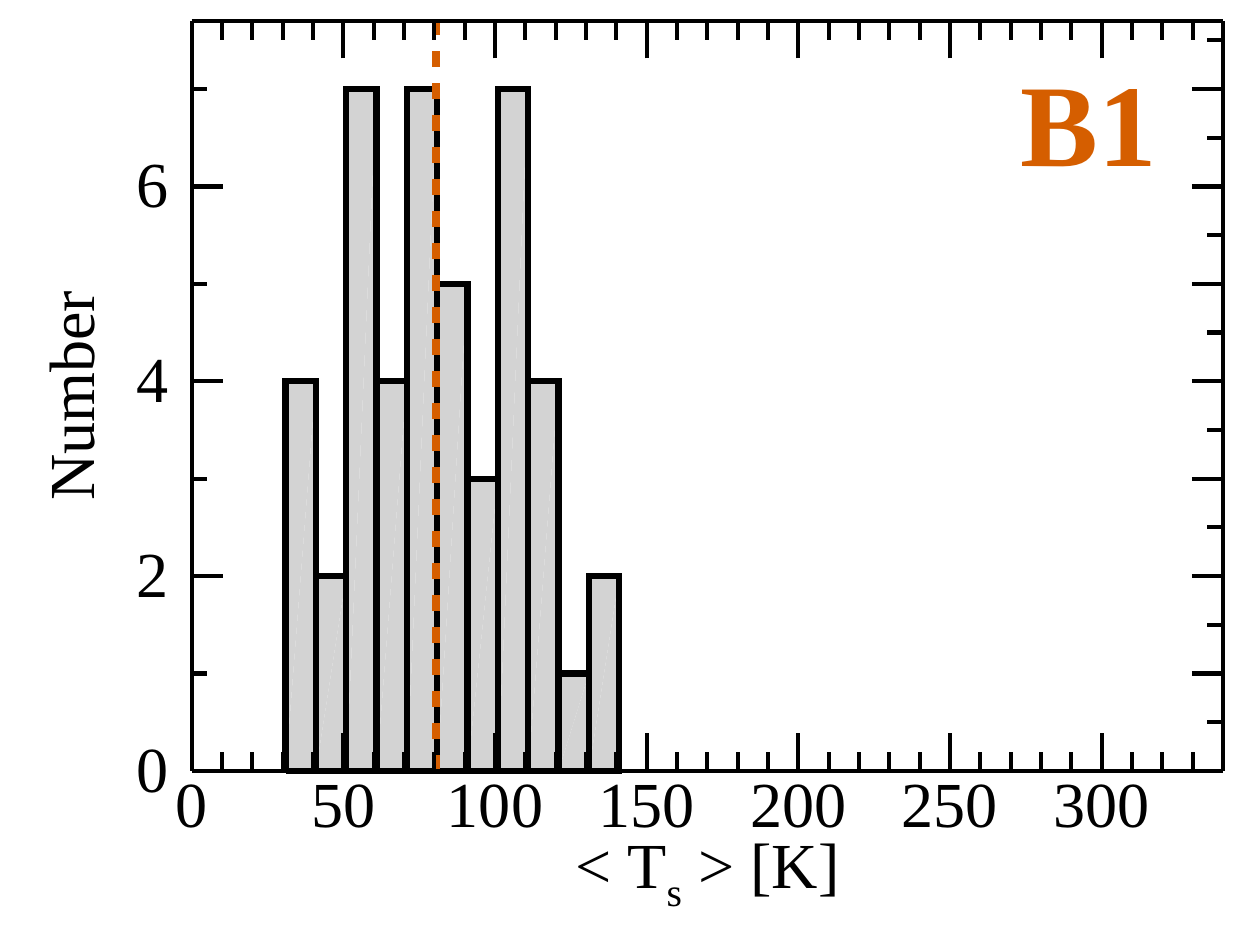}
    \includegraphics[scale=0.28]{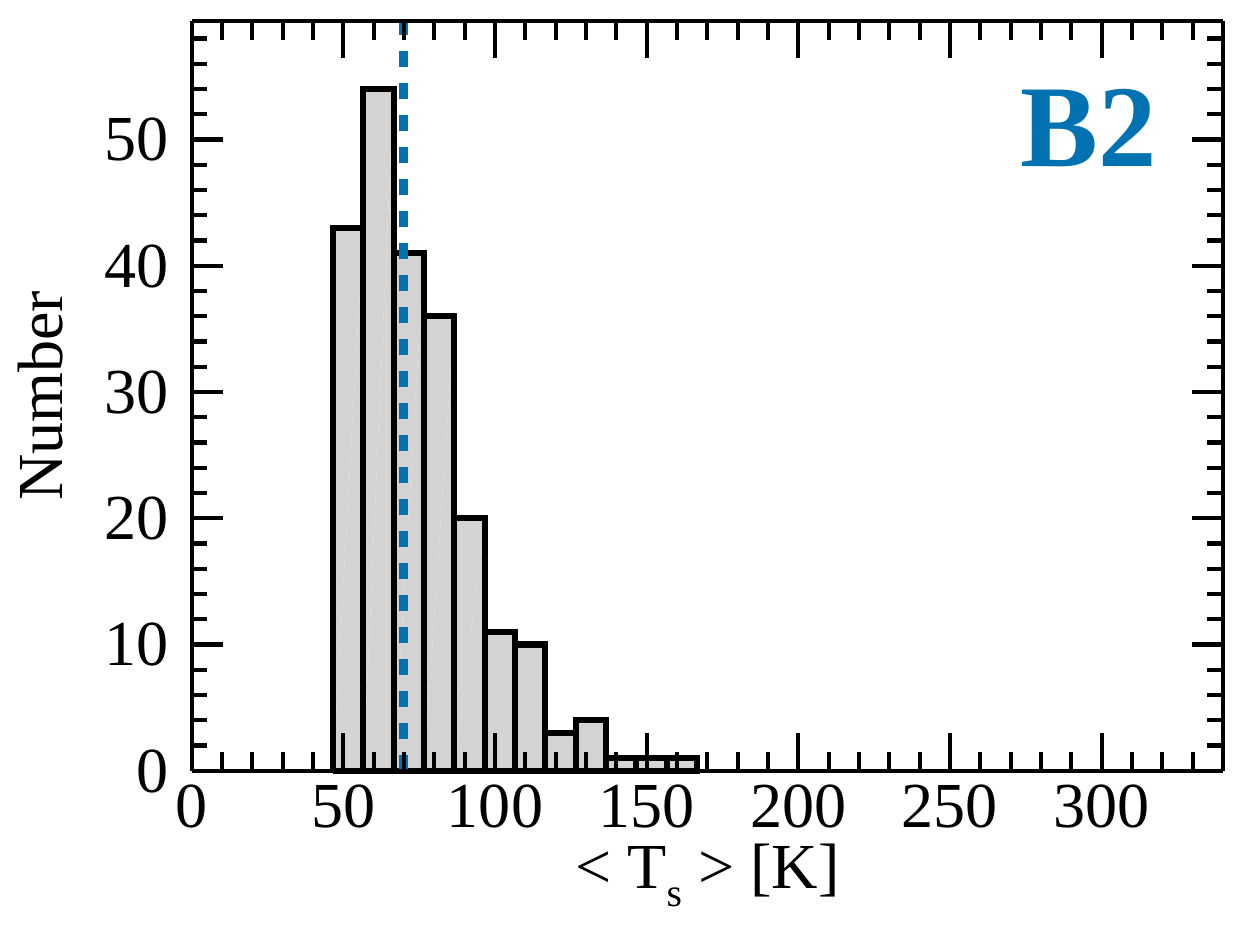}
    \includegraphics[scale=0.28]{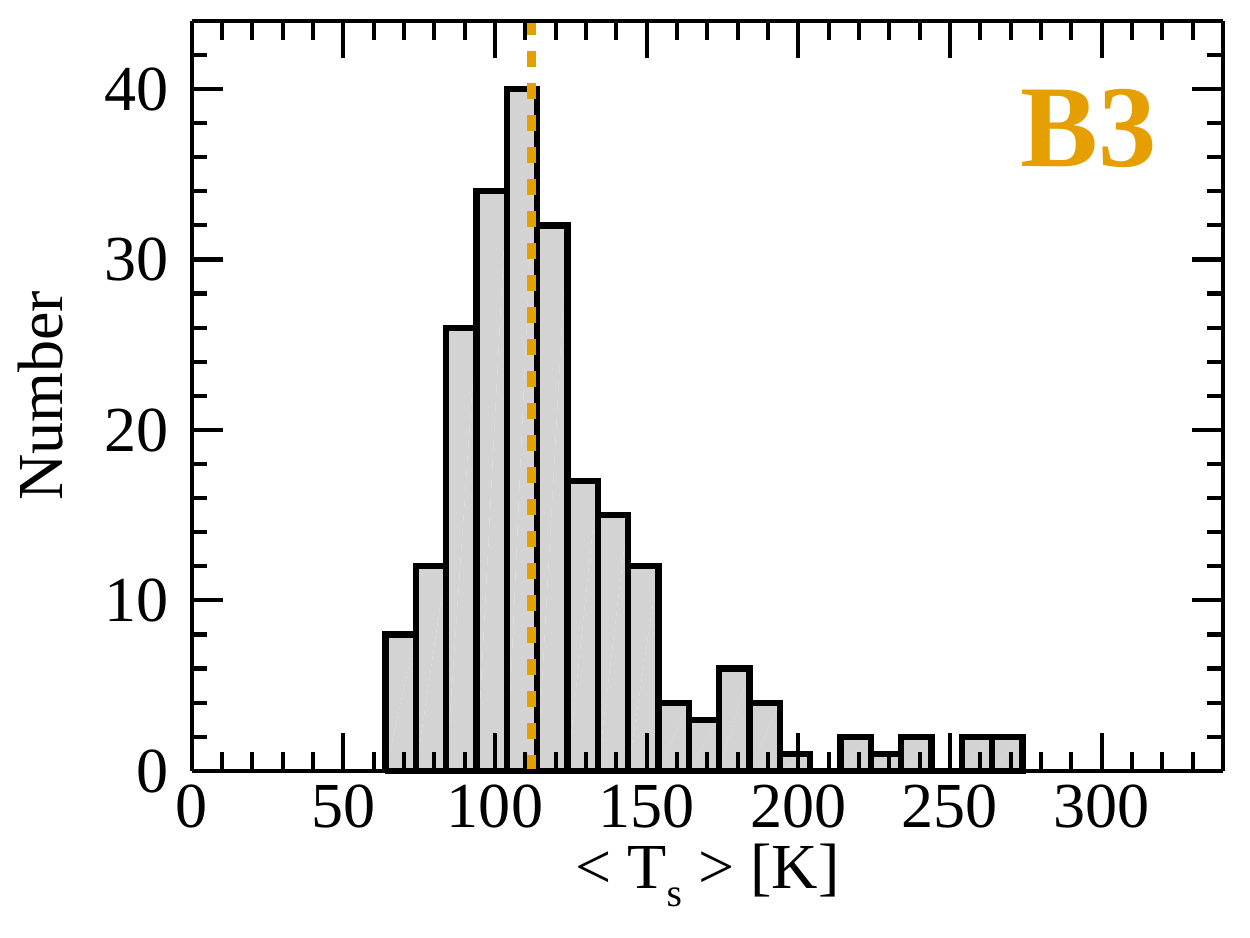}
    \includegraphics[scale=0.28]{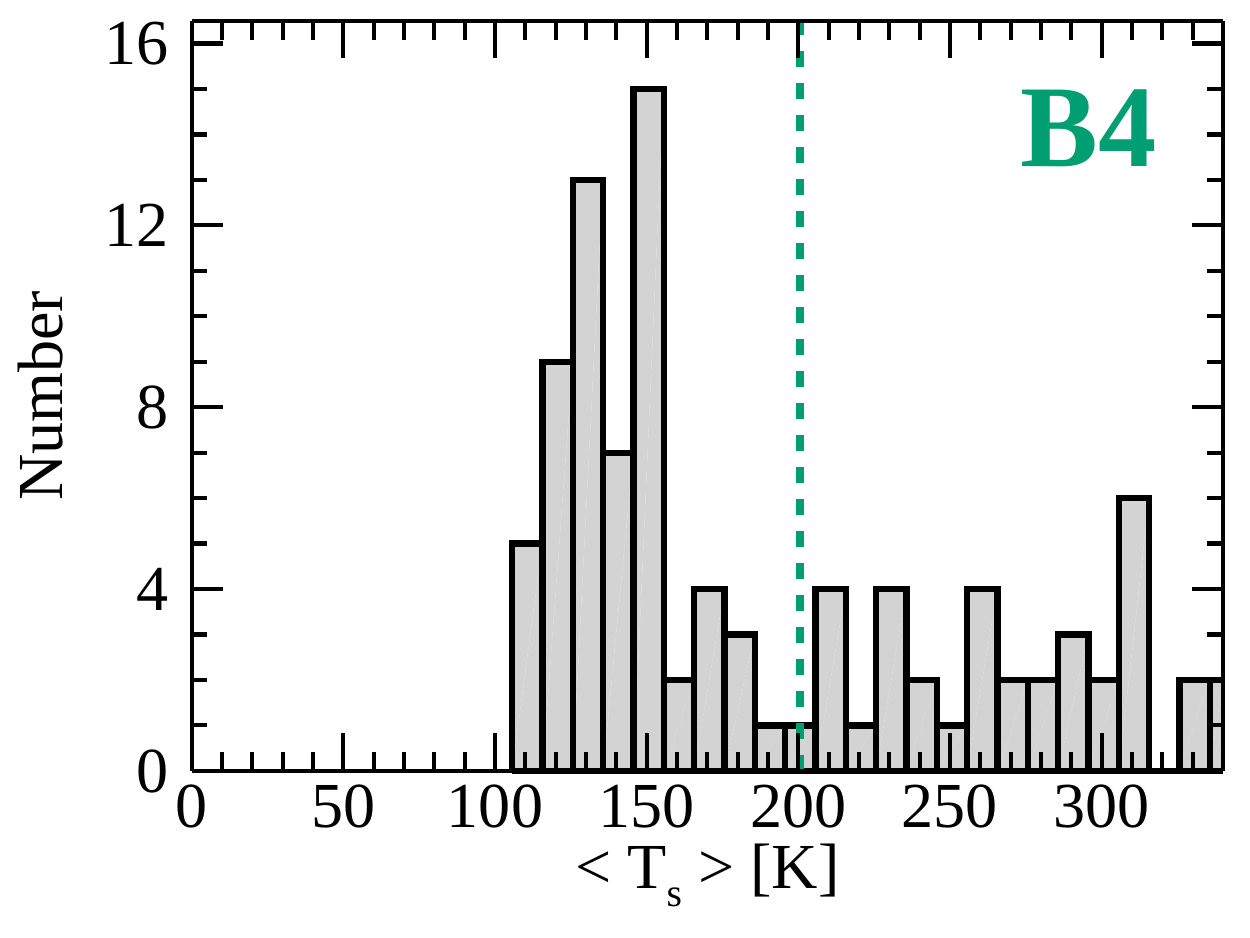}
    \caption{\label{f:hist_avg_tspin} Histograms of the LOS average spin temperatures
             that were calculated for the entire velocity range (Entire) and the individual bands (B1 to B4). 
             The median of each histogram is indicated as the dashed line.}
\end{figure*}

In addition to the Gaussian parameters, we calculated the LOS average spin temperature <$T_{\textrm{s}}$> on 30$''$ scales, 
assuming an isothermal single-phase approximation as outlined by \citet{dickey2000} 

\begin{equation} \label{eq:LOS_avg_Ts}
	<T_{\textrm{s}}>\,= q~\frac{N\textrm{(H~\textsc{i})}_{\textrm{unc}}}{\textrm{EW}} = q~\frac{\int T_{\textrm{em-only}}(\varv)~d\varv}{\int (1-e^{-\tau(\varv)})~d\varv}, 
\end{equation}

\noindent 
where $q$ is the fraction of gas in front of the continuum (e.g., $q = 1$ when all the gas is in front of the continuum), 
$N$(H~\textsc{i})$_{\textrm{unc}}$ is the \HI column density uncorrected for self-absorption, 
and EW is the equivalent width. 
For our calculation, we assumed a statistically probable value of $q = 0.5$ 
and used the entire velocity range (referred as ``entire''), as well as the individual bands (B1 to B4). 
To ensure a reliable analysis, we estimated the 1$\sigma$ uncertainties in $N$(H~\textsc{i})$_{\textrm{unc}}$, EW, and <$T_{\textrm{s}}$>
($\sigma_{N\textrm{(H~\textsc{i})}_{\textrm{unc}}}$, $\sigma_{\textrm{EW}}$, and $\sigma_{<T_{\textrm{s}}>}$) 
by propagating the error spectra of $T_{\textrm{em-only}}(\varv)$ and $1 - e^{-\tau(\varv)}$ 
and selected the pixels that meet $N$(H~\textsc{i})$_{\textrm{unc}} > 3\sigma_{N\textrm{(H~\textsc{i})}_{\textrm{unc}}}$,
$\textrm{EW} > 3\sigma_{\textrm{EW}}$, and <$T_{\textrm{s}}$>~$> 3\sigma_{<T_{\textrm{s}}>}$ thresholds. 
The calculated <$T_{\textrm{s}}$> values for the selected pixels are presented in Figure~\ref{f:hist_avg_tspin} and Table~\ref{t:band_info}. 

The <$T_{\textrm{s}}$> values for the entire velocities range from 64~K to 208~K (median of 92~K) 
and show roughly a Gaussian distribution with a long tail toward <$T_{\textrm{s}}$>~$\gtrsim 150$~K.
This distribution has a small standard deviation of 27~K,
suggesting that <$T_{\textrm{s}}$> does not change significantly on average 
as it is a complex quantity that depends on both $T_{\textrm{s}}$ and a fraction of the CNM. 
As for the individual bands, {B2 shows the coldest <$T_{\textrm{s}}$> distribution with a median of 70~K. 
B1 is slightly warmer than B2 with a median of 81~K, while having a non-negligible fraction of the pixels (13\%) 
with exceptionally low <$T_{\textrm{s}}$>~$\lesssim 50$~K.
Finally, B3 and B4 are systematically warmer than B1 and B2 with medians of 112~K and 201~K.}

\begin{figure*}
    \centering
    \includegraphics[scale=0.28]{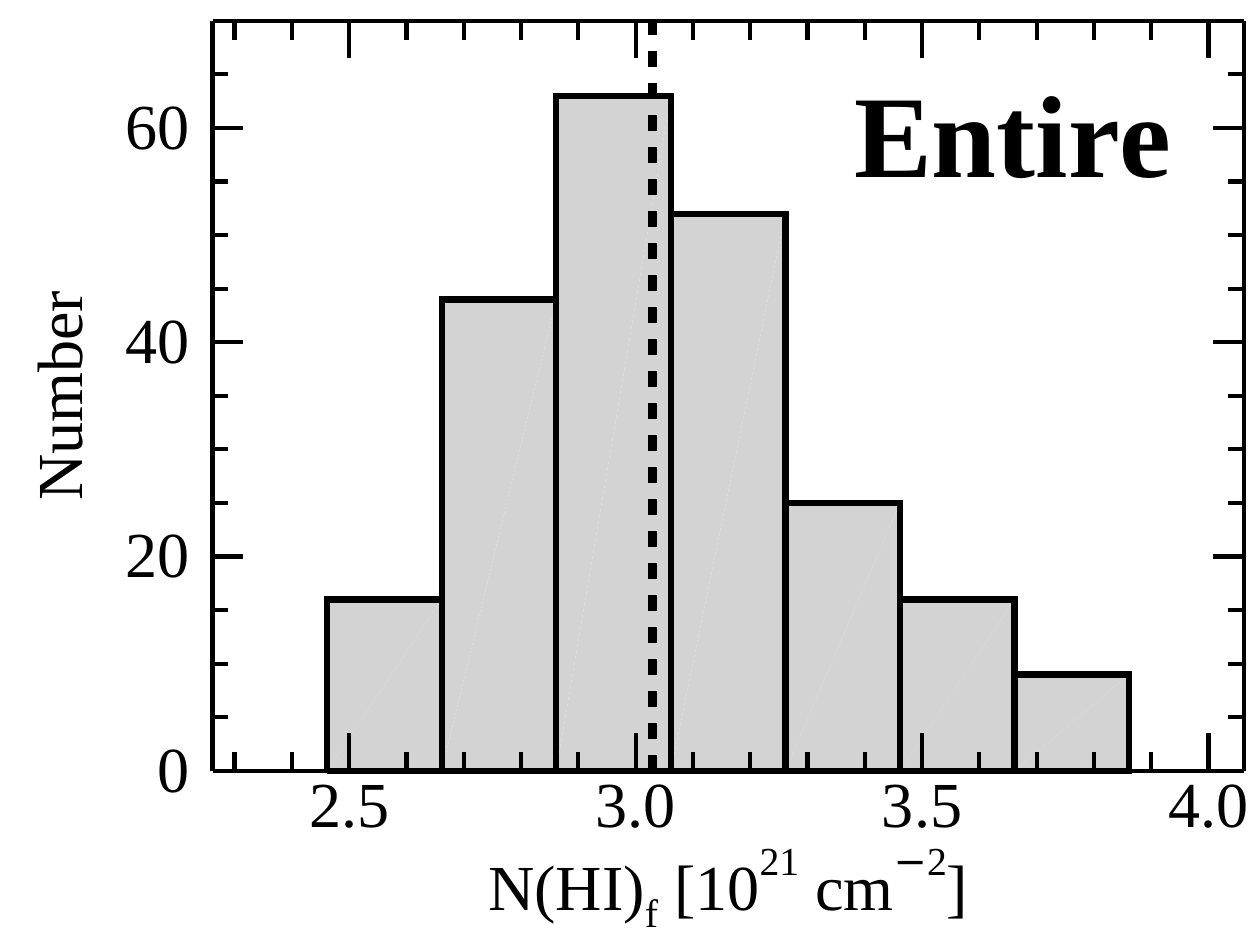} 
    \includegraphics[scale=0.28]{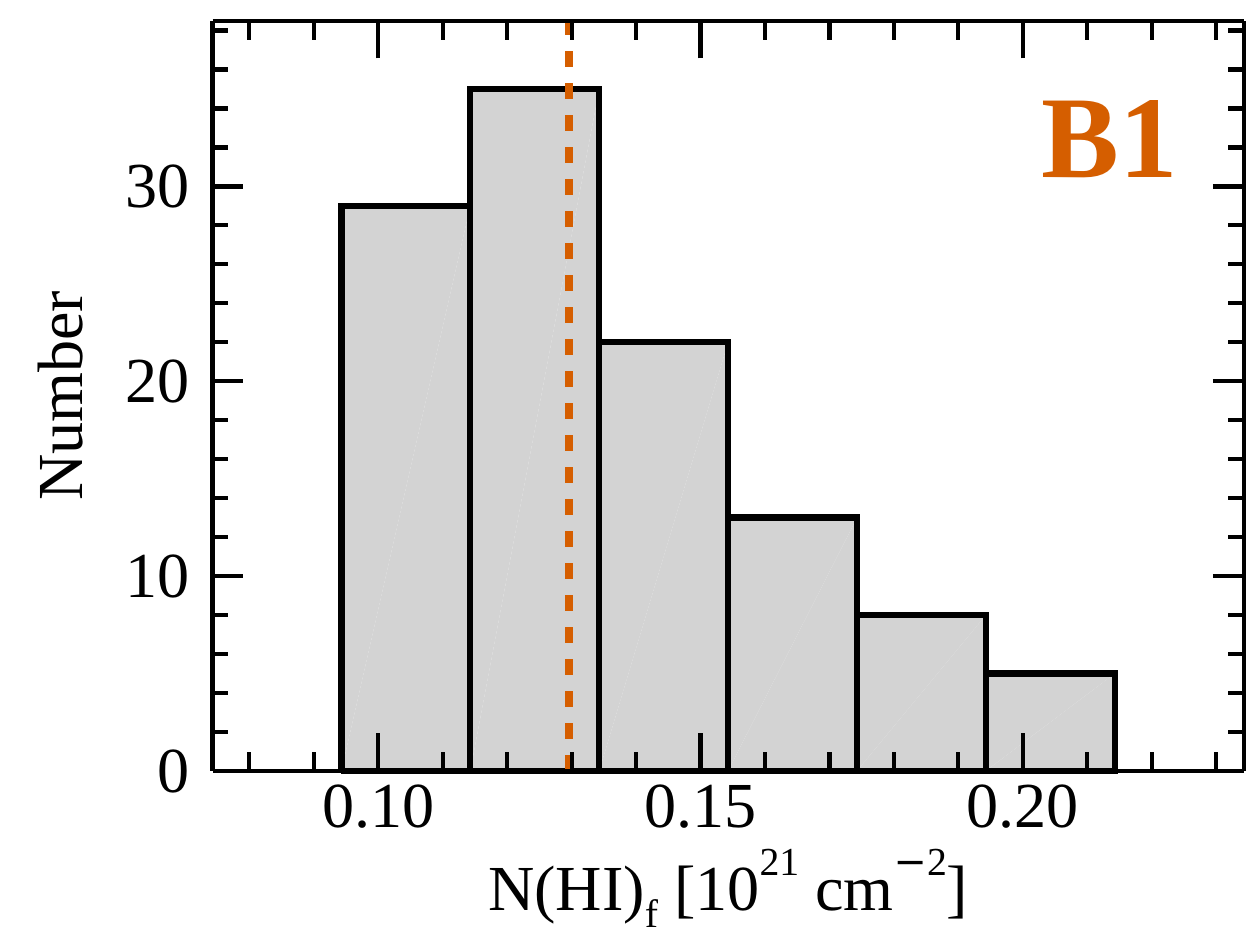}
    \includegraphics[scale=0.28]{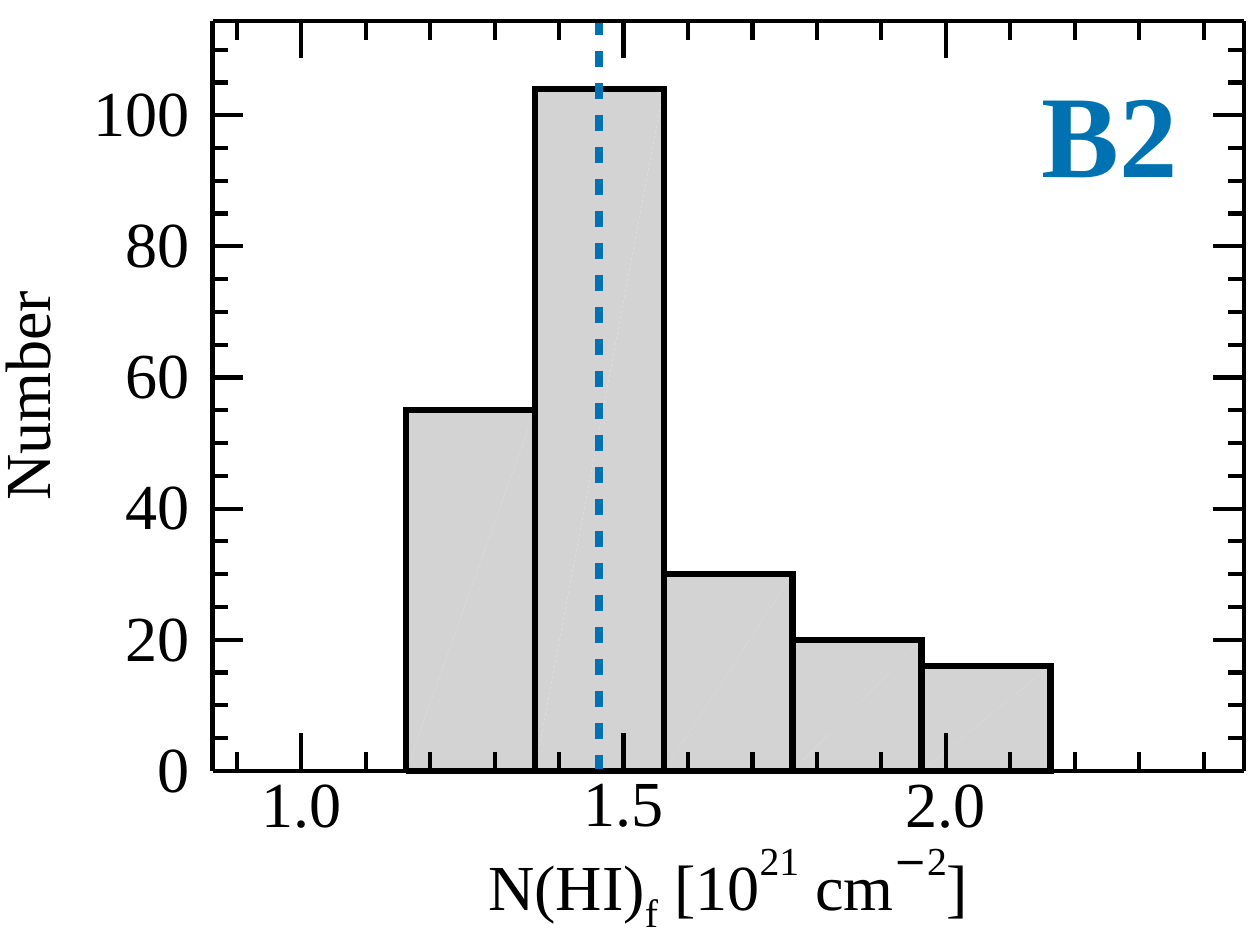}
    \includegraphics[scale=0.28]{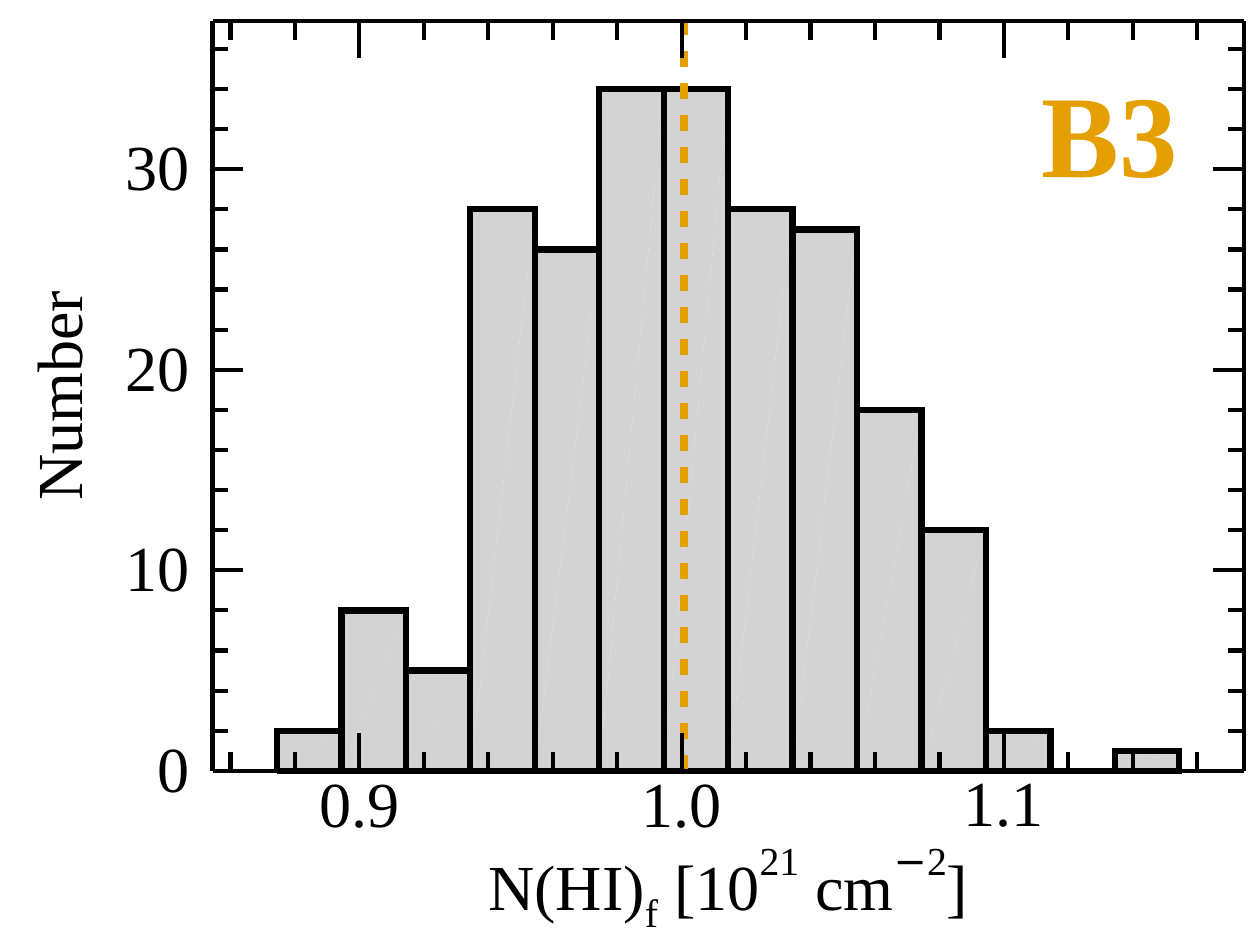}
    \includegraphics[scale=0.28]{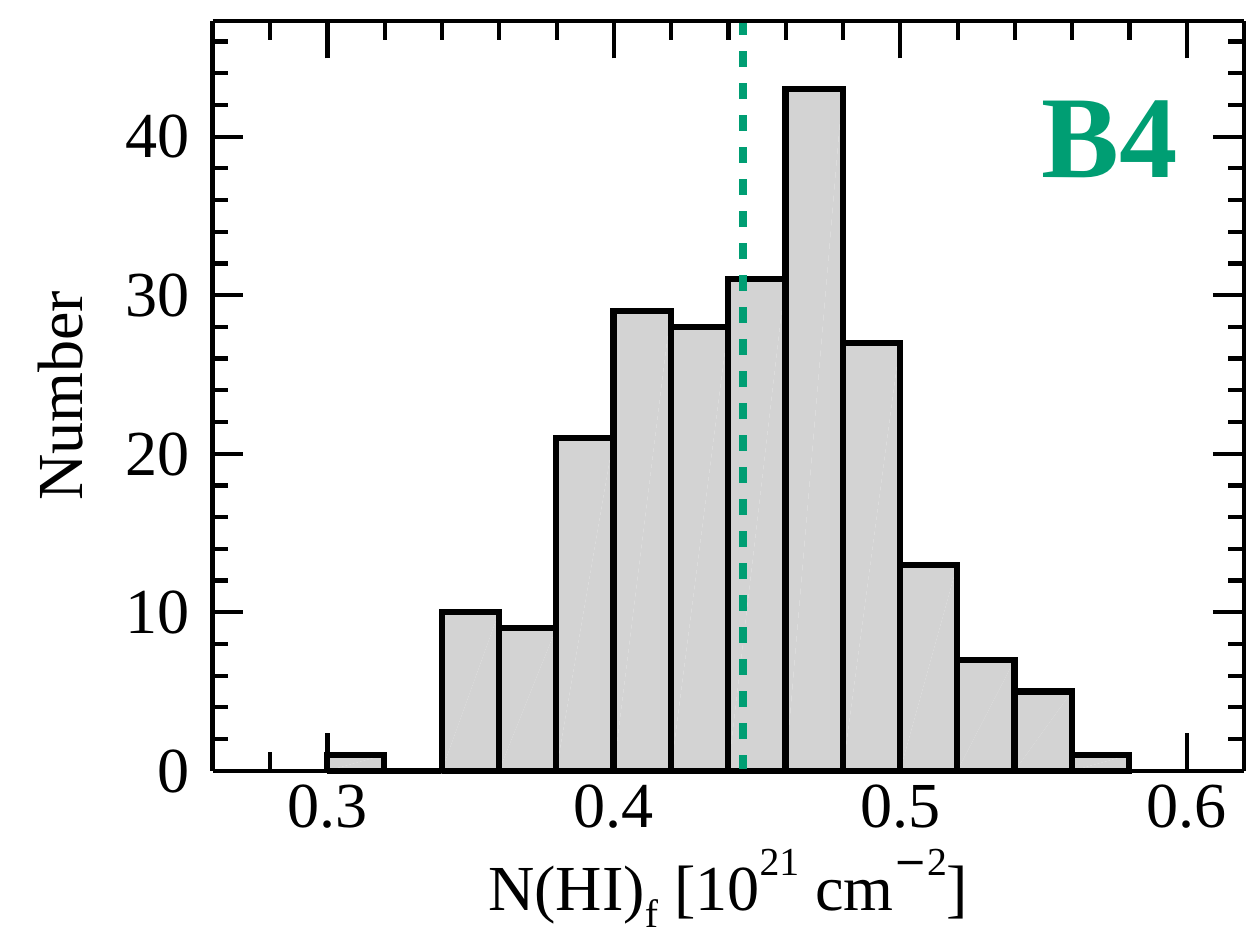}
    \caption{\label{f:hist_NHI} Same as Figure~\ref{f:hist_avg_tspin}, but for the opacity-corrected total \HI column densities.}
\end{figure*}

\subsubsection{Opacity-corrected Total \HI Column Density} \label{s:result_nhi}

To compute the opacity-corrected total \HI column density for the foreground gas that is seen in absorption, 
we first estimated the correction factor $f_{\textrm{chan}}(\varv)$ 
with the one-phase approximation (each channel represents gas at a single temperature) 

\begin{equation} \label{eq:f_chan}
    f_{\textrm{chan}}(\varv) = \frac{\tau(\varv)}{1-e^{-\tau(\varv)}}
\end{equation}

\noindent and applied it to the 30$''$-scale emission-only cube with $C = 1.823 \times 10^{18}$ and $q = 0.5$ as follows \citep{dickey2000}: 

\begin{equation} \label{eq:N_HI}
	N(\textrm{H~\textsc{i}})_{\textrm{f}}~({\textrm{cm}^{-2}})= C q \int f_{\textrm{chan}}(\varv)T_{\textrm{em-only}}(\varv)d\varv~\left({\textrm{K~km~s}^{-1}}\right). 
\end{equation}

\noindent 
We performed this calculation for the entire velocity range, as well as for the individual bands (B1 to B4), 
and selected the pixels whose $N$(H~\textsc{i})$_{\textrm{f}}$ values are larger than the 3$\sigma$ uncertainties in $N$(H~\textsc{i})$_{\textrm{f}}$. 
The results for the selected pixels are summarized in Figure~\ref{f:hist_NHI} and Table~\ref{t:band_info}.

The total \HI column densities calculated for the entire velocity range
vary from 2.5 to 3.8~$\times$~10$^{21}$~cm$^{-2}$ with a median of 3.0~$\times$~10$^{21}$~cm$^{-2}$. 
This small variation in $N$(H~\textsc{i})$_{\textrm{f}}$ is consistent with what \citet{roman-duval2014} found from 1$'$-scale \HI observations of the LMC 
and is mainly driven by B2 and B3 whose \HI abundances dominate the observed LOSs (median $N$(H~\textsc{i})$_{\textrm{f}}$ of (1.0--1.5)~$\times$~$10^{21}$~cm$^{-2}$). 
Similarly, B1 and B4 show a factor of two or so variation in $N$(H~\textsc{i})$_{\textrm{f}}$, 
although the values are systematically lower than that for B2 (median $N$(H~\textsc{i})$_{\textrm{f}}$ of (1.3--4.4)~$\times$~10$^{20}$~cm$^{-2}$). 
The saturation in $N$(H~\textsc{i})$_{\textrm{f}}$ for B1--B4 indicates the presence of \HI shielding layers for H$_{2}$ formation \citep[e.g.,][]{lee2012,lee2015}, 
and we will discuss further the conditions for the H~\textsc{i}-to-H$_{2}$ transition in 30 Dor in Section~\ref{s:HI-to-H2}.

In addition to $N$(H~\textsc{i})$_{\textrm{f}}$, 
we calculated the optically-thin \HI column density along the whole LOS that includes both the foreground and background 

\begin{equation} \label{eq:N_HI_unc}
	N(\textrm{H~\textsc{i}})_{\textrm{unc}}~({\textrm{cm}^{-2}})= C \int T_{\textrm{em-only}}(\varv)d\varv~\left({\textrm{K~km~s}^{-1}}\right)
\end{equation}

\noindent using the entire velocity range and the individual bands (B1--B4). 
While summarizing the results in Table~\ref{t:band_info} for the pixels whose values exceed the 3$\sigma$ uncertainties in $N$(H~\textsc{i})$_{\textrm{unc}}$, 
we do not discuss them further since these \HI column densities are not corrected for optically thick \HI 
and are therefore lower limits on the true total \HI column densities.

\subsection{Comparison of Global CNM, [C~\textsc{ii}], and CO Distributions} \label{s:CNM_CII_CO_spec} 
 
\begin{figure}
    \centering
    \includegraphics[scale=0.4]{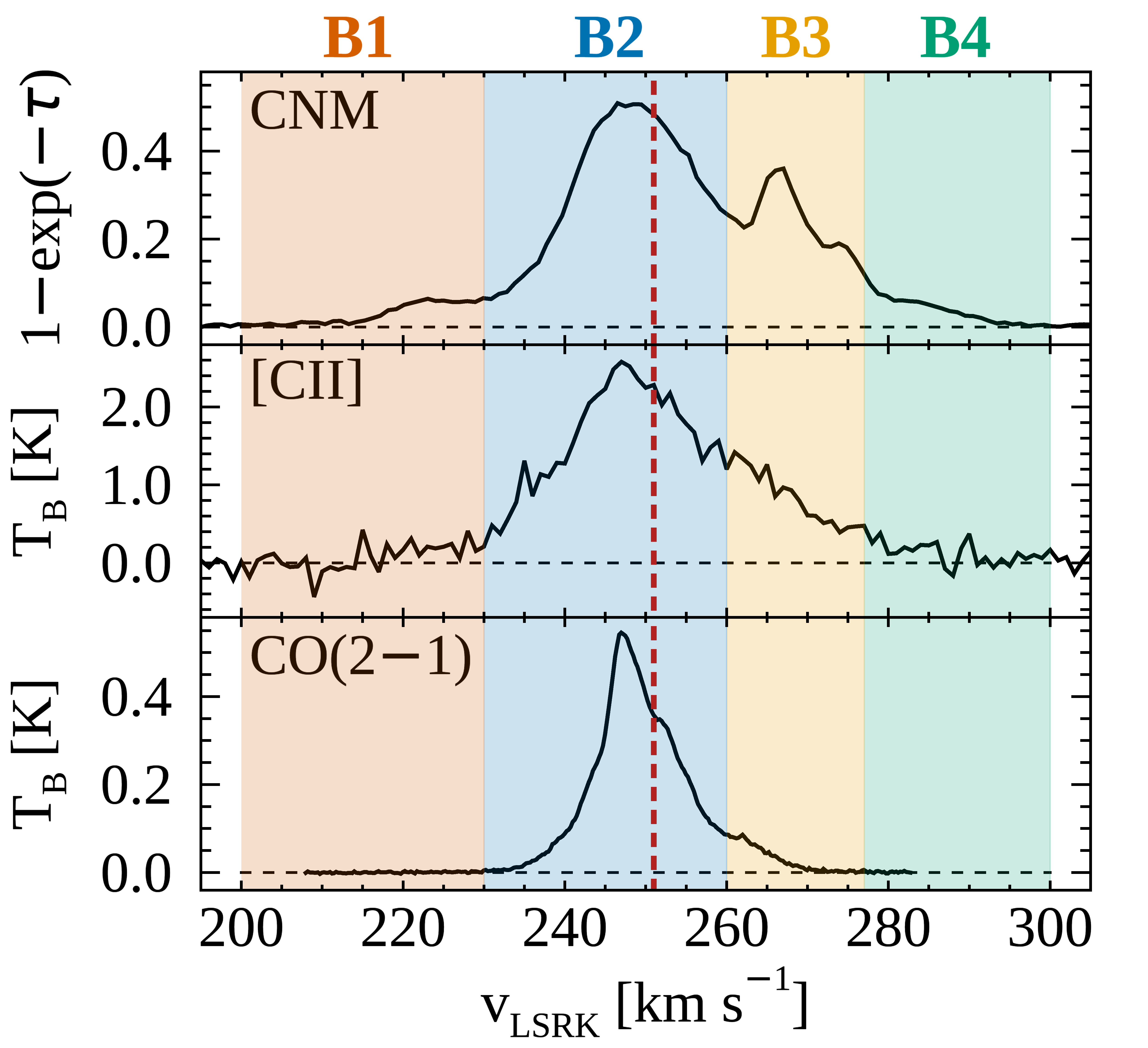}
	\caption{\label{f:HI_CO21_CII_spec} Comparison of the 1/$\sigma^{2}$-weighted average CNM, [C~\textsc{ii}], and CO spectra (30$''$ scales). 
	The four velocity bands are indicated in different colors, 
	{and the central velocity of H$\alpha$ emission is marked as the red dashed line (Section~\ref{s:multi-phase_structure}).}} 
\end{figure}

Finally, we compared the global distribution of the CNM to \CII and CO emission 
to glimpse into the environments traced by the CNM. 
To do so, we selected detections based on a 3$\sigma$ threshold {for each tracer}  
($\sigma$ = {pixel-by-pixel} statistical error from off-line channels)
and produced 1/$\sigma^{2}$-weighted average spectra for the CNM, [C~\textsc{ii}], and CO. 
A comparison of these average spectra is presented in Figure~\ref{f:HI_CO21_CII_spec}.

Figure~\ref{f:HI_CO21_CII_spec} shows that the three tracers are strongest in B2, 
suggesting that the main dense structure in 30 Dor is located at B2 velocities. 
Compared to the CNM, \CII and CO emission appear over a narrower range of velocities (B2 and B3) 
and share relatively simple spectral shapes.
This similarity between \CII and CO emission is consistent with what has been observed in the solar neighborhood \citep{hall2020}. 



\begin{deluxetable*}{l c c c c c c}
\centering
\tablecaption{\label{t:band_info} \HI Properties at the Individual Velocity Bands}
    \tablewidth{0pt}
    \setlength{\tabcolsep}{15pt}
    \tabletypesize{\small}
    \tablehead{
    \colhead{Band} & \colhead{$\varv_{\textrm{LSRK}}$} & \colhead{\# of Pixels} & \colhead{Fitted $1 - e^{-\tau}$} & 
    \colhead{<$T_{\textrm{s}}$>} & \colhead{$N(\textrm{H~\textsc{i}})_{\textrm{f}}$} & \colhead{$N(\textrm{H~\textsc{i}})_{\textrm{unc}}$} \\
    \colhead{} & \colhead{(km~s$^{-1}$)} & \colhead{} & \colhead{} & \colhead{(K)} & \colhead{(10$^{20}$~cm$^{-2}$)} & \colhead{(10$^{20}$~cm$^{-2}$)} \\
    \colhead{(1)} & \colhead{(2)} & \colhead{(3)} & \colhead{(4)} & \colhead{(5)} & \colhead{(6)} & \colhead{(7)} }  
    \startdata
     B1 & 200--230 & (116, 46, 112, 109) & 0.03--0.17, 0.08 (0.01) & 31.0--138.4, 80.6 (19.1) & 0.9--2.1, 1.3 (0.3) & 1.8--4.2, 2.6 (0.6) \\ 
     B2 & 230--260 & (225, 225, 225, 225) & 0.04--0.97, 0.39 (0.01) & 46.6--164.7, 69.9 (6.3) & 11.6--21.5, 14.6 (1.1) & 19.4--27.1, 23.4 (1.6) \\ 
     B3 & 260--277 & (210, 223, 225, 225) & 0.04--0.61, 0.26 (0.01) & 63.9--265.5, 112.1 (11.0) & 8.7--11.3, 10.0 (0.6) & 15.3--19.5, 17.3 (1.0) \\
     B4 & 277--300 & (99, 118, 225, 225) & 0.02--0.14, 0.05 (0.01) & 105.6--558.7, 200.5 (49.8) & 3.0--5.6, 4.4 (0.4) & {6.0--10.5, 8.7 (0.8)} \\ 
     Entire & 200--300 & (225, 225, 225, 225) & 0.02--0.97, 0.24 (0.01) & 63.9--207.8, 92.2 (5.6) & 24.6--38.3, 30.3 (1.3) & {42.2--59.7, 51.6} (2.2) \\ 
    \enddata
    \tablecomments{\small (1) Division of the velocities for our analyses; 
                                    (2) Velocity range;
                                    (3) Numbers of the pixels that were used for the calculation of the peak $1 - e^{-\tau}$, 
				    <$T_{\textrm{s}}$>, $N$(H~\textsc{i})$_{\textrm{f}}$, and $N$(H~\textsc{i})$_{\textrm{unc}}$ are presented in parentheses;
				    (4) Peak value in the fitted $1 - e^{-\tau}$; 
				    (5) LOS average spin temperature; 
				    (6) Opacity-corrected total \HI column density for the foreground gas; 
				    (7) Optically-thin \HI column density for the whole LOS; 
				    (4, 5, 6, 7) Range and median value are listed.   
				    The number in each parentheses corresponds to the median value in the 1$\sigma$ uncertainties. 
				    (5, 6) $q = 0.5$ was adopted. 
                                   }
\end{deluxetable*}
                         
\section{Analyses} \label{s:analyses}

Our results so far suggest that the CNM structures at B1 to B4 velocities are physically distinct and trace different environments. 
In terms of the multi-phase structures {toward} 30 Dor, the CNM is overall more extended than \CII and CO emission, probing relatively diffuse gas. 
In this section, we will analyze closely the observed properties of the CNM along with supplementary data 
to investigate how the CNM changes in different environments.

\subsection{Spatial Distributions of the CNM Properties} \label{s:spatial_dist}

\begin{figure*}
    \centering
    \includegraphics[scale=0.18]{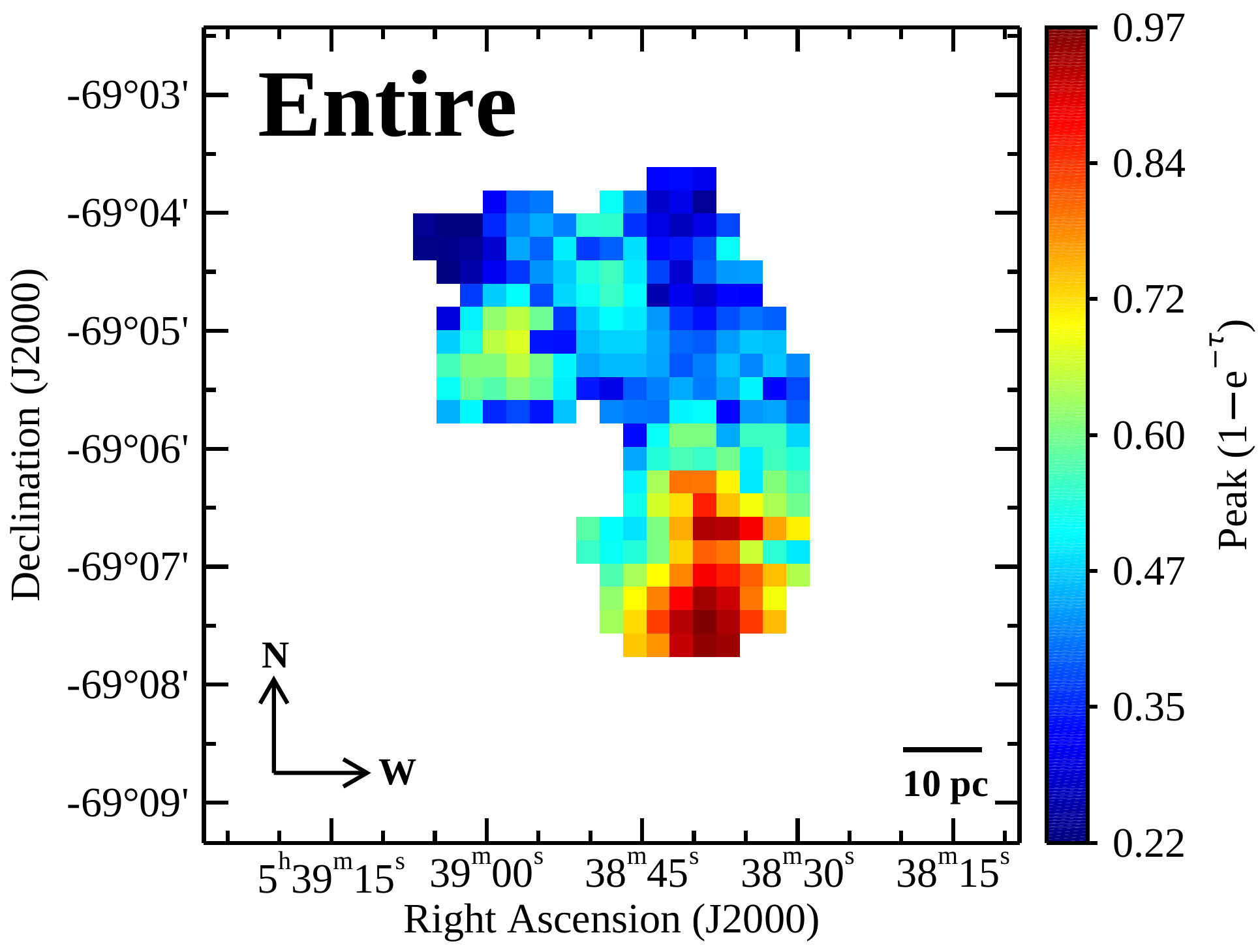}
    \includegraphics[scale=0.18]{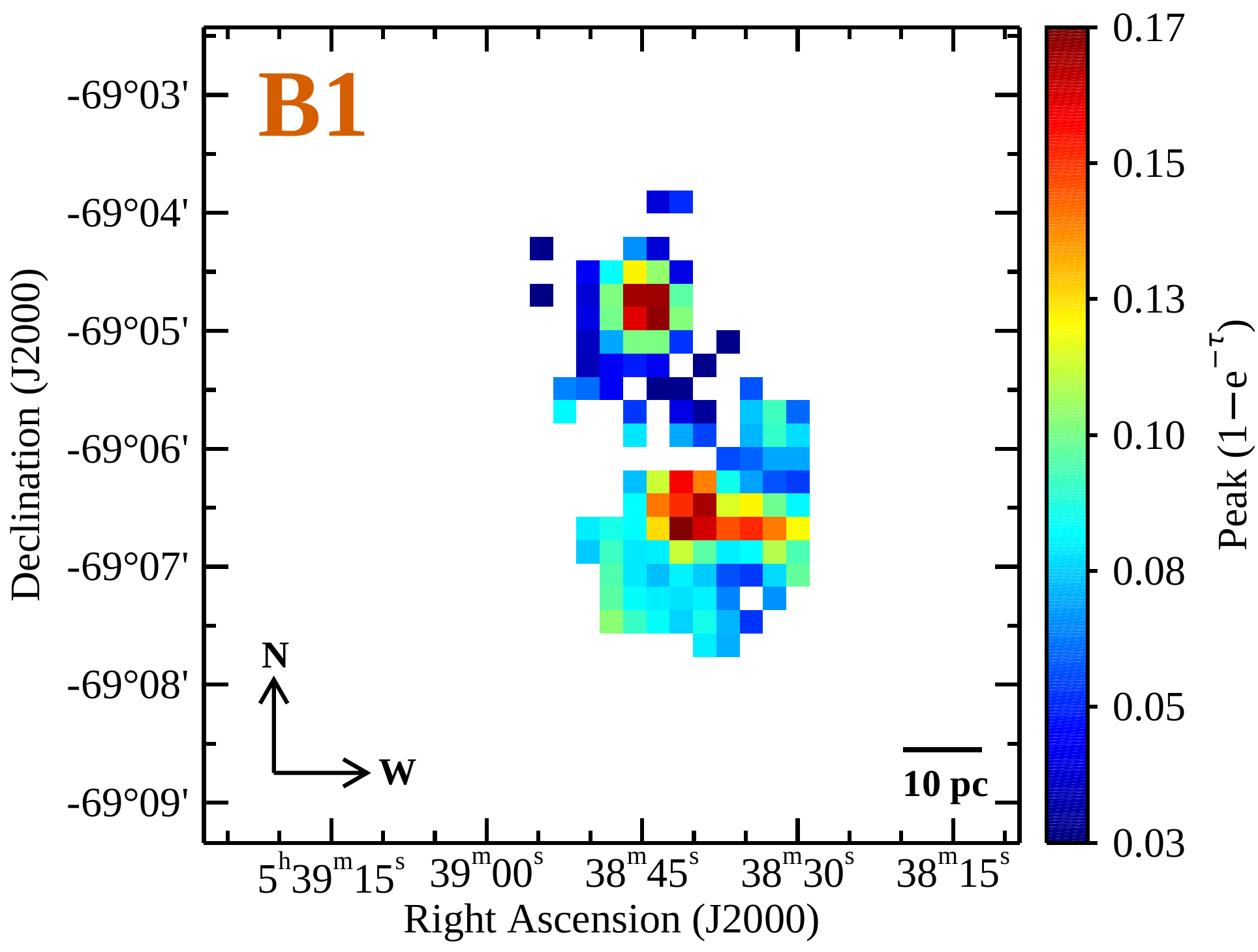}
    \includegraphics[scale=0.18]{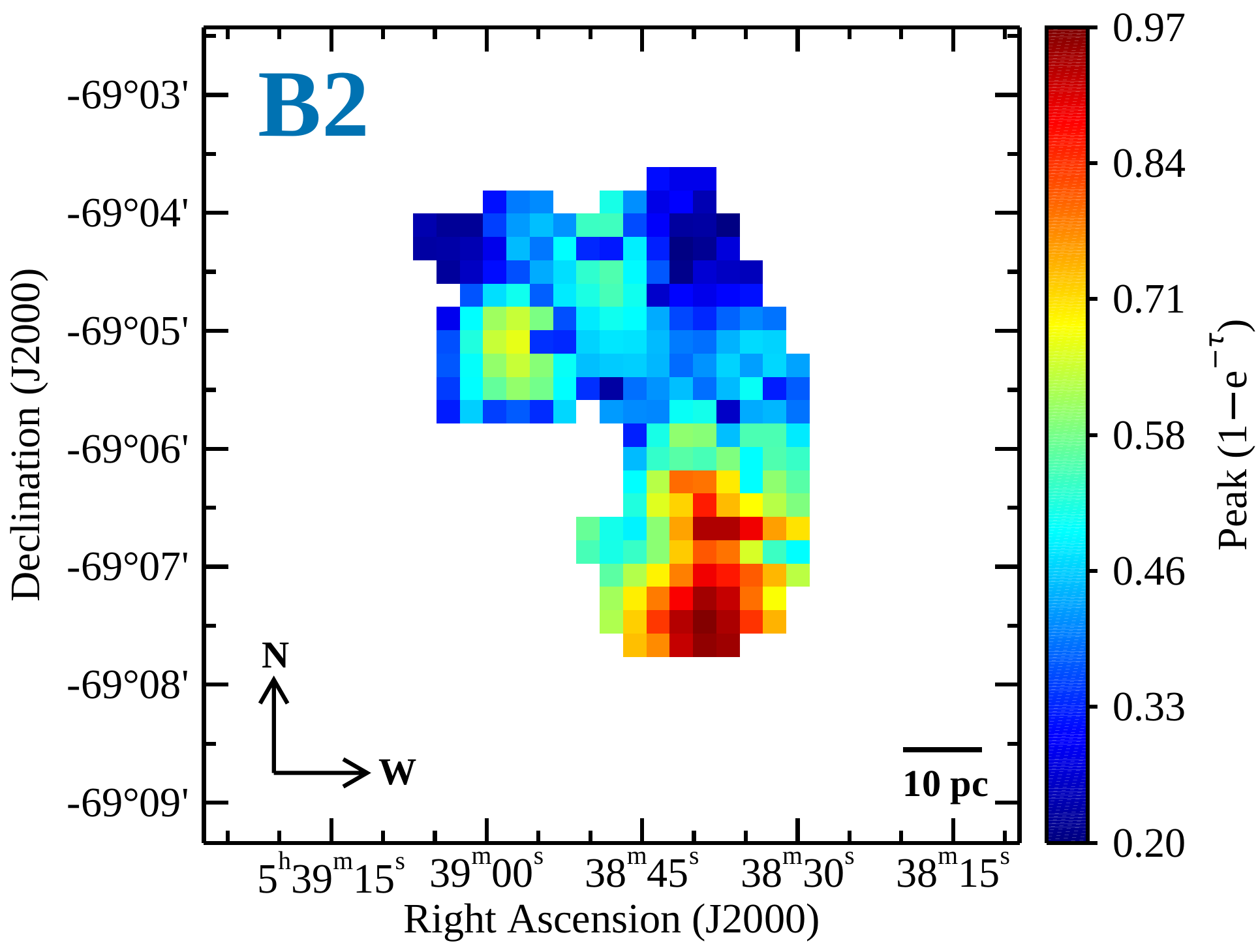}
    \includegraphics[scale=0.18]{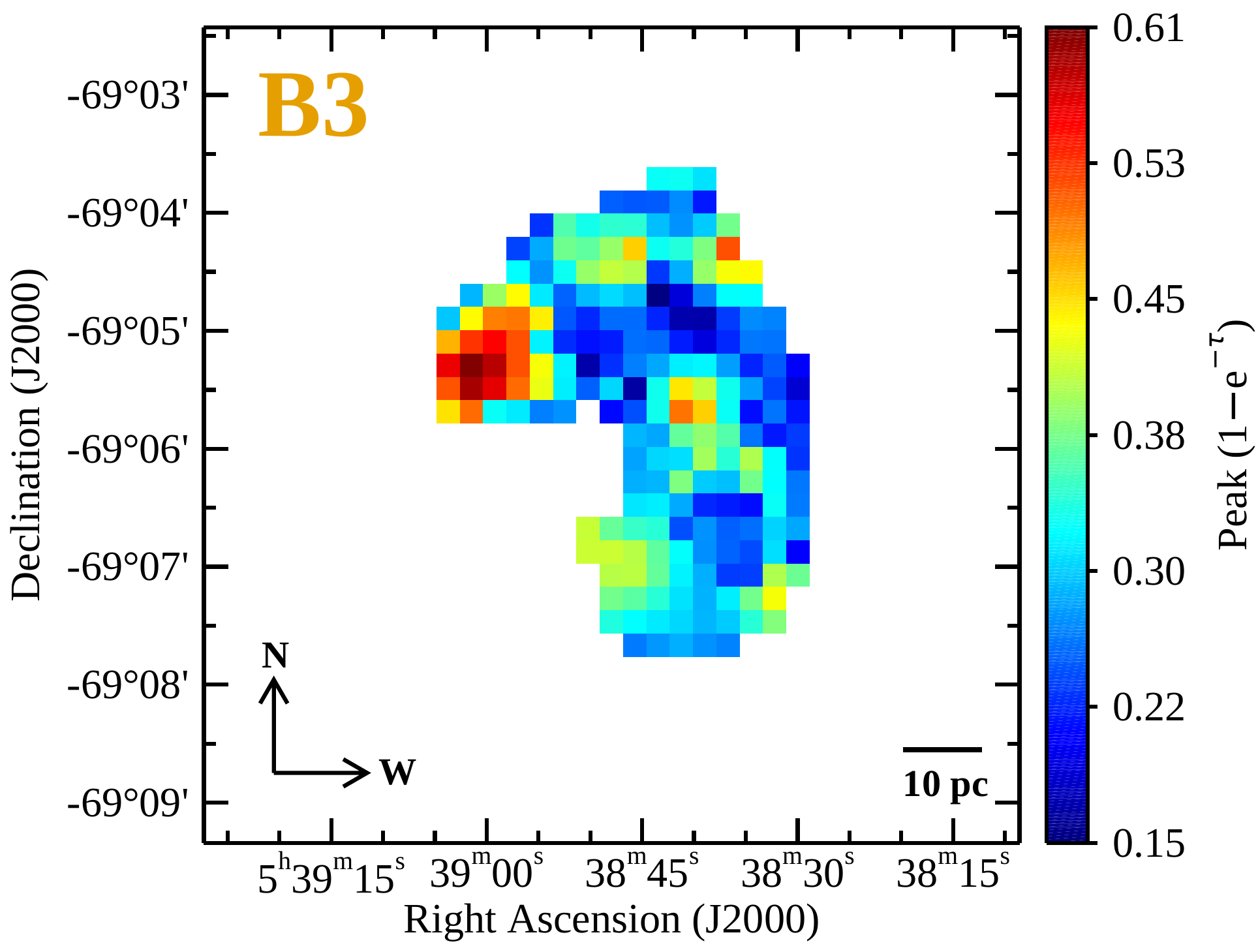}
    \includegraphics[scale=0.18]{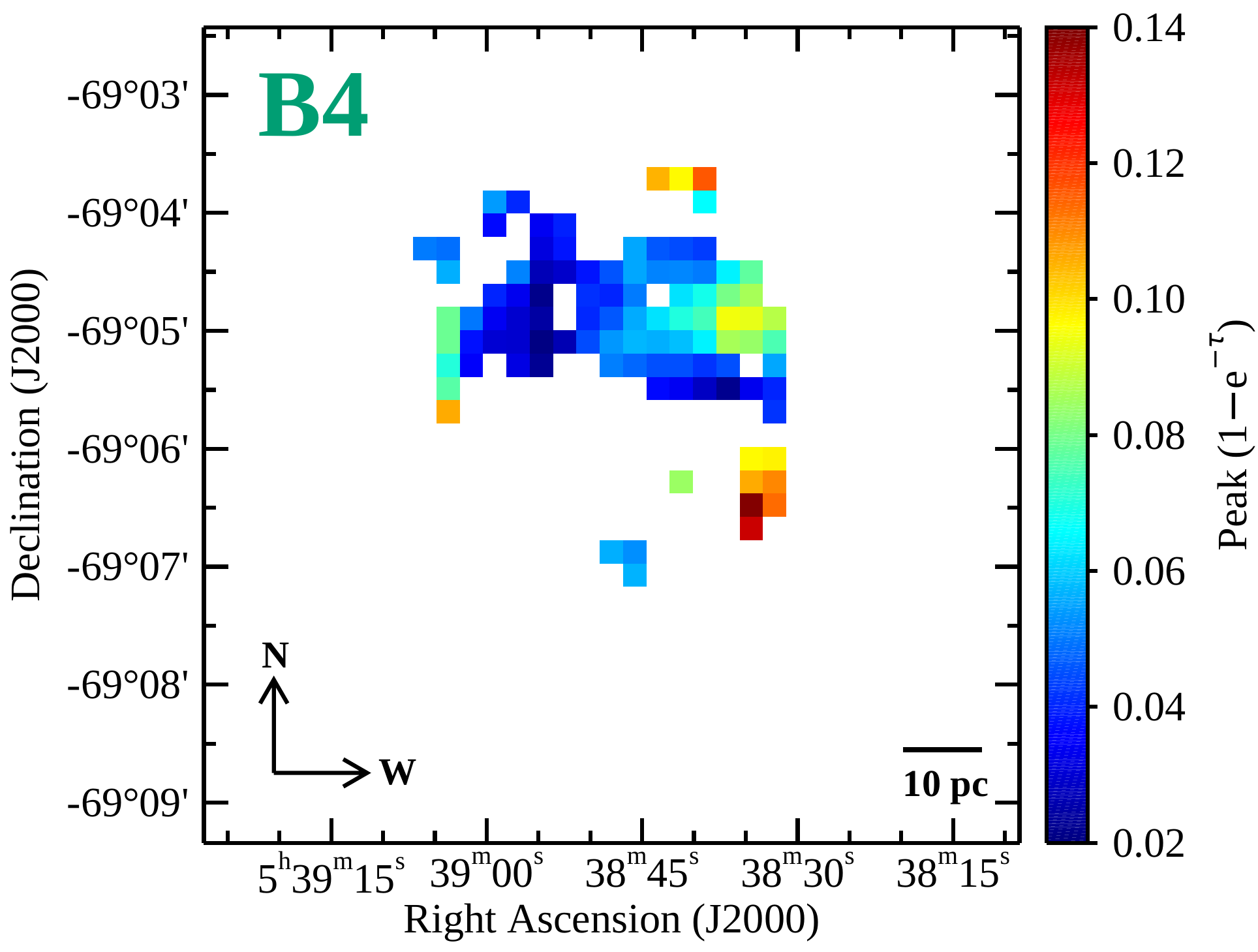}
    \caption{\label{f:peak_tau_per_band} Spatial distributions of the peak $1 - e^{-\tau}$ values for the entire velocity range (Entire) 
             and the individual bands (B1 to B4). The angular resolution of these maps is 30$''$.} 
\end{figure*}

\begin{figure*}
    \centering
    \includegraphics[scale=0.18]{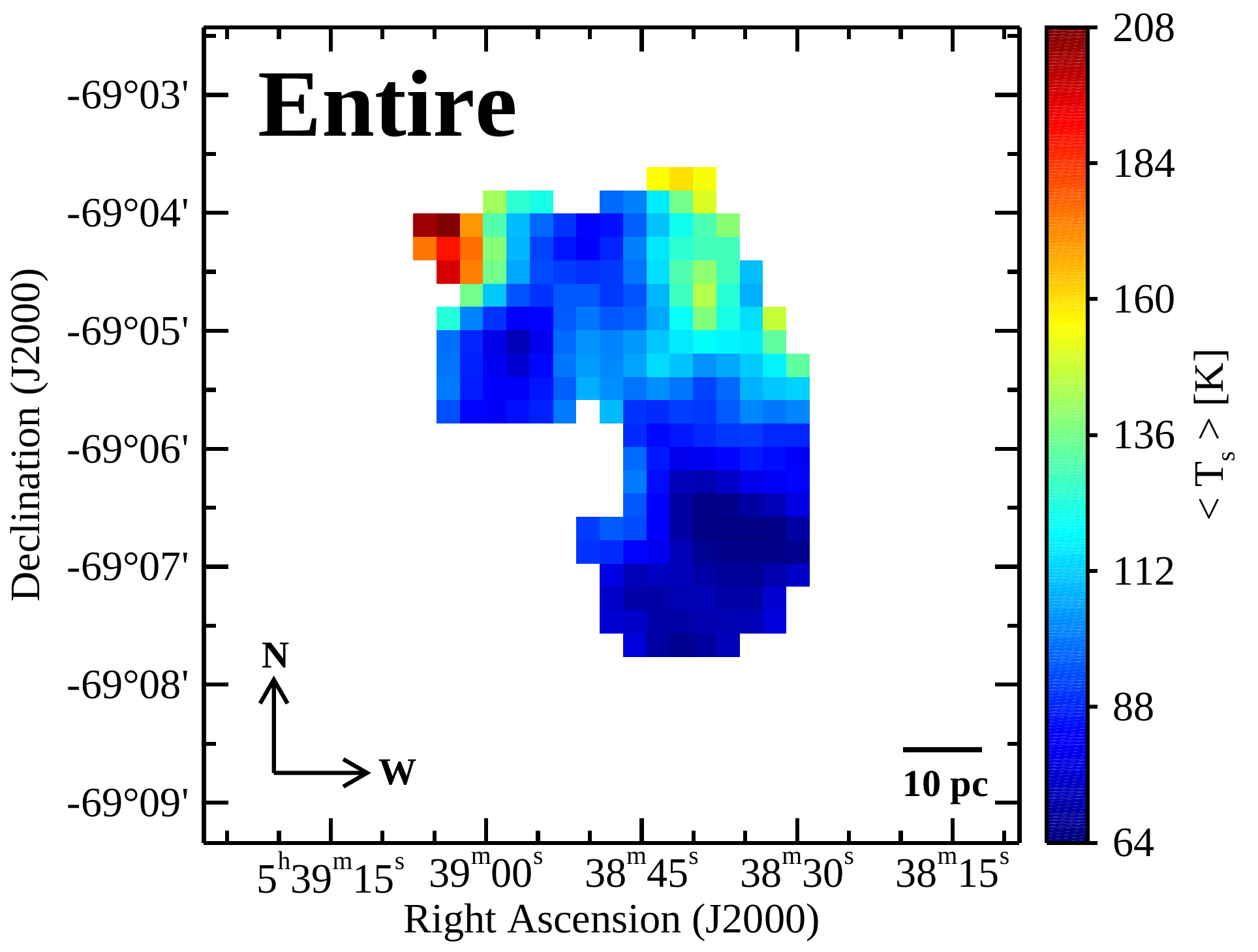}
    \includegraphics[scale=0.18]{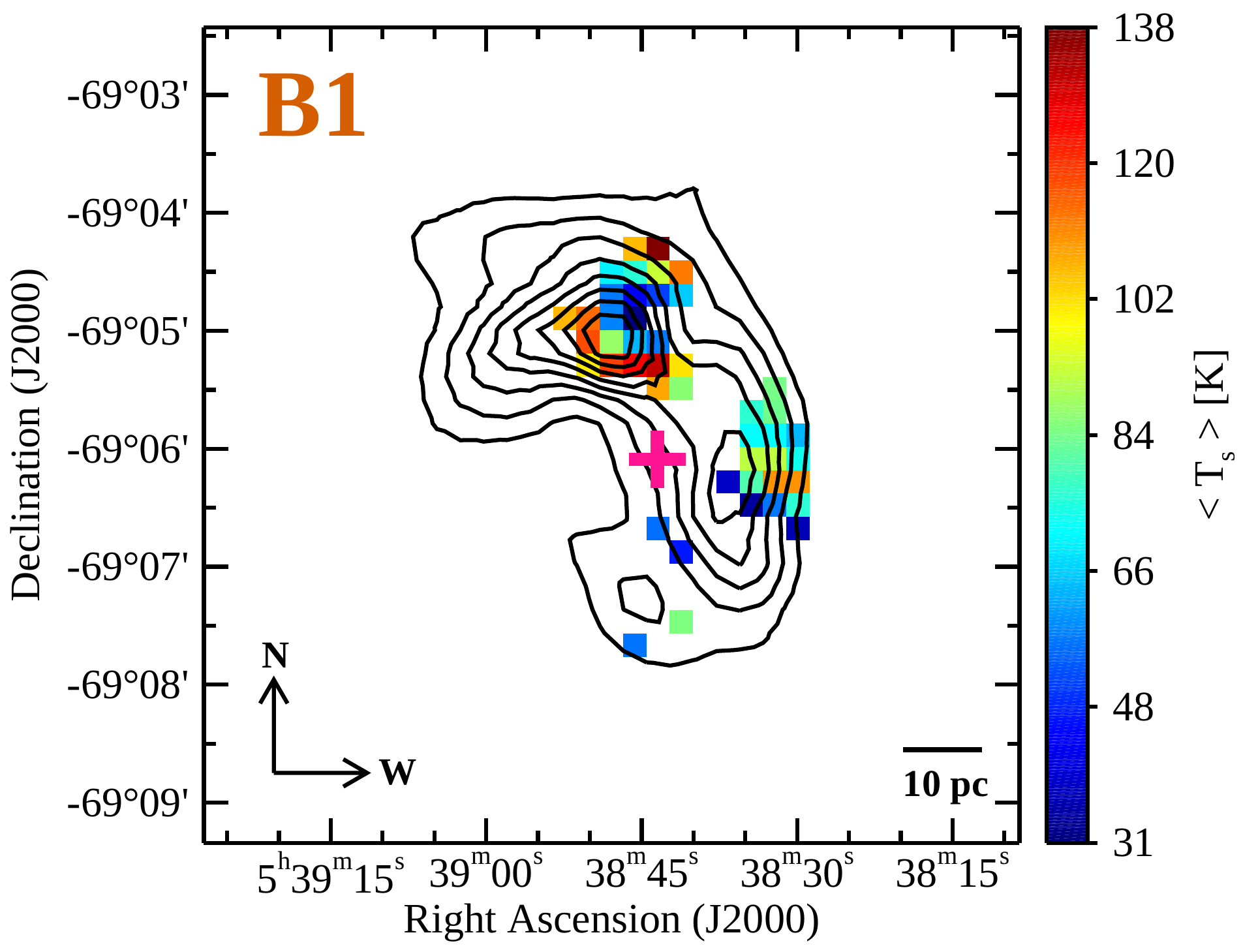} 
    \includegraphics[scale=0.18]{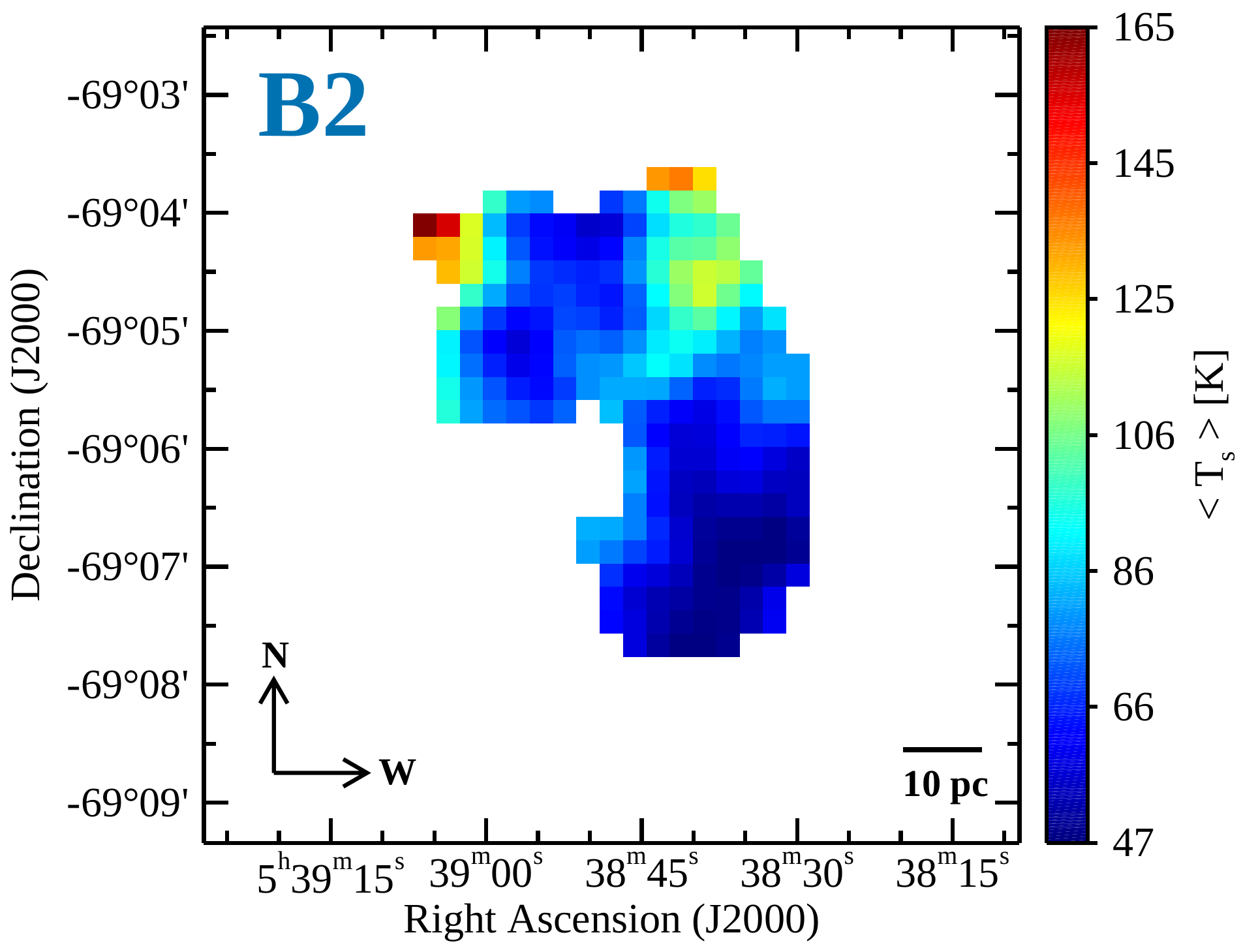}
    \includegraphics[scale=0.18]{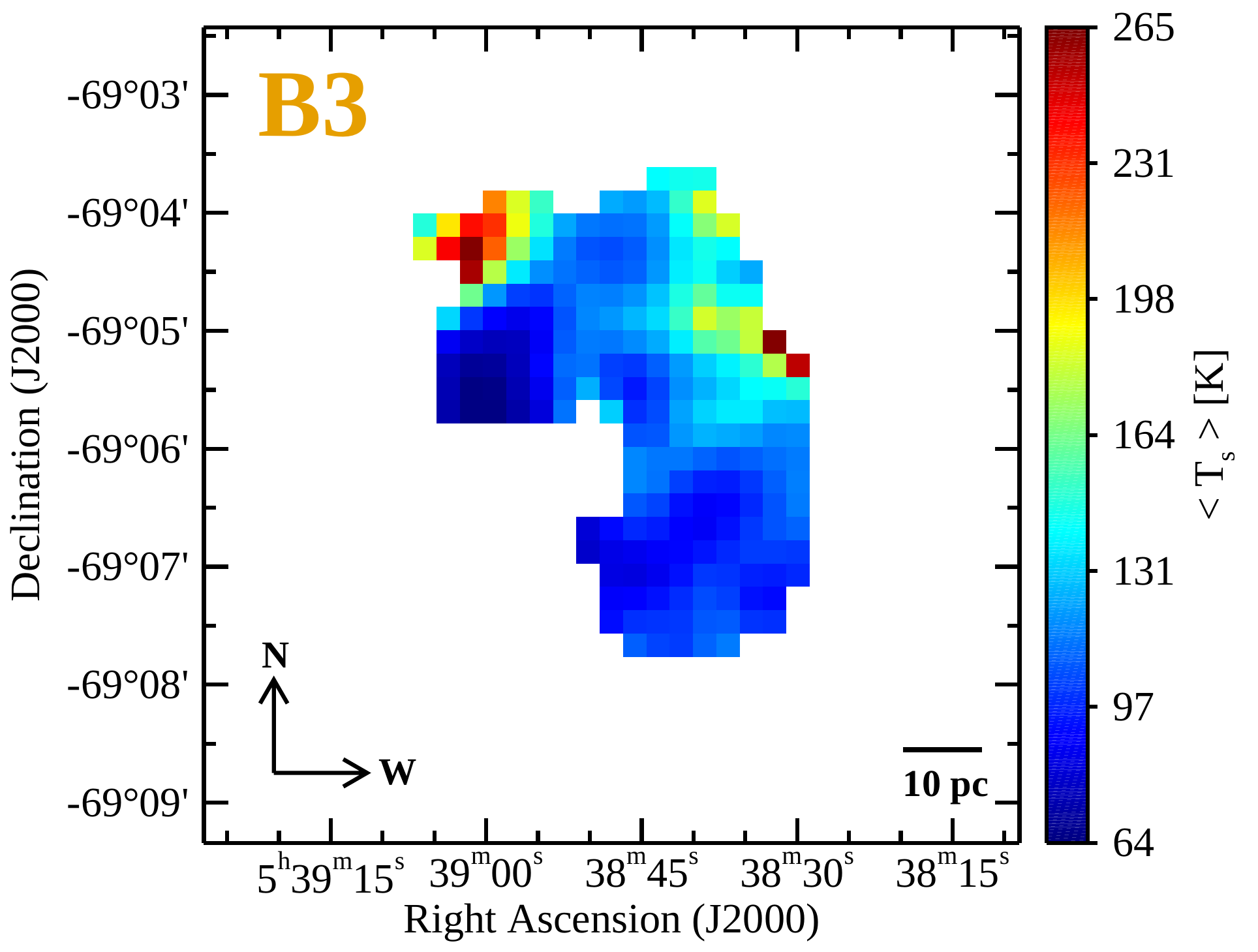}
    \includegraphics[scale=0.18]{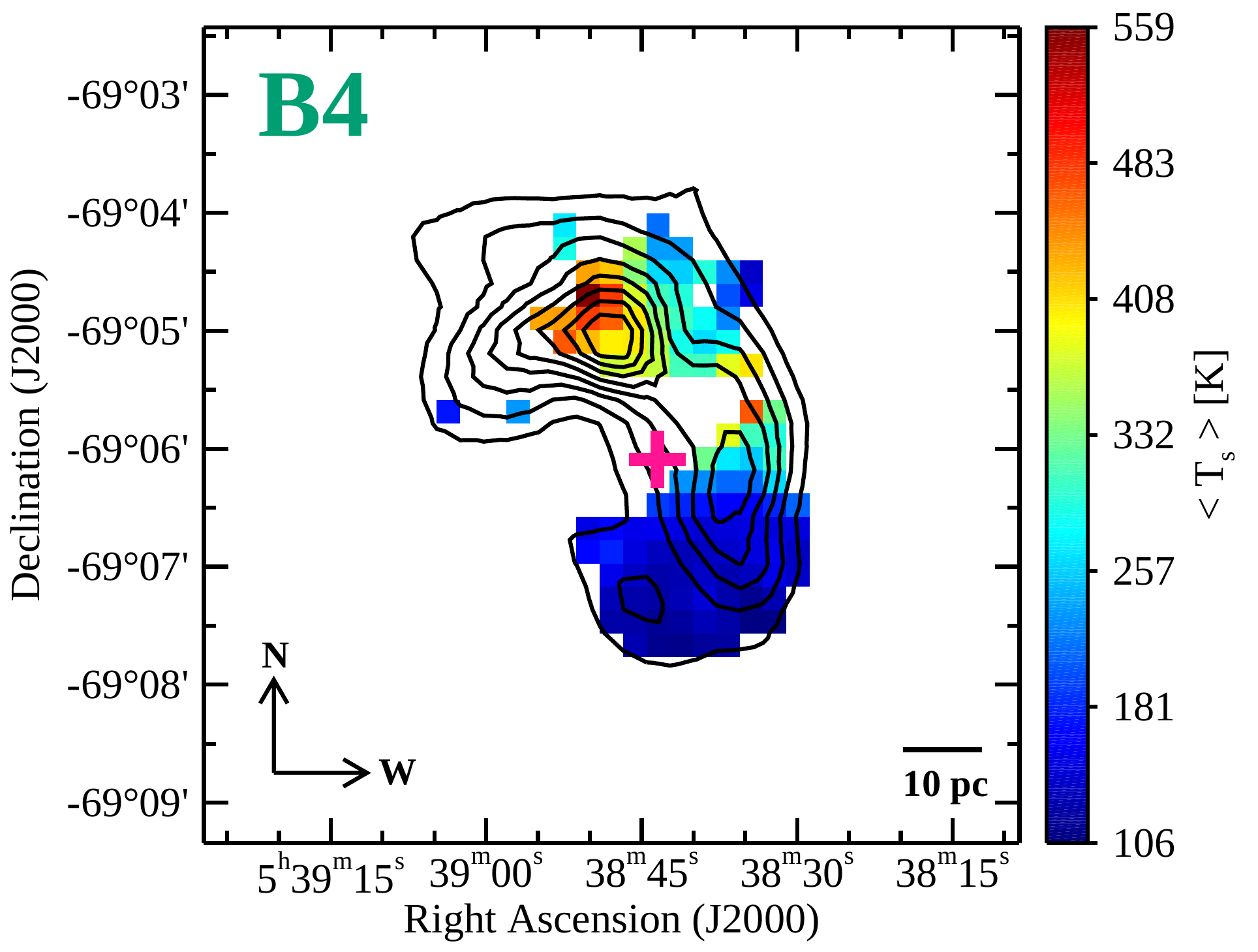} 
    \caption{\label{f:avg_tspin_map_all_pixels_per_band} Same as Figure~\ref{f:peak_tau_per_band}, but for <$T_{\textrm{s}}$>.
             For B1 and B4, the 1.4 GHz continuum data from the {high}-resolution cube are overlaid as the contours
             with levels ranging from 20\% to 90\% of the peak value ({448}~K) in steps of 10\%.
	     The location of R136 is indicated as the red cross.}
\end{figure*}

First, we probed the spatial distributions of the CNM properties
that were calculated for the entire velocity range, as well as for the B1--B4 bands 
(Figures~\ref{f:peak_tau_per_band}, \ref{f:avg_tspin_map_all_pixels_per_band}, and \ref{f:NHI_map_per_band}). 
For the three properties in our examination, peak $1 - e^{-\tau}$, <$T_{\textrm{s}}$>, and $N$(H~\textsc{i})$_{\textrm{f}}$,  
we found that the distributions for the entire velocity range resemble those for B2, 
hinting that the CNM structures at B2 velocities are the most prominent in 30 Dor.
In the following discussions, we focus on the comparison between the B1--B4 distributions. 


For B2 and B3, the absorption is detected over almost the entire region of the strong continuum,    
with the peak $1 - e^{-\tau}$ changing significantly from 0.04 to 0.97 for B2 and from 0.04 to 0.61 for B3. 
In terms of the spatial distribution, B2 shows the highest optical depths in the south at 
$(\alpha, \delta)_{\textrm{J2000}}$~$\sim$~(05$^{\textrm{h}}$38$^{\textrm{m}}$39$^{\textrm{s}}$, $-$69$\degree$07$'$32$''$), 
while $1 - e^{-\tau} < 0.7$ in the north at $\delta > -69\degree06\arcmin$. 
On the other hand, B3 has the highest $1 - e^{-\tau}$ values in the northeastern region at 
$(\alpha, \delta)_{\textrm{J2000}}$~$\sim$~(05$^{\textrm{h}}$39$^{\textrm{m}}$01$^{\textrm{s}}$, $-$69$\degree$05$'$20$''$). 
The rest of B3 shows moderate absorption with $1 - e^{-\tau}$~$\sim$~0.3, 
with a few spots where $1 - e^{-\tau}$ reaches 0.5, e.g.,
$(\alpha, \delta)_{\textrm{J2000}}$~$\sim$~(05$^{\textrm{h}}$38$^{\textrm{m}}$41$^{\textrm{s}}$, $-$69$\degree$05$'$44$''$). 
In contrast to B2 and B3, the absorption is only partially detected for B1 and B4 with much less strength. 
Specifically, B1 shows mostly $1 - e^{-\tau} < 0.1$, 
except for two spots at $(\alpha, \delta)_{\textrm{J2000}}$~$\sim$~(05$^{\textrm{h}}$38$^{\textrm{m}}$43$^{\textrm{s}}$, $-$69$\degree$04$'$56$''$)
and (05$^{\textrm{h}}$38$^{\textrm{m}}$41$^{\textrm{s}}$, $-$69$\degree$06$'$44$''$) where $1 - e^{-\tau}$~$\sim$~0.17. 
In B4, the absorption is primarily detected in the north at $\delta > -69\degree05\arcmin40\arcsec$ with $1 - e^{-\tau}$~$\sim$~0.02--0.1, 
except for a spot at $(\alpha, \delta)_{\textrm{J2000}}$~$\sim$~(05$^{\textrm{h}}$38$^{\textrm{m}}$34$^{\textrm{s}}$, $-$69$\degree$06$'$32$''$)
where $1 - e^{-\tau}$~$\sim$~0.14.


The above-mentioned structure in absorption mainly determines 
the distributions of the spin temperature and opacity-corrected total \HI columm density, 
since the emission varies much more smoothly than the absorption. 
In general, B2 has lower spin temperatures than B3, 
i.e., <$T_{\textrm{s}}$> primarily in the range of 50--75~K for B2 and 70--120~K for B3. 
Both B2 and B3 reveal higher spin temperatures along the northern edge of the continuum at 
$(\alpha, \delta)_{\textrm{J2000}}$~$\sim$~(05$^{\textrm{h}}$39$^{\textrm{m}}$04$^{\textrm{s}}$, $-$69$\degree$04$'$20$''$), 
with <$T_{\textrm{s}}$> reaching 165~K and 265~K for B2 and B3, respectively. 
Compared to {B3}, B1 and B4 have systematically lower and higher spin temperatures. 
For example, while <$T_{\textrm{s}}$> is measured only for a handful number of pixels, 
it mostly ranges from 30~K to 80~K for B1. 
On the other hand, <$T_{\textrm{s}}$> changes from 106~K to 559~K for B4, 
with the lowest values in the south at $\delta < -69\degree06\arcmin15\arcsec$. 
Interestingly, the lowest and highest spin temperatures for B1 and B4 are found around the continuum peak, 
implying that the expanding \HII region somehow plays a role in the heating and cooling of the cold \HI. 


In terms of the opacity-corrected \HI column density, B2 is the most dominant structure 
with $N$(H~\textsc{i})$_{\textrm{f}}$ increasing from 1.2 to 2.2 $\times$ 10$^{21}$ cm$^{-2}$ 
toward the south at $\delta < -69\degree06\arcmin$. 
In B3, the \HI column density distribution becomes much more uniform with $N$(H~\textsc{i})$_{\textrm{f}}$ $\sim$ (0.9--1.1) $\times$ 10$^{21}$ cm$^{-2}$, 
and its highest value is found in a small region at 
$(\alpha, \delta)_{\textrm{J2000}}$~$\sim$~(05$^{\textrm{h}}$38$^{\textrm{m}}$46$^{\textrm{s}}$, $-$69$\degree$05$'$08$''$)
that corresponds to the peak of the continuum. 
The remaining B1 and B4 have systematically lower \HI column densities than B2 and B3. 
Specifically, B4 shows about half the \HI column density of B3, 
i.e., $N$(H~\textsc{i})$_{\textrm{f}}$ $\sim$ (0.3--0.6) $\times$ 10$^{21}$ cm$^{-2}$, 
with the higher values in the south at $\delta < -69\degree06\arcmin$. 
Similarly, B1 shows low \HI column densities of (0.1--0.2) $\times$ 10$^{21}$ cm$^{-2}$ for a small number of pixels. 
The spatial coverage for B1 is limited mainly because B1 is weak in both emission and absorption. 

\subsection{Multi-phase Structures {in and around} 30 Dor} \label{s:multi-phase_structure}

\begin{deluxetable}{l c c c}[t!]
\centering
\tablecaption{\label{t:CII_CO_table} \CII and CO Properties at B2 and B3 Velocities}
    \tablewidth{0pt}
    \setlength{\tabcolsep}{5pt}
    \tabletypesize{\small}
    \tablehead{
    \colhead{Band} & \colhead{\# of Pixels} & \colhead{$W$([C~\textsc{ii}]) (K~km~s$^{-1}$)} & \colhead{$W$(CO) (K~km~s$^{-1}$)} \\ 
    \colhead{(1)} & \colhead{(2)} & \colhead{(3)} & \colhead{(4)}}
    \startdata
     B2 & (376, 675) & $-$4.9, 226.9, 33.8 & 0.1, 40.5, 2.0 \\
     B3 & (376, 675) & $-$7.4, 63.9, 10.5 & 0.0, 3.6, 0.3 \\ 
    \enddata
    \tablecomments{\small(1) Velocity band; 
	(2) Numbers of the detected pixels that were considered for the $W$([C~\textsc{ii}]) and $W$(CO) statistics are presented in parentheses;
	(3, 4) Minimum, maximum, and median values are summarized.}
\end{deluxetable}
             
\begin{figure*}
    \centering
    \includegraphics[scale=0.18]{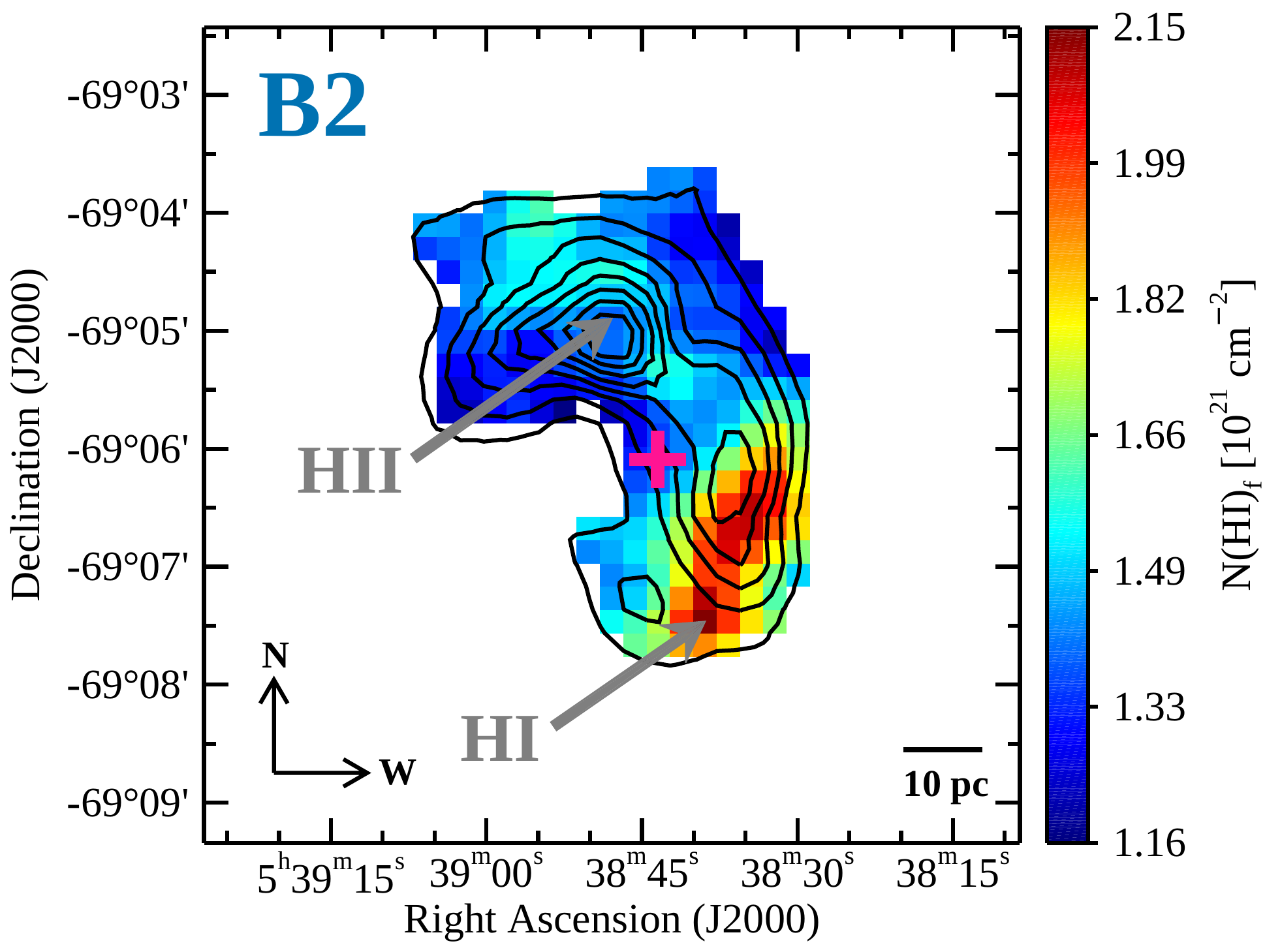}
    \includegraphics[scale=0.18]{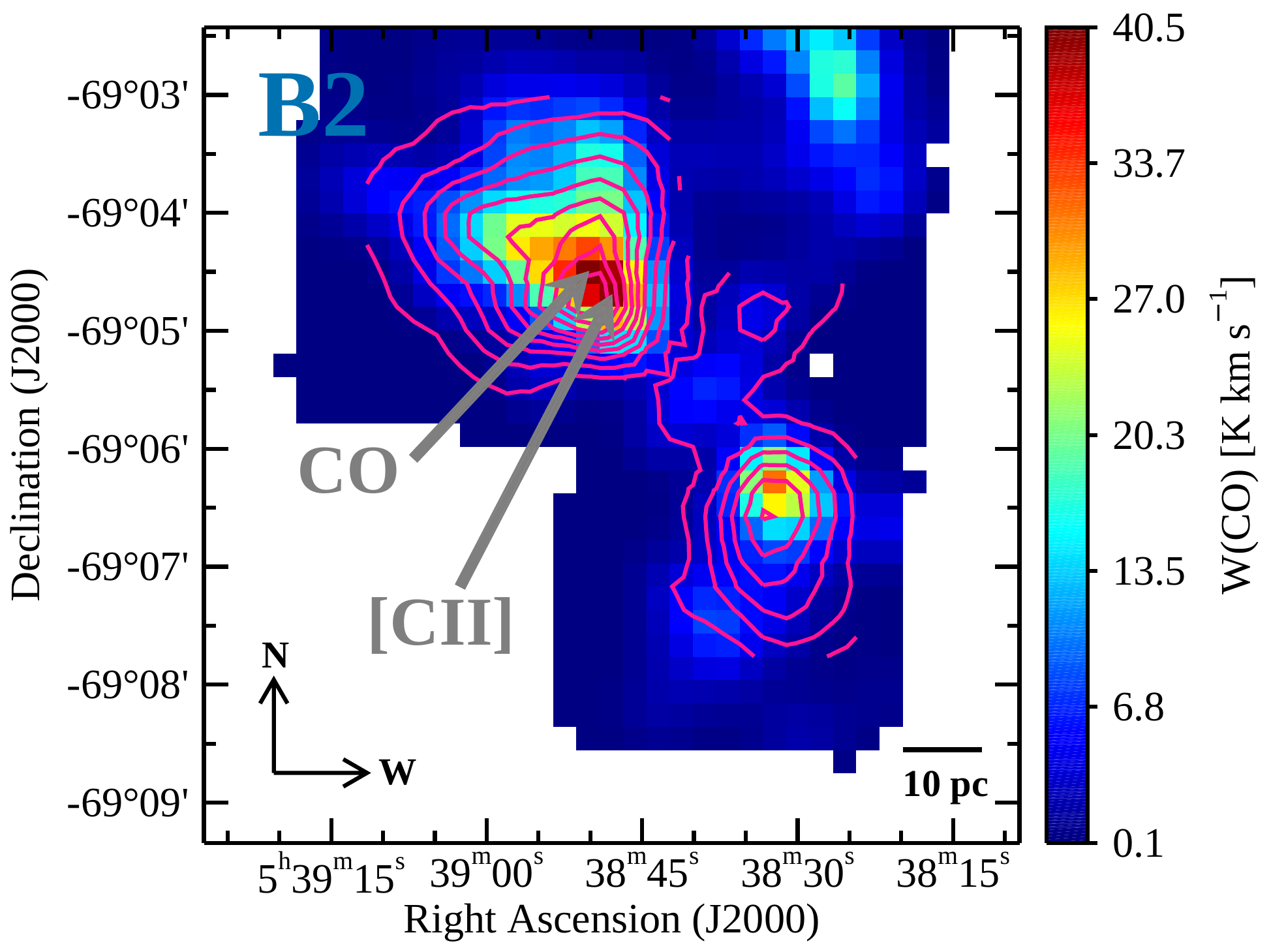}
    \includegraphics[scale=0.18]{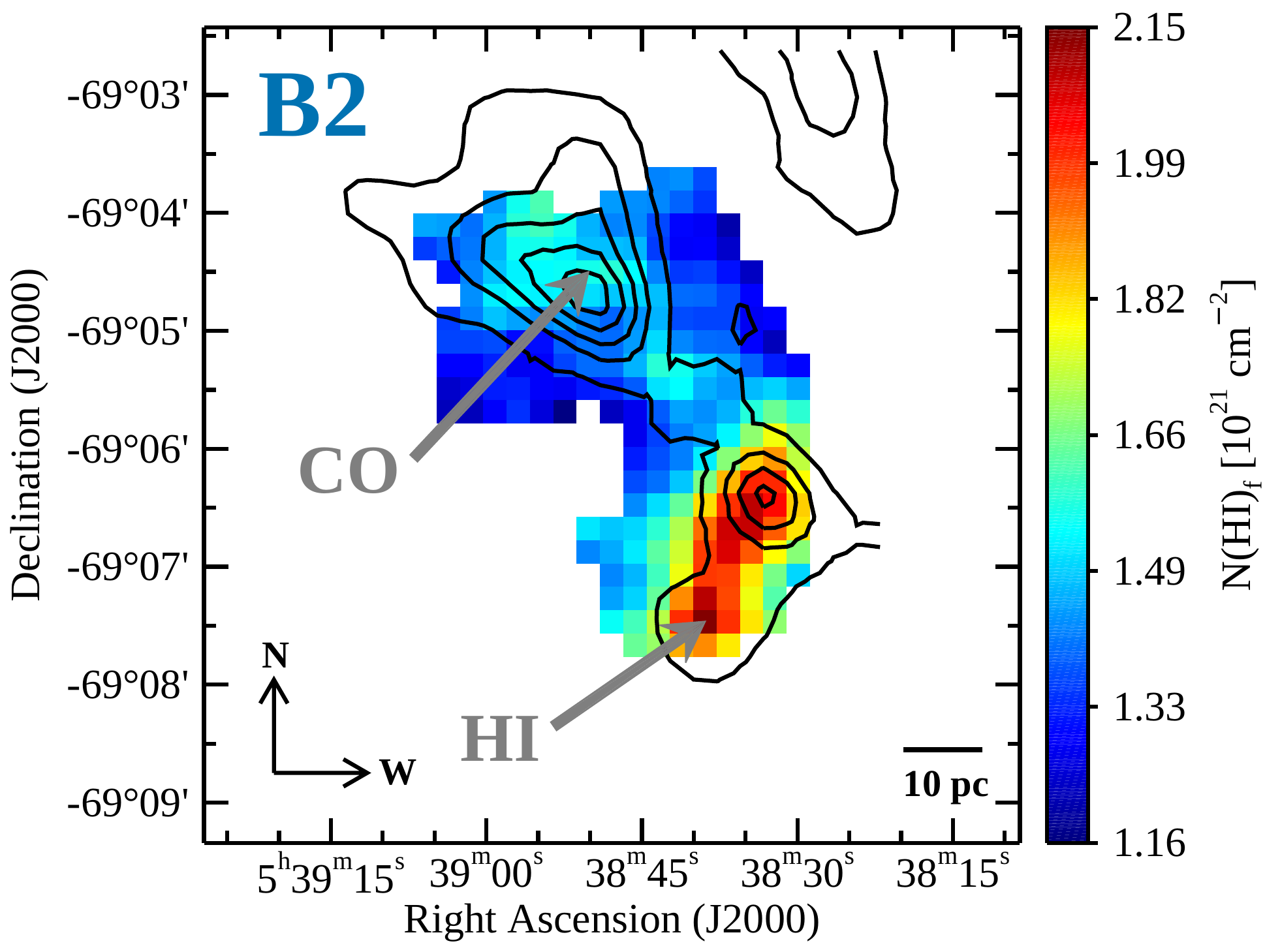}
    \includegraphics[scale=0.18]{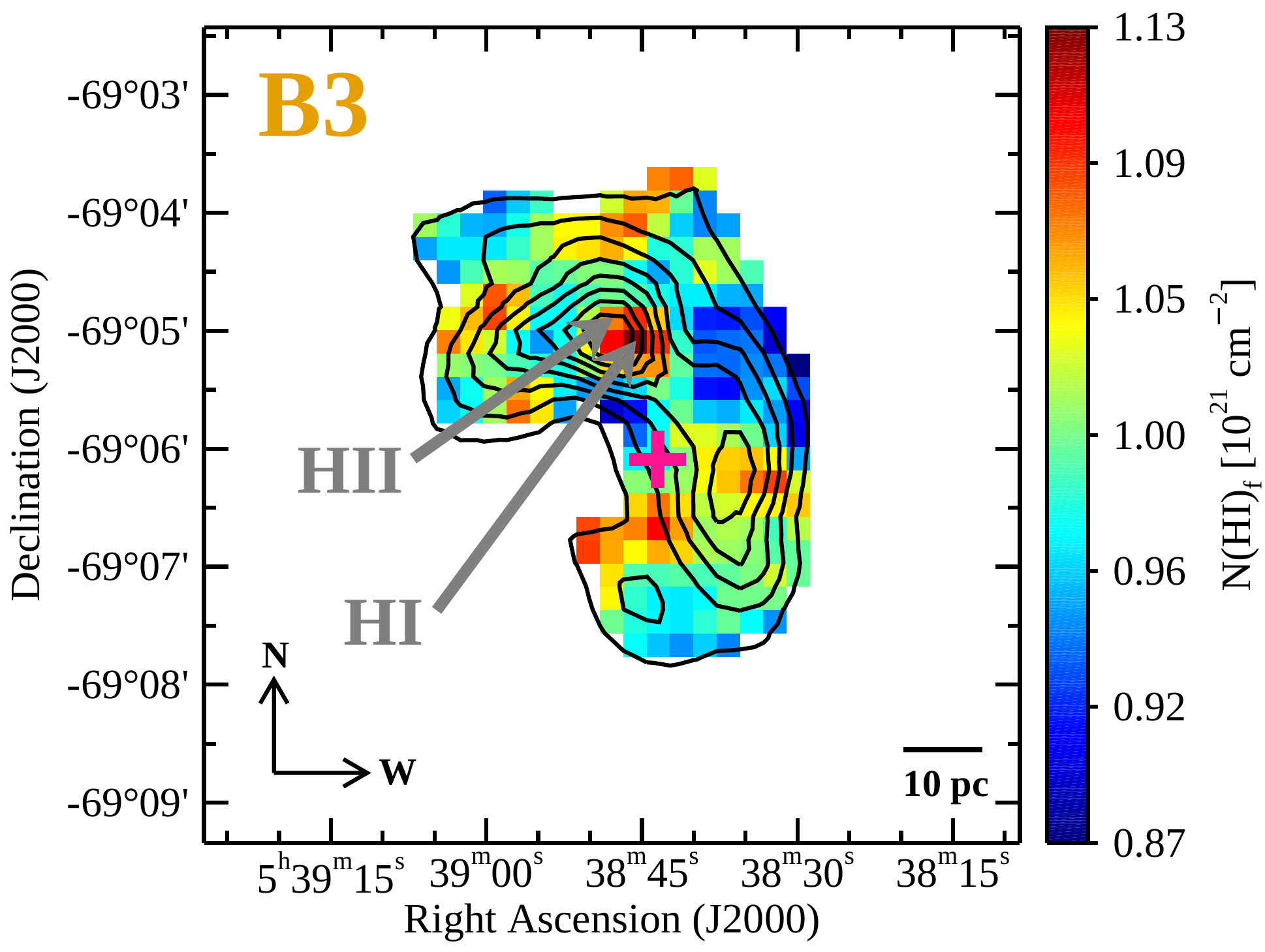}
    \includegraphics[scale=0.18]{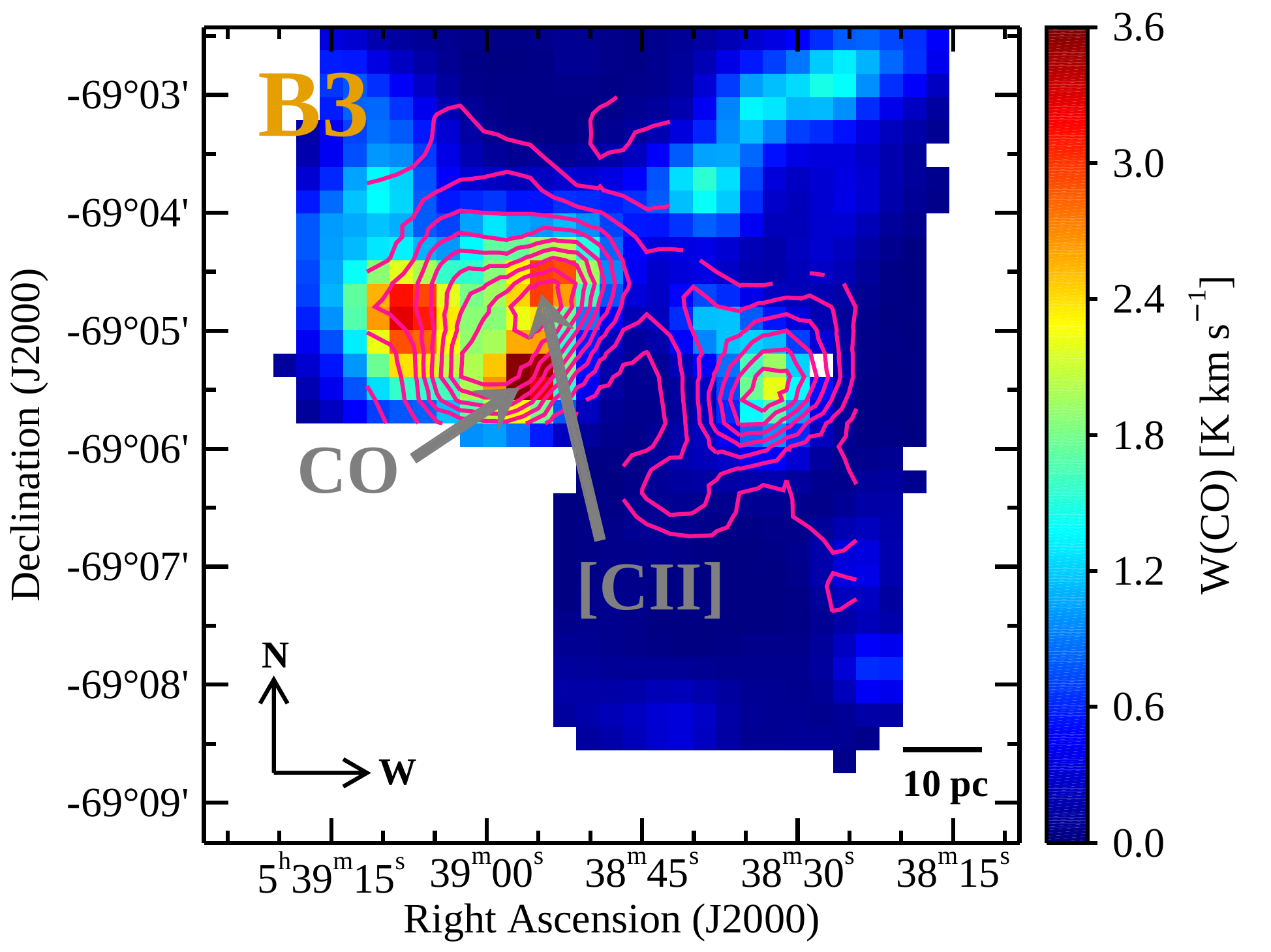}
    \includegraphics[scale=0.18]{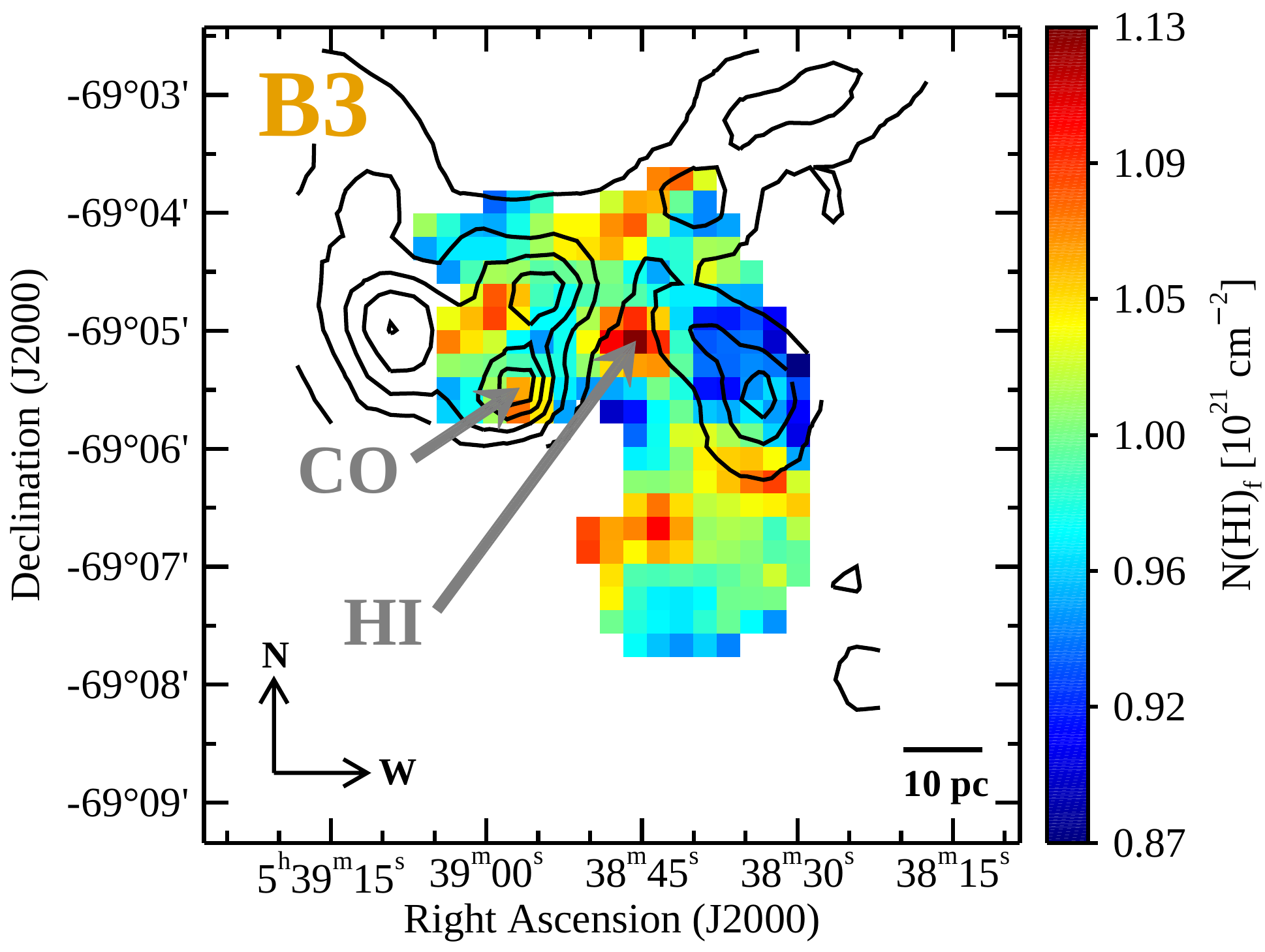}
	\caption{\label{f:multi-phase_compare} {Comparison of the multi-phase tracers for B2 and B3 (all tracers on 30$''$ scales).
	     The colorscale images show $N$(H~\textsc{i})$_{\textrm{f}}$ in the left and right panels and $W$(CO) in the middle panel.
             The 1.4 GHz continuum data are overlaid as the contours in the left panel with levels ranging from 20\% to 90\% of the peak value ({448}~K) in 10\% steps.
             Similarly, $W$([C~\textsc{ii}]) and $W$(CO) are shown as the contours in the middle and right panels, respectively,
             with levels ranging from 10\% to 90\% of the peak values in 10\% steps for [C~\textsc{ii}] and 20\% steps for CO.
             The peak values are 227~K~km~s$^{-1}$ and 64~K~km~s$^{-1}$ for [C~\textsc{ii}] in B2 and B3, respectively, 
	     and 40~K~km~s$^{-1}$ and 4~K~km~s$^{-1}$ for CO in B2 and B3, respectively.
             The location of R136 is shown as the red cross, and 
	     the peak of each tracer is indicated with the gray arrow.}}
\end{figure*}

To understand the context of the observed CNM structures, 
we examined the spatial distributions of multi-phase tracers {in and around} 30 Dor. 
For our examination, we focused on B2 and B3 and compared the H~\textsc{i}, [C~\textsc{ii}], and CO distributions 
along with the 1.4 GHz continuum (tracing the \HII region).
To derive the [C~\textsc{ii}] and CO distributions for B2 and B3, 
we computed the integrated intensities, $W$([C~\textsc{ii}]) and $W$(CO), by integrating the data cubes over the corresponding velocity ranges.
The comparison of the individual tracers is presented in Figure~\ref{f:multi-phase_compare} and Table~\ref{t:CII_CO_table}. 

{We first start our discussion from B2. 
Given that H$\alpha$ emisson peaks at 251~km~s$^{-1}$ in 30 Dor
(266~km~s$^{-1}$ in the heliocentric frame; \citealt{chu1994}), 
we probe the structures associated with the main \HII complex at B2 velocities. 
These structures are strong in H~\textsc{i}, CO, and [C~\textsc{ii}] as well (Section~\ref{s:CNM_CII_CO_spec}). 
When examined closely, H~\textsc{i}, CO, [C~\textsc{ii}], and 1.4 GHz continuum show largely two structures in the north and south. 
CO, [C~\textsc{ii}], and 1.4 GHz continuum are strongest in the north, 
and the CO and [C~\textsc{ii}] peaks are slightly offset from the 1.4 GHz continuum peak. 
On the other hand, the \HI peak surrounds the weaker CO and [C~\textsc{ii}] clumps in the south. 
These differences between the northern and southern structures in terms of the relative strengths in the ionized, atomic, and molecular gas tracers 
likely result from the variations in local conditions, such as the gas density and strength of UV radiation \citep[e.g.,][]{chevance2016}.}

{The multi-phase structures at B3 velocities are drastically different from those at B2 velocities.
For example, the \HI and 1.4 GHz continuum distributions do not have much similarity, except that their peaks coincide well. 
In addition, CO and [C~\textsc{ii}] show overall two structures in the west and east, which tend to surround multiple \HI clumps. 
While both B3 and B4 are inflows, B3 is systematically colder and denser than B4, 
as evidenced by the lower spin temperatures and higher \HI column densities (Section~\ref{s:tspin_calculation}), 
as well as by the presence of CO and [C~\textsc{ii}] emission. 
This suggests that B3 may be able to fuel star formation in 30 Dor and 
possbily has a different origin compared to B4 (Section~\ref{s:outflow_inflow} for details).} 

\subsection{Comparison to the Milky Way and the LMC} \label{s:MW_vs_30Dor}

\begin{figure*} 
    \centering 
    \includegraphics[scale=0.35]{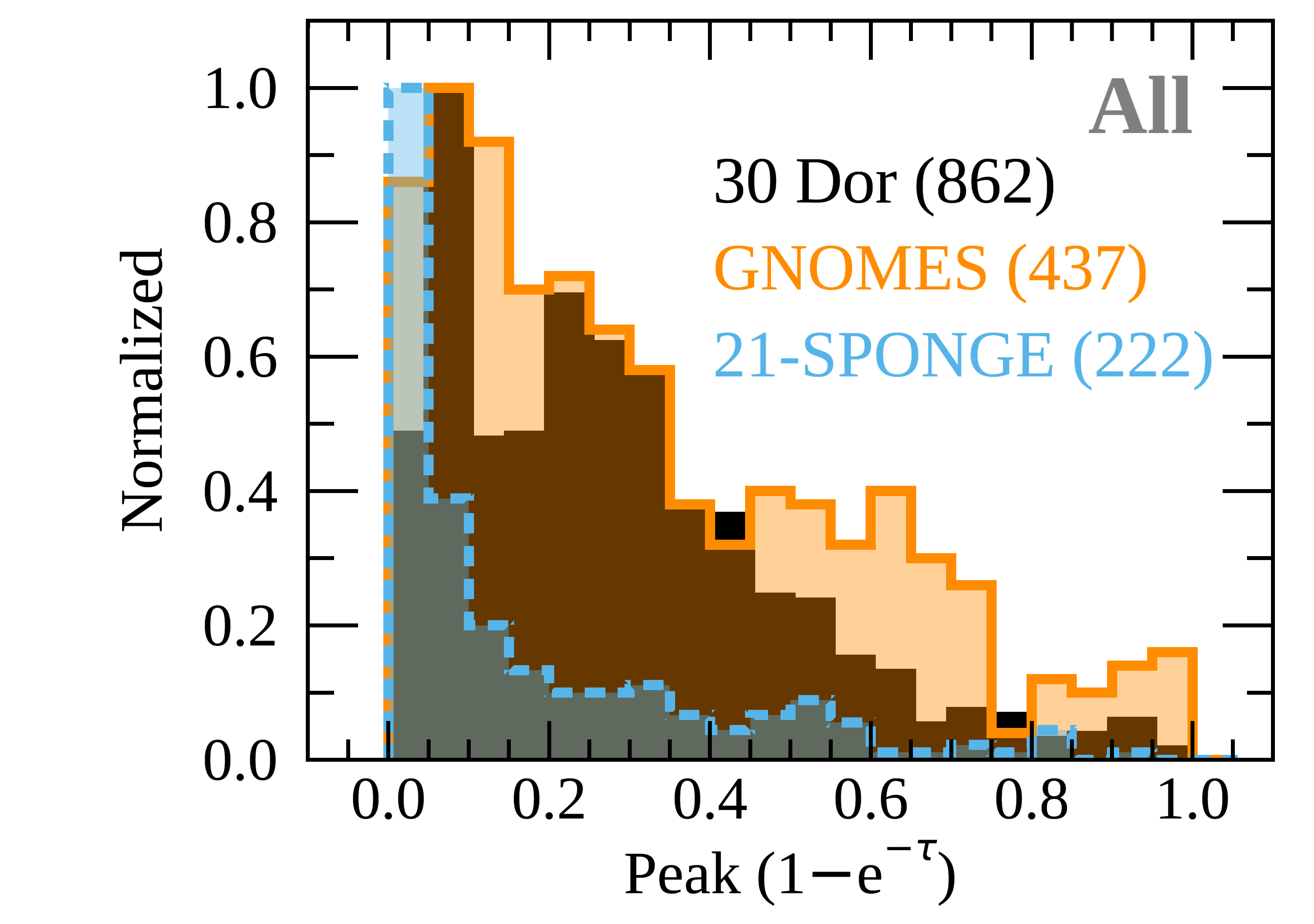} 
    \includegraphics[scale=0.35]{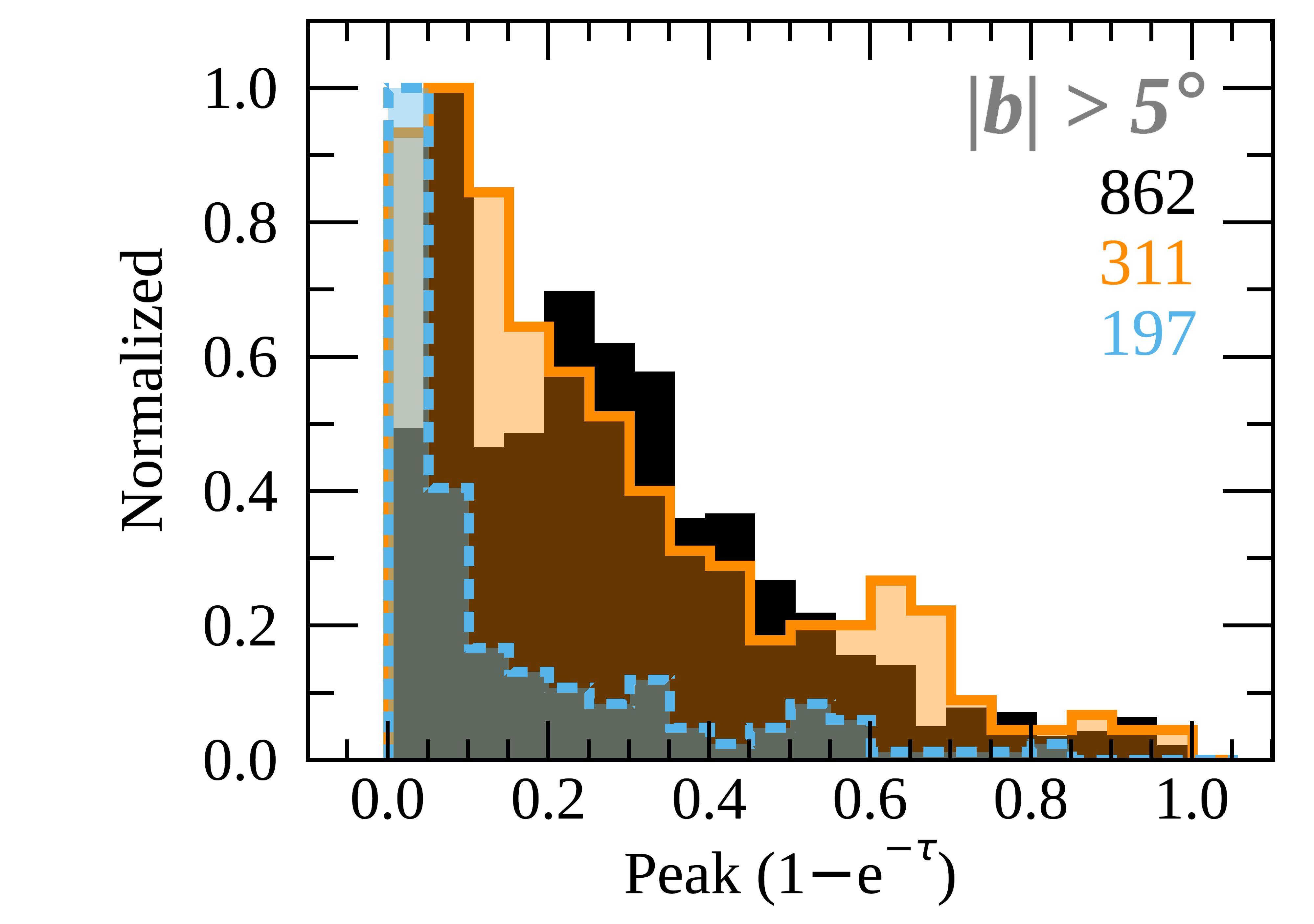}
    \includegraphics[scale=0.35]{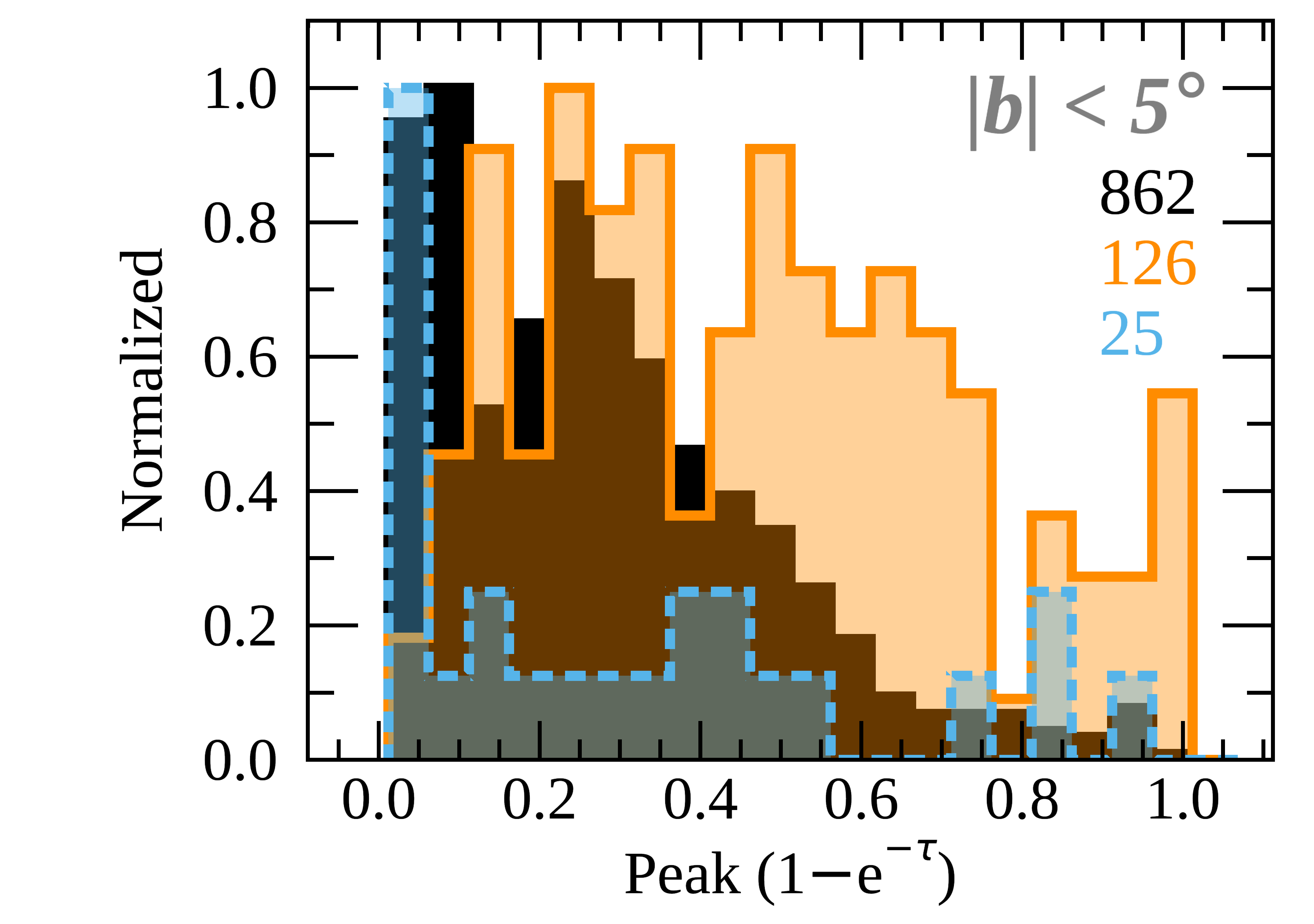} 
    \includegraphics[scale=0.35]{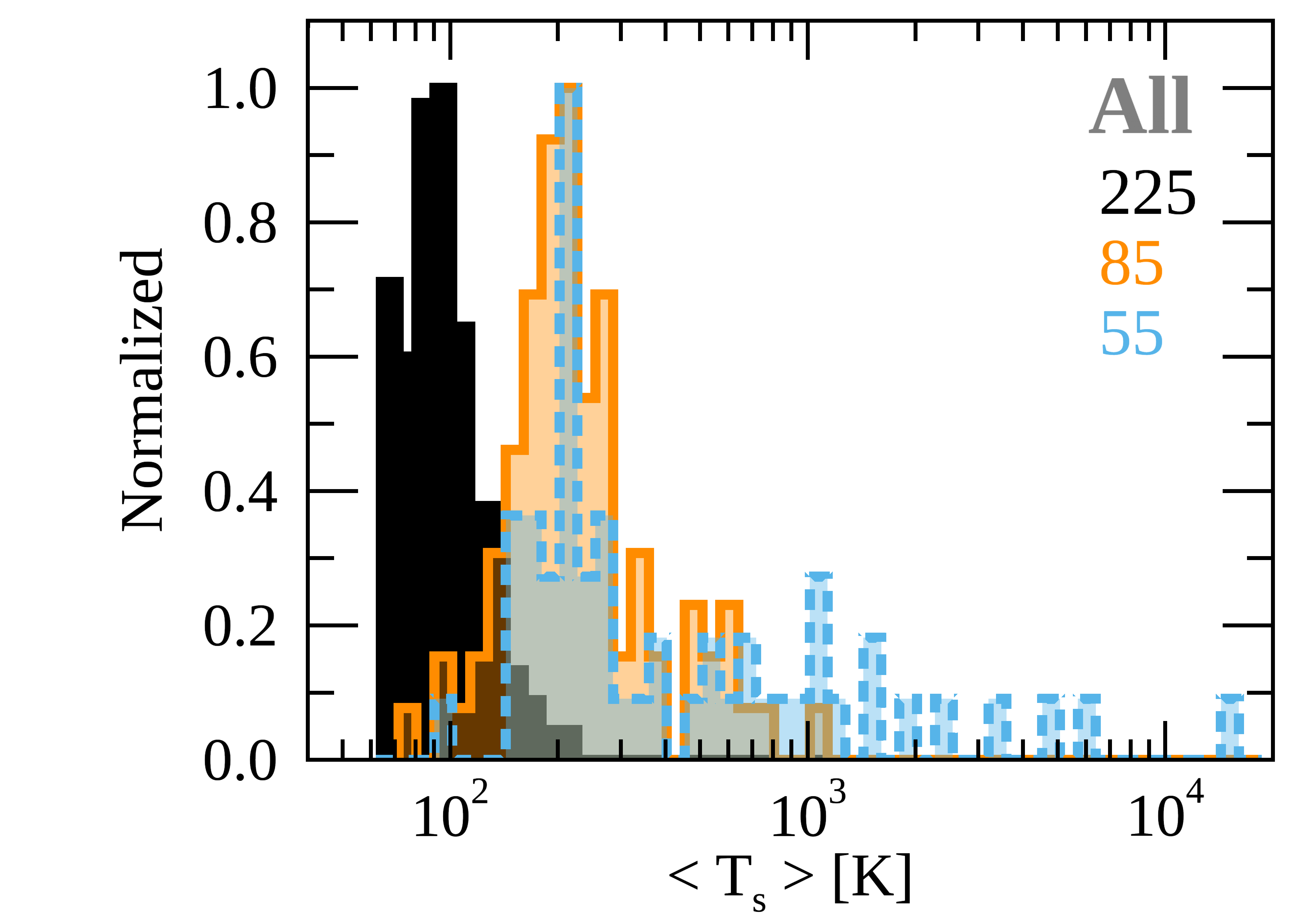} 
    \includegraphics[scale=0.35]{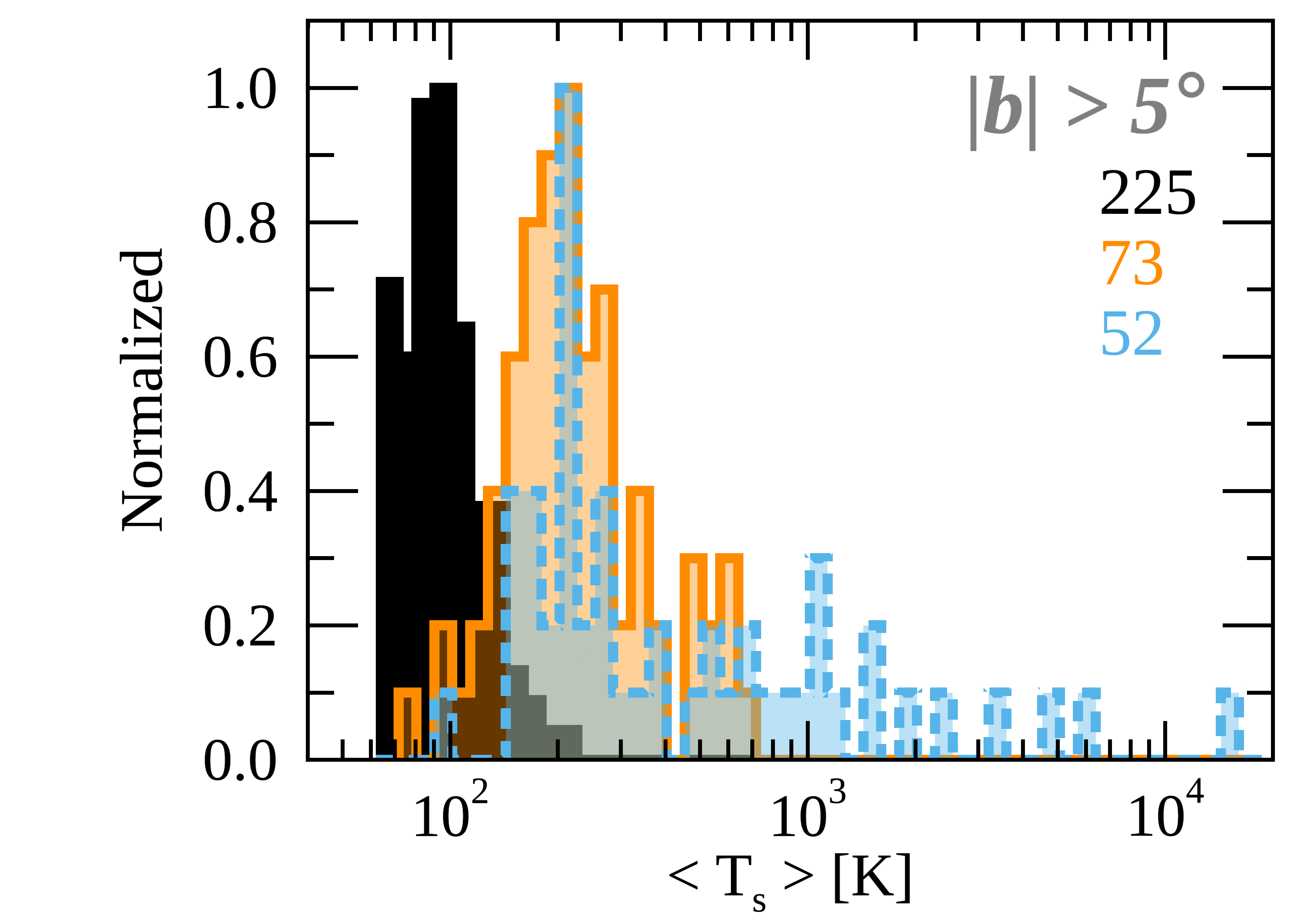}
    \includegraphics[scale=0.35]{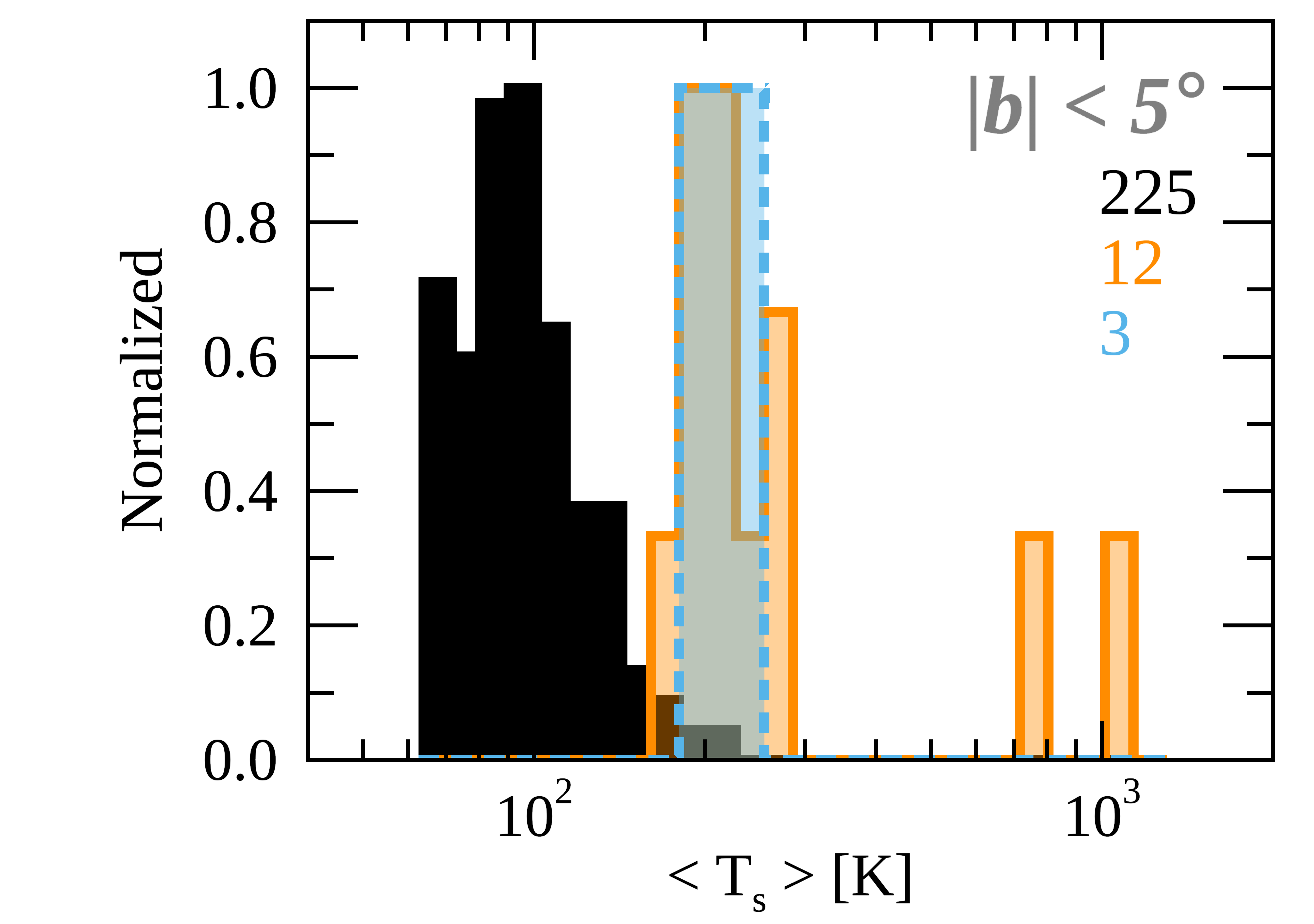}
    \includegraphics[scale=0.35]{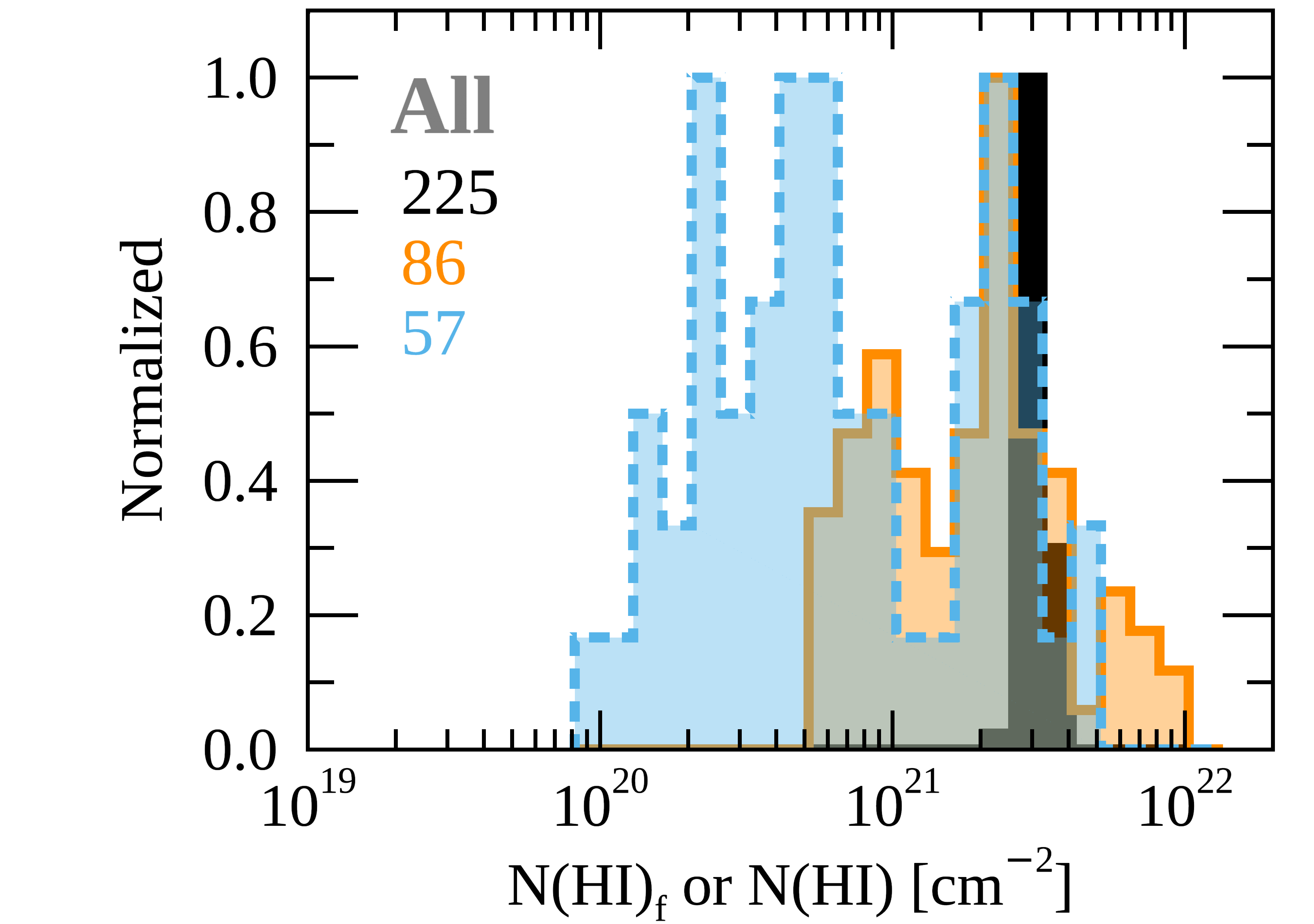}
    \includegraphics[scale=0.35]{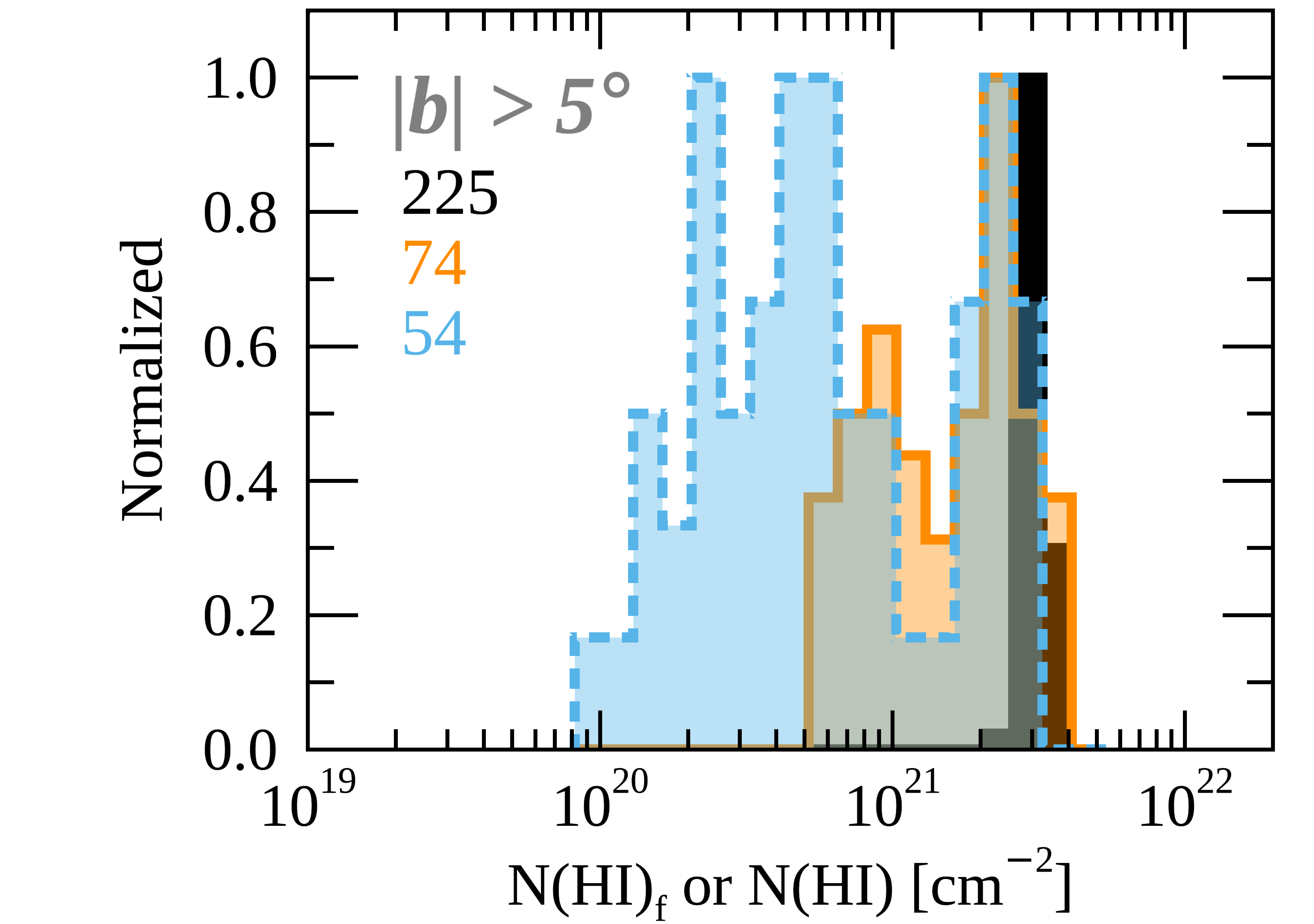}
    \includegraphics[scale=0.35]{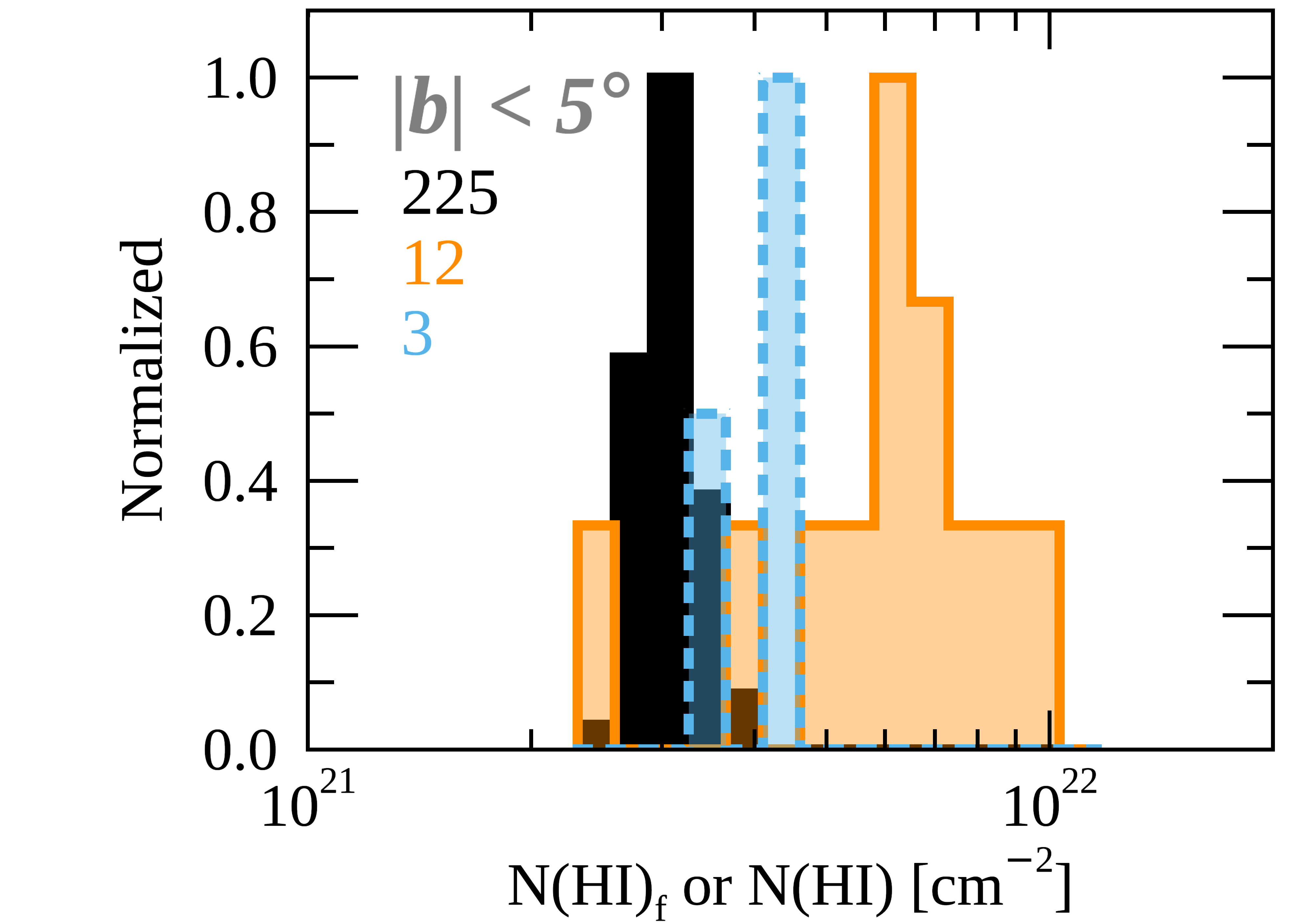}
    \caption{\label{f:MW_30Dor_hist} Normalized histograms of the peak $1 - e^{-\tau}$ (top), <$T_{\textrm{s}}$> (middle), 
	     and $N$(H~\textsc{i})$_{\textrm f}$/$N$(H~\textsc{i}) (bottom) from the 21-SPONGE, GNOMES, and 30 Dor observations. 
             The values of the individual Gaussian components were used for the peak $1 - e^{-\tau}$, 
	     while those of the entire velocities were employed for <$T_{\textrm{s}}$>, $N$(H~\textsc{i})$_{\textrm f}$, and $N$(H~\textsc{i}). 
	     Different $q$ values of 0.5 and 1 were adopted for 30 Dor and the Milky Way, respectively, 
	     and the $N$(H~\textsc{i}) histograms hence show $N$(H~\textsc{i})$_{\textrm{f}}$ for 30 Dor (foreground) 
	     and $N$(H~\textsc{i}) for the Milky Way (all gas along the observed LOSs). 
	     We note that $N$(H~\textsc{i})$_{\textrm{f}}$ for 30 Dor is a lower limit on the total \HI column density along the whole LOSs, 
	     since it includes the foreground gas only.
             Finally, the numbers of the Gaussian components and LOSs considered for the histograms are indicated in sky blue, orange, and black, respectively.
             (Left) Entire population of the observed LOSs is included for the comparison.
             (Middle) LOSs at $|b|~> 5^{\circ}$ for the Milky Way are selected to compare with the 30 Dor results. 
	     (Right) Same as the middle panel, but with the LOSs at $|b|~< 5^{\circ}$ for the Milky Way.}
\end{figure*}

Finally, we compared the observed properties of \HI in 30 Dor to those in the Milky Way and the LMC 
to examine how \HI changes in different environments.
For the comparison to the Milky Way, we selected the 21-SPONGE \citep{murray2018} and GNOMES \citep{stanimirovic2014, nguyen2019} surveys 
and calculated <$T_{\textrm{s}}$> and $N$(H~\textsc{i}) by applying Equations~(\ref{eq:LOS_avg_Ts}), (\ref{eq:f_chan}), and (\ref{eq:N_HI}) 
to the published \HI emission and absorption spectra with $q = 1$ (appropriate since their background sources are extragalactic).
These two surveys are distinctly different:  
21-SPONGE probes mostly diffuse environments at high and intermediate latitudes with an exceptional optical depth sensitivity of $\sigma_{\tau} < 10^{-3}$, 
while GNOMES focuses on the surroundings of MCs where total column densities are relatively high. 
To follow the measurements for the Milky Way as much as possible, 
we used the peak $1 - e^{-\tau}$ values of the individual Gaussian components, 
as well as the <$T_{\textrm{s}}$> and $N$(H~\textsc{i})$_{\textrm f}$ values
that were estimated for the entire velocity range with $q = 0.5$, for 30 Dor. 
The histograms of the peak $1 - e^{-\tau}$, LOS average spin temperature, and opacity-corrected total \HI column density 
are presented in Figure~\ref{f:MW_30Dor_hist}.

A comparison of the 21-SPONGE and GNOMES distributions shows that they are distinctly different even for the same latitude ranges, 
implying that the region of interest and sensitivity of a survey affect the observed properties of H~\textsc{i}. 
With this caveat in mind, we conclude that \HI in 30 Dor is dissimilar from that in the Milky Way at high latitudes ($|b|~>5^{\circ}$) 
in a way that the LOS average spin temperature is systematically lower and the total \HI column density is higher. 
The LOS average spin temperature in 30 Dor is still lower when it is compared to that in the Milky Way at low latitudes ($|b|~<5^{\circ}$).

In addition to the Milky Way, we compared our 30 Dor observations to the latest survey of \HI absorption in the LMC by Liu et al. (2025, in preparation; L25 hereafter). 
L25 measured \HI absorption toward 92 extragalactic background sources with the Australia Telescope Compact Array (ATCA) 
and calculated <$T_{\textrm{s}}$> and $N$(H~\textsc{i}) for all LOSs
by applying Equations~(\ref{eq:LOS_avg_Ts}), (\ref{eq:f_chan}), and (\ref{eq:N_HI}) with $q = 1$ 
to their absorption spectra and the emission data from \citet{kim1999}. 
A qualitative comparison shows lower LOS average spin temperatures and higher total \HI column densities in our observations 
than the L25 LOSs that are located far away from 30 Dor (angular separation $>$ 1$^{\circ}$). 
For a small number of L25 LOSs closer to 30 Dor, on the other hand, the LOS average spin temperature is comparable 
and the optical depth is higher compared to what we measured. 

In summary, we found that \HI in 30 Dor is distinctly different from that in the high-latitude Milky Way, 
as well as from that in the less star-forming part of the LMC, 
with lower LOS average spin temperatures and higher total \HI column densities. 
The lower LOS average spin temperatures suggest an intrinsically colder temperature of the CNM and/or a higher fraction of the CNM in 30 Dor. 
According to the theoretical model by \citet{bialy2019}, 
the density and fraction of the CNM increases, as the thermal pressure increases in a half-solar metallicity environment.
{Our approximate derivation of the thermal pressure based on the H~\textsc{i}, [C~\textsc{ii}], and dust data
indeed revealed that 30 Dor and its surrounding region have thermal pressures of (0.7--36.7)~$\times$~10$^{4}$~cm$^{-3}$~K (Appendix \ref{appendix_a}), 
which are systematically higher than solar neighborhood values \citep[e.g.][]{jenkins2011, goldsmith2018}.}
Higher CNM fractions may be symptomatic of starburst conditions, 
as the \HI observations of the Galactic mini-starburst region W43 also hinted \citep{bihr2015}.

\begin{figure*}
    \centering
    \includegraphics[scale=0.18]{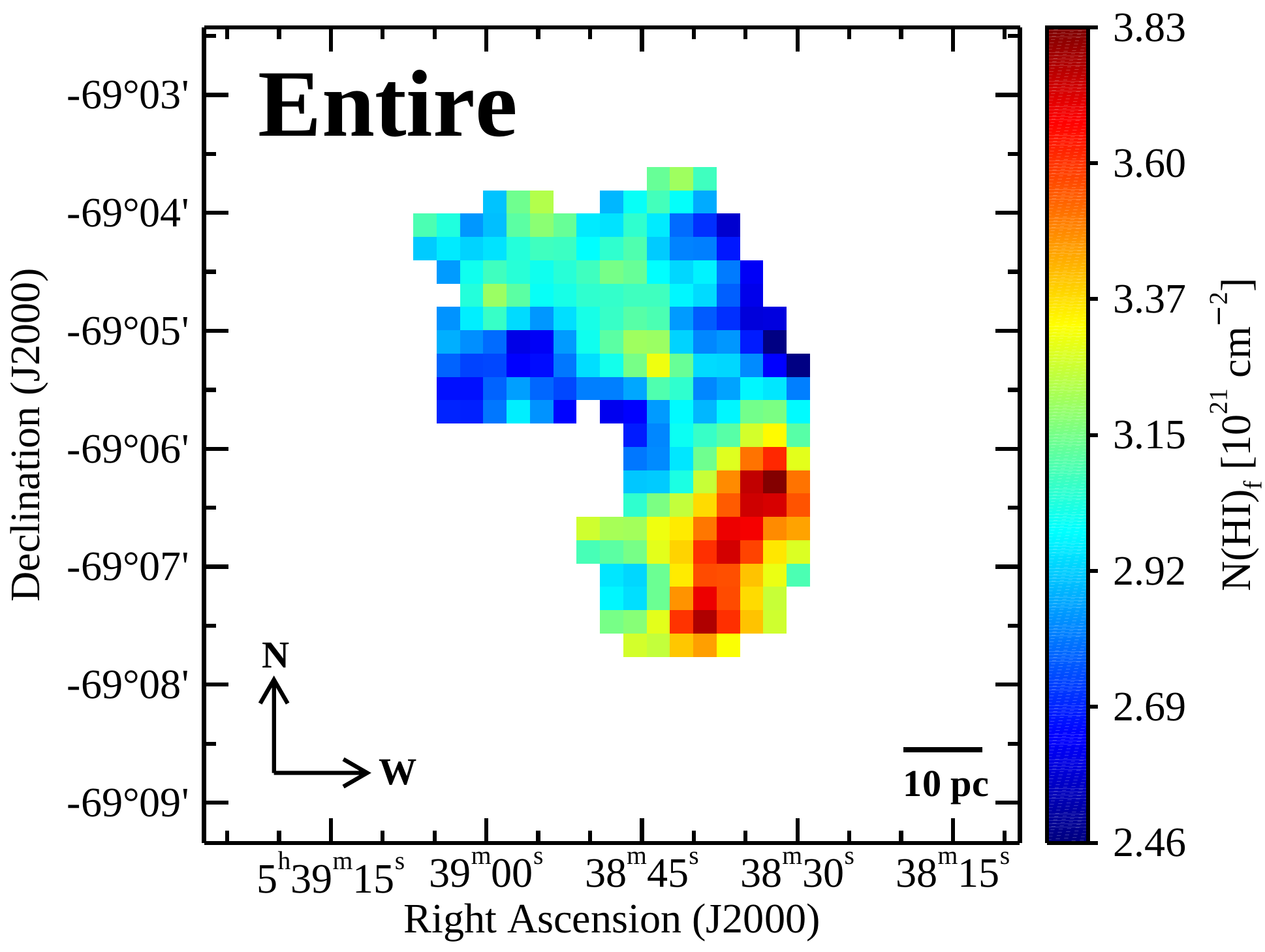}
    \includegraphics[scale=0.18]{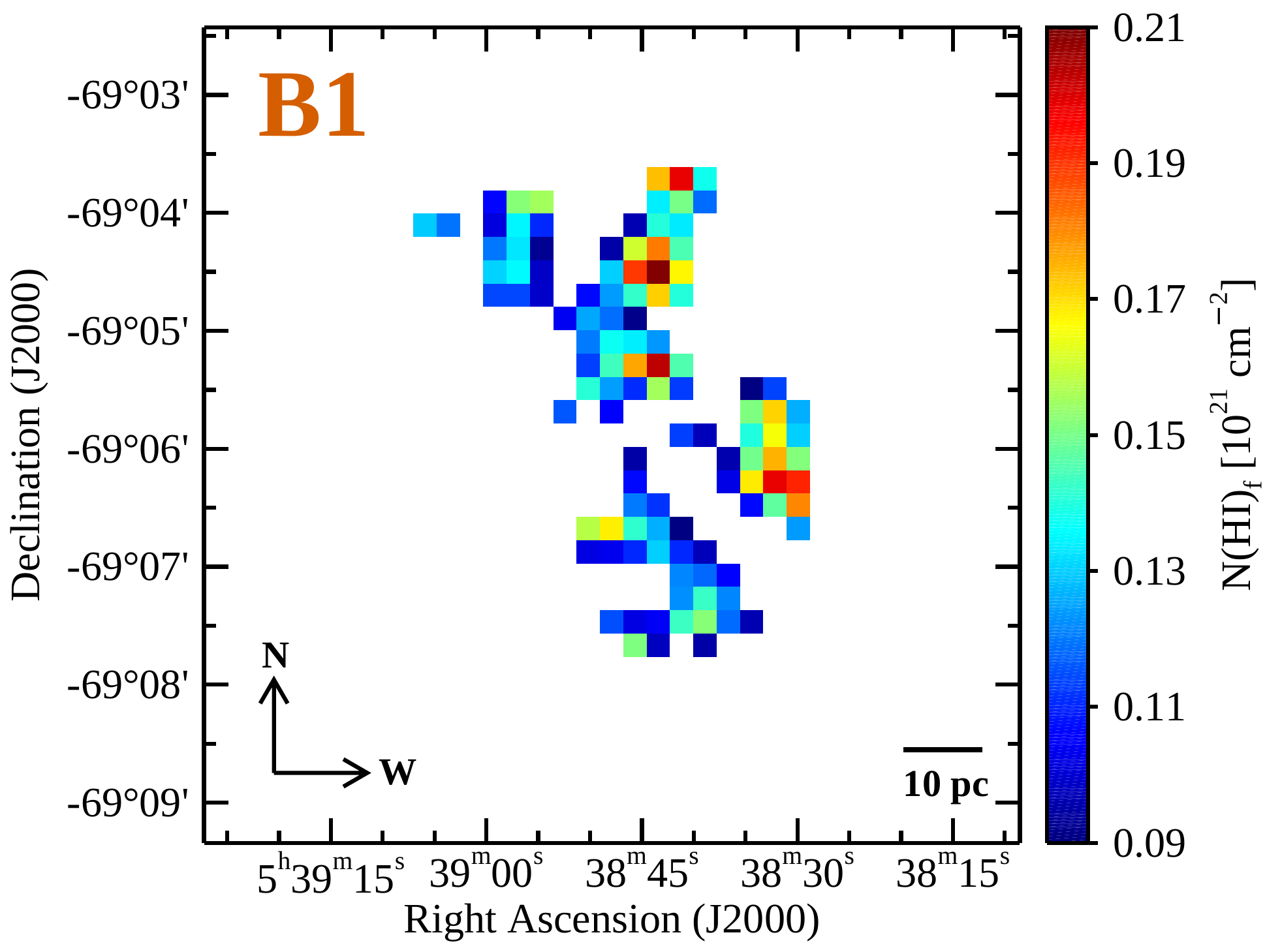} 
    \includegraphics[scale=0.18]{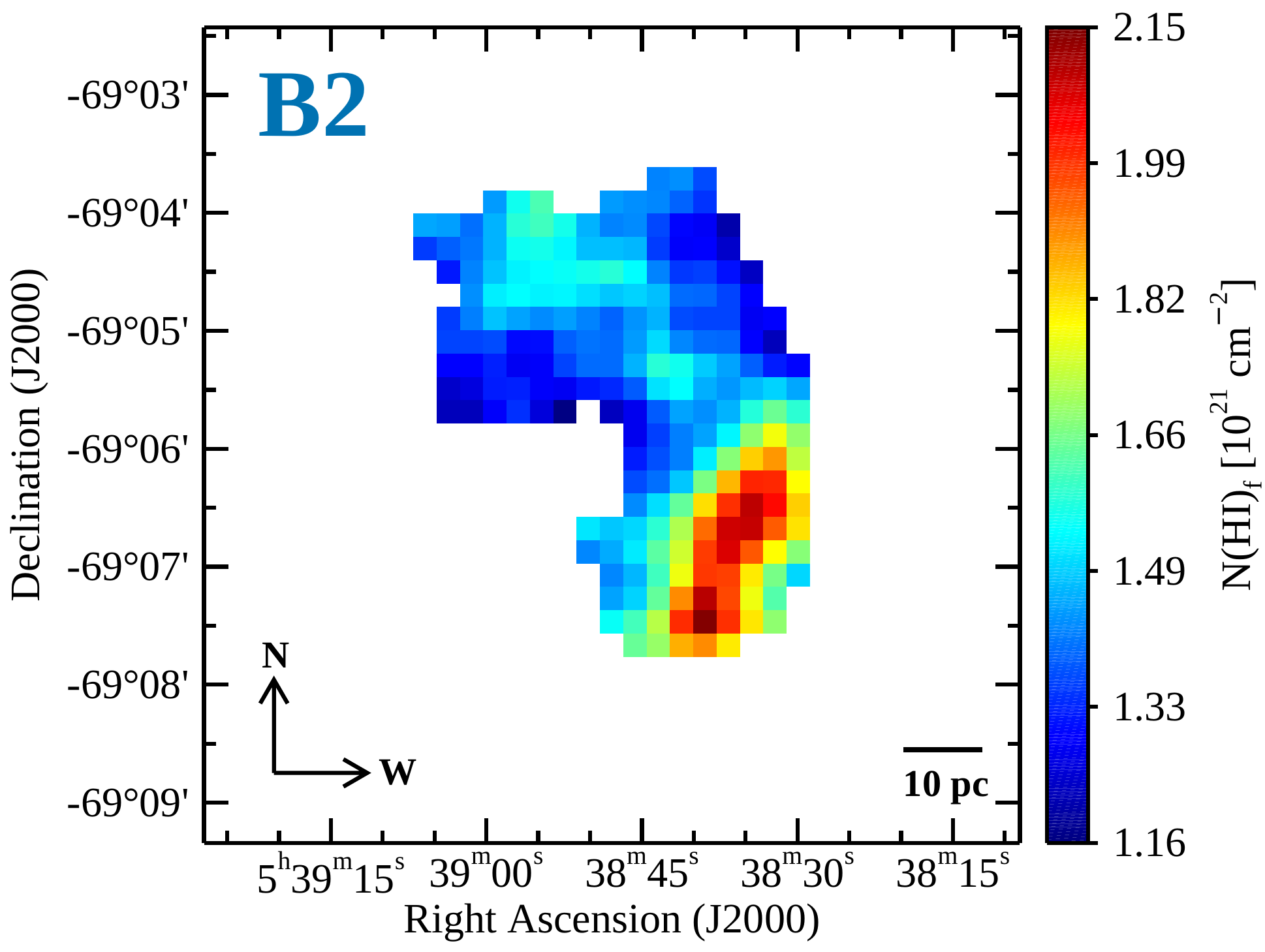}
    \includegraphics[scale=0.18]{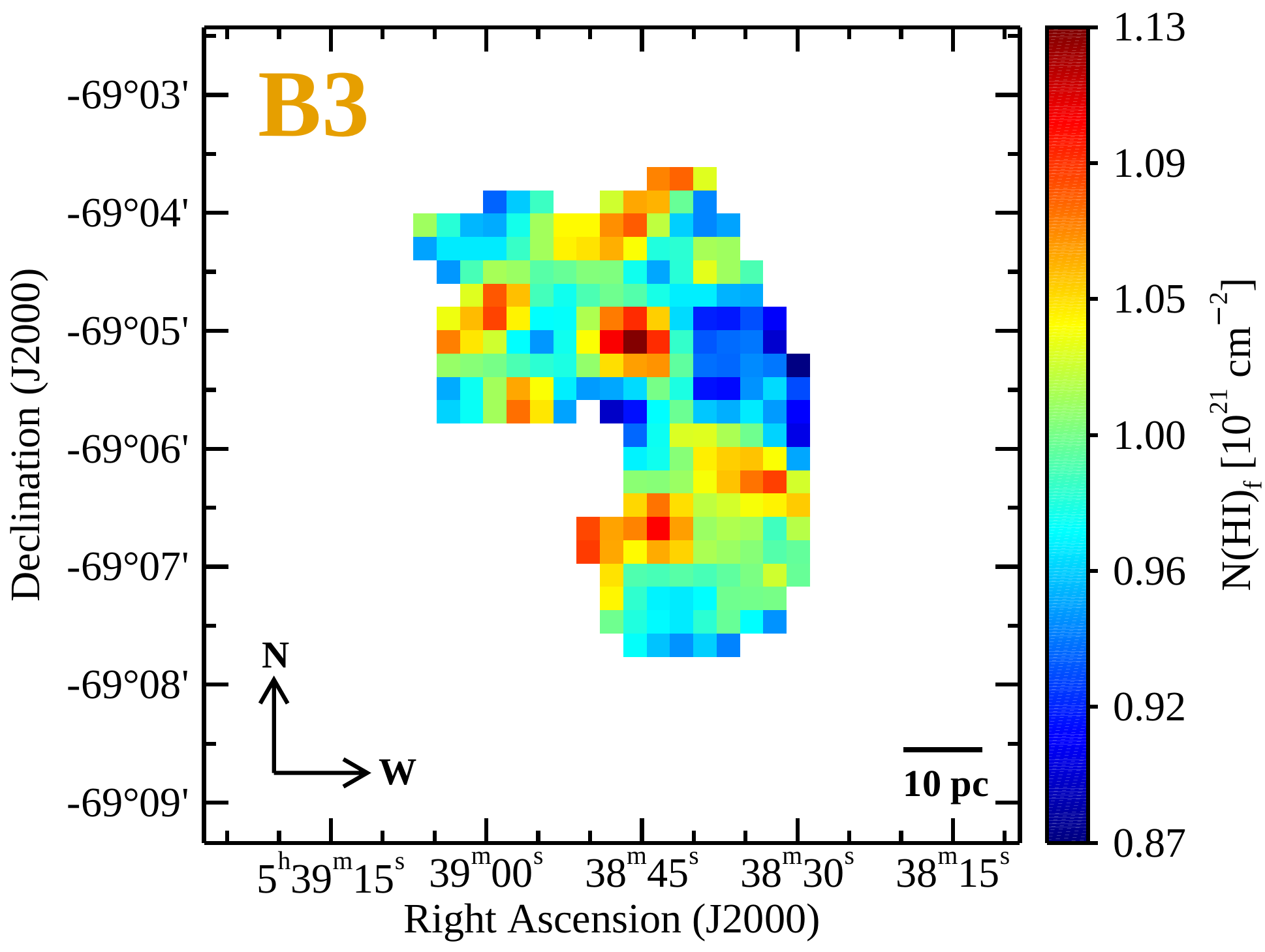}
    \includegraphics[scale=0.18]{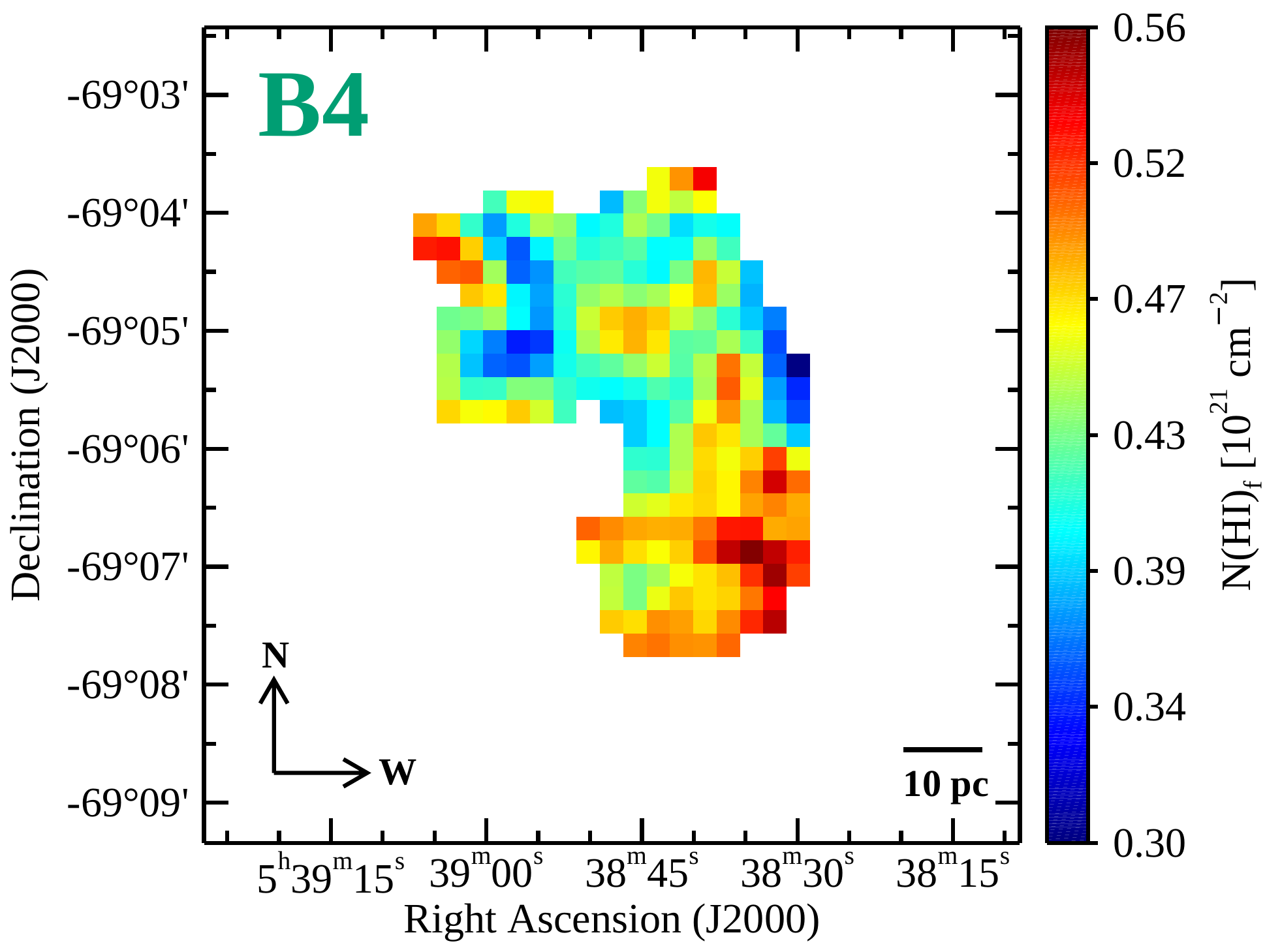}
    \caption{\label{f:NHI_map_per_band} Same as Figure~\ref{f:peak_tau_per_band}, but for $N$(H~\textsc{i})$_{\textrm{f}}$.}
\end{figure*}

\section{Discussion} \label{s:discussion} 

\subsection{H~\textsc{i}-to-H$_{2}$ Transition in 30 Dor} \label{s:HI-to-H2}

In Section~\ref{s:result_nhi}, we showed that the \HI column densities at B1--B4 velocities 
change by a factor of two or so only across $\sim$60 pc scale regions. 
Considering that our $N$(H~\textsc{i}) estimates are corrected for high optical depth,  
this result strongly indicates the presence of \HI shielding layers in 30 Dor.
Steady-state H$_{2}$ formation models have predicted such layers in star-forming regions 
where a certain amount of \HI column density is required to shield H$_{2}$ against dissociating UV photons.
For example, \citet{sternberg2014} formulated an analytic model of the H~\textsc{i}-to-H$_{2}$ transition in a one-dimensional plane-parallel slab of gas  
and predicted the total \HI column density for one-sided beamed UV radiation as 

\begin{equation}
    \label{eq:S14_NHI}
    N\textrm{(H~\textsc{i})}~(\textrm{cm}^{-2}) = \frac{5.3\times10^{20}}{Z'\phi_{\textrm{g}}}\textrm{ln}\left(\frac{\alpha G}{2}+1\right)
\end{equation}

\noindent where $Z'$ is the metallicity relative to the solar value, and $\phi_{\textrm{g}}$ is the order unity grain composition factor. 
Here one-sided irradiation by a beam radiation field is adopted, 
considering that the \HI column densities were measured for foreground structures 
that are primarily illuminated by R136 \citep[e.g.,][]{chevance2016, lee2019}. 

For the dimensionless parameter $\alpha G$, \citet{bialy2016} provided the following formula 
based on an updated expression of the total H$_{2}$-dust-limited effective dissociation bandwidth: 
\begin{equation}
    \label{eq:alphaG}
    \alpha G = 0.59 I_{\textrm{UV}} \left(\frac{100~\textrm{cm}^{-3}}{n}\right) \left(\frac{9.9}{1 + 8.9 \phi_{\textrm{g}} Z'}\right)^{0.37}
\end{equation}

\noindent where $I_{\textrm{UV}}$ is the strength of UV radiation normalized to the Draine field \citep{draine1978, bialy2020}, 
and $n$ is the total gas number density.
 




Equation~\ref{eq:S14_NHI} shows that $\alpha G$ solely determines 
the total \HI column density at the H~\textsc{i}-to-H$_{2}$ transition point for given environmental conditions ($Z'$ and $\phi_{\textrm{g}}$). 
The dimensionless parameter $\alpha G$ has the physical meaning of the ratio of the effective H$_{2}$ photodissociation rate (accounting for UV shielding) 
to the H$_{2}$ formation rate and can range from small to large values. 
For example, when $\alpha G \ll 1$ (``weak-field limit''), 
H$_{2}$ self-shielding is mainly responsible for the absorption of dissociating UV photons 
and results in a gradual H~\textsc{i}-to-H$_{2}$ transition where most of the \HI column density is accumulated past the transition point. 
On the other hand, when $\alpha G \gtrsim 1$ (``strong-field limit''),
dust absorption becomes dominant and makes the H~\textsc{i}-to-H$_{2}$ transition sharp due to the exponential reduction of UV photons. 
In this case, the \HI column density is primarily built up prior the transition point, producing distinct \HI shielding layers. 

When $Z' = 0.5$ and $\phi_{\textrm{g}} = 1$ are assumed, 
the measured \HI column densities of (1.3--14.6) $\times$ 10$^{20}$ cm$^{-2}$ at B1--B4 velocities (median values in Table~\ref{t:band_info}) 
correspond to $\alpha G$ $\sim$ {0.3--5.9}. 
An interesting region to compare with is the low-mass star-forming complex in the solar neighborhood, Perseus, 
where $\alpha G$ $\sim$ 5--26 was estimated \citep{bialy2015}. 
Perseus is significantly different from 30 Dor in terms of environmental conditions
\citep[e.g., solar metallicity, weak UV radiation of 1--2 Draine fields, and low thermal pressure of a few $\times$ 10$^{3}$ cm$^{-3}$~K;][]{lee2012,goldsmith2018} 
and \HI properties \citep[e.g., low CNM fraction of 0.3;][]{stanimirovic2014}. 
Yet Perseus and 30 Dor are comparable as to the nature of the H~\textsc{i}-to-H$_{2}$ transition (``strong-field limit''). 
This could be because dust grains, the site for H$_{2}$ formation and the primary source for the reduction of UV photons, 
play a key role in the H~\textsc{i}-to-H$_{2}$ transition over a wide range of environmental conditions 
and the decrease of metallicity in 30 Dor is not so significant as to change the nature of the H~\textsc{i}-to-H$_{2}$ transition. 
Perhaps the most surprising result from our analysis would be that 
\HI shielding layers predicted by simple steady-state H$_{2}$ formation models exist in extreme environments like 30 Dor 
where various energetic processes (e.g., stellar outflows and winds and expanding \HII regions) constantly influence the surrounding ISM. 
We will revisit the H~\textsc{i}-to-H$_{2}$ transition in and around 30 Dor in our forthcoming paper, 
confronting the model by \citet{sternberg2014} more rigorously with the observed ISM properties (e.g., $I_{\textrm{UV}}$, $n$, and $\phi_{\textrm{g}}$).

\subsection{Outflows and Inflows} \label{s:outflow_inflow}

As discussed in Section~\ref{s:gfit}, we interpret B1 as outflows {and B3 and B4 as inflows}.    
To derive the outflow and inflow rates $\dot{M}_{\textrm{outflow}}$ and $\dot{M}_{\textrm{inflow}}$, 
we assumed spherically symmetric flows that are moving at a constant radial velocity of $\delta \varv_{\textrm{r}}$ 
with a covering factor of $f_{\textrm{c}}$. 
These flows start from the surface of the \HII region whose radius is $R$,   
and the density decreases radially as $1/r^{2}$.  
The mass flux rate $\dot{M}$ can then be calculated as

\begin{equation} \label{eq:mass_rates}
\begin{split}
	\dot{M}~(M_{\odot}~\textrm{yr}^{-1})& = 4 \pi \mu m_{\textrm{H}} f_{\textrm{c}} \delta \varv_{\textrm{r}} R N(\textrm{H~\textsc{i}}) \\
		& = 1.4 \times 10^{-5} f_{\textrm{c}} \left(\frac{\delta \varv_{\textrm{r}}}{\textrm{km~s}^{-1}}\right)
		                                      \left(\frac{R}{\textrm{pc}}\right)
						      \left(\frac{N(\textrm{H~\textsc{i}})}{10^{20}~\textrm{cm}^{-2}}\right) 
\end{split}
\end{equation}

\noindent 
where $\mu = 1.4$ is the average particle mass in units of the hydrogen mass $m_{\textrm{H}}$,
$f_{\textrm{c}}$ is the covering factor, 
$\delta \varv_{\textrm{r}}$ is the relative radial velocity of the flow with respect to the \HII region, 
$R$ is the radius of the \HII region, 
and $N$(H~\textsc{i}) is the \HI column density that is measured up to the \HII region. 
For our calculation, we adopted $f_{\textrm{c}}$~=~0.5 for B1 and {$f_{\textrm{c}}$~=~1 for B3 and B4, 
based on the covering fraction of \HI in Figure \ref{f:NHI_map_per_band}, 
and $\delta \varv_{r}$~=~$-$25~km~s$^{-1}$, 18~km~s$^{-1}$, and 34~km~s$^{-1}$ for B1, B3, and B4, respectively,  
where we took $\delta \varv_{r}$ to be the median central velocity of the fitted Gaussians minus the \HII region velocity of 251~km~s$^{-1}$. 
In addition, we used $N$(H~\textsc{i})~=~1.3~$\times$~10$^{20}$~cm$^{-2}$, 1.0~$\times$~10$^{21}$~cm$^{-2}$, and 4.4~$\times$~10$^{20}$~cm$^{-2}$ 
(median \HI column densities) for B1, B3, and B4, respectively, as well as $R$ = 2$'$ or 30 pc}. 

{While keeping in mind that our estimates are approximate based on simplified spherical flows for the front side only 
and could change depending on the actual geometry that is currently unknown (e.g., biconic flows), 
we measured $\dot{M}_{\textrm{outflow}}$~=~0.007~$M_{\odot}$~yr$^{-1}$ for B1 
and $\dot{M}_{\textrm{inflow}}$~=~0.08~$M_{\odot}$~yr$^{-1}$ and 0.06~$M_{\odot}$~yr$^{-1}$ for B3 and B4, respectively.
The total inflow rate of 0.14~$M_{\odot}$~yr$^{-1}$ is comparable to the star formation rate of 0.18~$M_{\odot}$~yr$^{-1}$  
that was estimated by \citet{nayak2023} based on the number of young stellar objects (YSOs), 
suggesting that the accreted atomic gas could sustain star formation in 30 Dor. 
The outflow rate, on the other hand, is only 5\% of the total inflow rate.} 


Recently, \citet{poudel2025} analyzed ULLYSES UV absorption spectra toward eight LOSs probing 30 Dor and its surrounding region 
and found blueshifted absorption components at 150--230~km~s$^{-1}$ in atomic (O~\textsc{i} $\lambda$1302) and 
ionized (Si~\textsc{ii} $\lambda$1526, Fe~\textsc{ii} $\lambda$2344, C~\textsc{iv} $\lambda$1550, etc.) gas tracers. 
B1 lies within the observed velocities of these absorption components, 
suggesting that B1 is associated with large-scale outflows driven by active star formation in 30 Dor. 
In terms of mass, the \HI outflow rate of {0.007}~$M_{\odot}$~yr$^{-1}$ is a fraction of the ionized outflow rate of $\gtrsim$~0.02~$M_{\odot}$~yr$^{-1}$. 
One of the interesting features about B1 is that it is exceedingly weak in \HI emission (e.g., Figure~\ref{f:avg_spec}). 
This implies that the observed outflows consist of ionized gas along with the CNM and very little of the WNM. 
As discussed in Section~\ref{s:MW_vs_30Dor}, the high fraction of the CNM could result from high thermal pressures. 
For example, the WNM no longer exists at thermal pressures higher than 10$^{4}$~cm$^{-3}$~K in the solar neighborhood condition \citep{bialy2019}. 
While it is currently unclear exactly how stellar activities drive the observed outflows, 
high thermal pressures are expected from various radiative and mechanical feedback processes related to star formation in 30 Dor, 
e.g., strong UV radiation, stellar outflows and winds, etc.
{The thermal pressure of (2.6--20.7)~$\times$~10$^{4}$~cm$^{-3}$~K for B1 is indeed systematically higher than 
the solar neighborhood \citep[e.g.,][]{jenkins2011, goldsmith2018}, as well as B3 and B4 (Appendix \ref{appendix_a}).} 

In addition to the blueshifted absorption, \citet{poudel2025} found redshifted absorption from 290~km~s$^{-1}$ to 320~km~s$^{-1}$ toward six of the eight LOSs.
Considering that B4 is seen in \HI emission and absorption, 
the \citet{poudel2025} result suggests that the observed inflows consist of the CNM, WNM, and ionized gas. 
These multi-phase inflows have different properties compared to the outflows, 
i.e., lower thermal pressures (so that the CNM and WNM can co-exist), higher LOS average spin temperatures, 
and higher \HI column densities (Section~\ref{s:ts_derivation}).  
As for the origin of the inflows, \citet{poudel2025} suggested that the redshifted absorbers might represent 
inflowing materials that are recycling back from the outflows. 
{Or the inflows could be produced by the tidal interaction between the LMC and SMC \citep[e.g.,][]{olsen2011,fukui2017}.
Among these two possibilities, the colder and denser B3 might be ``galactic fountain'' flows, 
while B4 could originate from outside the LMC}. 

\section{Summary} \label{s:summary}

In this paper, we present a detailed study of the properties of the CNM 
in and around the extreme star-forming region 30 Dor. 
To probe the CNM across 30 Dor, we made use of the high-resolution GASKAP-H~\textsc{i} emission and absorption observations
and spatially mapped various CNM properties 
(e.g., central velocities, optical depths, and average spin temperatures) on 7~pc scales.
In addition, we compared the observed CNM properties to supplementary data 
to evaluate how the CNM changes in different environments and 
showed that the CNM is an important constituent of the baryon cycle in and around 30 Dor.  
Our key results are as follows.

\begin{enumerate}
\item \HI absorption is clearly detected across $\sim$60~pc scale regions,
      indicating the pervasiveness of the CNM in 30 Dor. 
\item Four distinct CNM structures with systematically different characteristics exist 
      at 200--230~km~s$^{-1}$, 230--260~km~s$^{-1}$, 260--277~km~s$^{-1}$, and 277--300~km~s$^{-1}$ (B1, B2, B3, and B4, respectively).
\item {The CNM at B2 velocities belongs to the main dense structure of 30 Dor, where ionized, atomic, and molecular gases are concentrated. 
      On the other hand, the CNM at B1 velocities traces outflows, while B3 and B4 are inflows with respect to 30 Dor. 
      Compared to the CNM at B2 velocities, B1, B3, and B4 are warmer and have lower \HI column densities.}  
\item Assuming simple spherically symmetric flows moving at a constant velocity, 
      the outflow and inflow rates were estimated to be {0.007}~$M_{\odot}$~yr$^{-1}$ and {0.14}~$M_{\odot}$~yr$^{-1}$, respectively. 
      The observed outflows and inflows are associated with a large amount of ionized gas, 
      and hence the total outflow and inflow rates would be much higher than what we measured in H~\textsc{i}.  
      As for the origin of the gas flows, the outflows likely originate from active star formation in 30 Dor,
      {while the inflows could be galactic fountain gas or be accreted from outside the LMC}. 
\item Compared to the high-latitude Milky Way and the less star-forming portion of the LMC, 
      30 Dor has systematically lower LOS average spin temperatures and higher \HI column densities. 
      The lower LOS average spin temperatures could result from a higher fraction of the CNM,  
      which is expected in the extreme surroundings of 30 Dor where UV photons are abundant and thermal pressures are high. 
\item The relatively uniform spatial distribution of the \HI column densities at B1--B4 velocities 
      indicates that \HI shielding layers for H$_{2}$ formation exist in 30 Dor. 
      Considering that the solar neighborhood and 30 Dor have significantly different ISM conditions, 
      the presence of \HI shielding layers in both environments suggests that 
      dust grains play an important role in the H~\textsc{i}-to-H$_{2}$ transition over a wide range of ISM conditions. 
\end{enumerate}

\begin{acknowledgements}
This scientific work uses data obtained from Inyarrimanha Ilgari Bundara/the Murchison Radio-astronomy Observatory. 
We acknowledge the Wajarri Yamaji People as the Traditional Owners and native title holders of the Observatory site. 
CSIRO's ASKAP radio telescope is part of the Australia Telescope National Facility (\url{https://ror.org/05qajvd42}). 
Operation of ASKAP is funded by the Australian Government with support from the National Collaborative Research Infrastructure Strategy.
ASKAP uses the resources of the Pawsey Supercomputing Research Centre. 
Establishment of ASKAP, Inyarrimanha Ilgari Bundara, the CSIRO Murchison Radio-astronomy Observatory and the Pawsey Supercomputing Research Centre 
are initiatives of the Australian Government, with support from the Government of Western Australia, Australia 
and the Science and Industry Endowment Fund, Australia. 
{GASKAP-\HI is partially funded by the Australian Government through an Australian Research Council Australian Laureate Fellowship (project number FL210100039 awarded to NMc-G).}
{The data to produce the low-resolution \HI emission were imaged using CHTC services at the University of Wisconsin's Center for High Throughput Computing \citep{GNT1-HW21}.}
{B.L. acknowledges support by the NRF, grant Nos. RS-2022-NR069020.}
This paper includes archived data obtained through the CSIRO ASKAP Science Data Archive, CASDA (\url{http://data.csiro.au})
\end{acknowledgements}

\appendix
\restartappendixnumbering

{\section{Thermal Pressures in and around 30 Dor} \label{appendix_a}}

{Our \HI observations along with the existing \CII data provide us with a unique opportunity 
to measure the thermal pressure of the neutral ISM in and around 30 Dor. 
For our measurement of the thermal pressure, we followed the methodology of \citet{goldsmith2018}
and focused on B1, B3, and B4 where the contribution from ionized gas to \CII emission is insignificant. }

{\citet{goldsmith2018} analyzed the H~\textsc{i}, [C~\textsc{ii}], and dust data toward four Galactic LOSs 
by assuming that \CII emission is optically thin. 
In this case, the \CII integrated intensity $W$([C~\textsc{ii}]) can be expressed as }

{\begin{equation} \label{eq:CII_int}
W(\textrm{[C~\textsc{ii}]})~(\textrm{K~km~s}^{-1}) = 3.43 \times 10^{-16} N(\textrm{C}^{+}) 
		                                    \left[1 + 0.5 e^{91.21/T_{\textrm{k}}} \left(1 + \frac{2.4 \times 10^{-6}}{R_{\textrm{ul}} n}\right)\right]^{-1}
\end{equation} }

\noindent 
{where $N$(C$^{+}$) is the C$^{+}$ column density, 
$T_{\textrm{k}}$ is the gas kinetic temperature, 
$R_{\textrm{ul}}$ is the de-excitation rate coefficient, 
$n$ is the density of collisional partners, 
91.21~K is the equivalent temperature, 
and 2.4~$\times$~10$^{-6}$~s$^{-1}$ is the spontaneous decay rate of the \CII line.
In addition, all carbon was assumed to be singly ionized with a fixed fractional abundance  
so that the C$^{+}$ column density can be directly estimated from the total hydrogen column density. }

{When the \CII excitation is done purely by collisions with atomic hydrogen, 
the following de-excitation rate coefficient can be adopted from \citet{barinovs2005} and \citet{goldsmith2012}: }

{\begin{equation} \label{eq:R_ul_HI} 
R_{\textrm{ul;H}^{0}} = 7.6 \times 10^{-10} (T_{\textrm{k}}/100)^{0.14}. 
\end{equation} }

\noindent 
{If the ratio of the C$^{+}$ column density to the integrated intensity is defined as $X = N(\textrm{C}^{+})/W(\textrm{[C~\textsc{ii}]})$,  
Equation~\ref{eq:CII_int} can be re-written as }

{\begin{equation} \label{eq:n_H0} 
n(\textrm{H}^{0})~(\textrm{cm}^{-3}) = \frac{3.16 \times 10^{3} (100 / T_{\textrm{k}})^{0.14}}{2 e^{-91.21 / T_{\textrm{k}}} (3.43 \times 10^{-16} X - 1) - 1}. 
\end{equation} }

{The \CII transition can also arise from the regions where both atomic and molecular hydrogens exist. 
In this case, the following effective de-excitation rate coefficient can be adopted }

{\begin{equation} \label{eq:R_ul_mix} 
R_{\textrm{ul;mix}} = f_{\textrm{n}}(\textrm{H}^{0})R_{\textrm{ul;H}^{0}} + f_{\textrm{n}}(\textrm{H}_{2})R_{\textrm{ul;H}_{2}} 
\end{equation} }

\noindent 
{where }

{\begin{equation} \label{eq:R_ul_H2} 
R_{\textrm{ul,H}_{2}} = 10^{-10} \left[4.9 + 0.22 (T_{\textrm{k}} / 100)\right] 
\end{equation} }

\noindent 
{and $f_{\textrm{n}}$ is the fractional number densities of the two forms of hydrogen. 
With this effective de-excitation rate coefficient, Equation~\ref{eq:CII_int} can be re-written as }

{\begin{equation} \label{eq:total_n} 
n(\textrm{H}^{0} + \textrm{H}_{2})~(\textrm{cm}^{-3}) = \frac{2.4 \times 10^{-6} R_{\textrm{ul;mix}}^{-1}}
	                                                {2 e^{-91.21 / T_{\textrm{k}}} (3.43 \times 10^{-16} X - 1) - 1}.
\end{equation} }

\begin{figure*}
    \centering
    \includegraphics[scale=0.25]{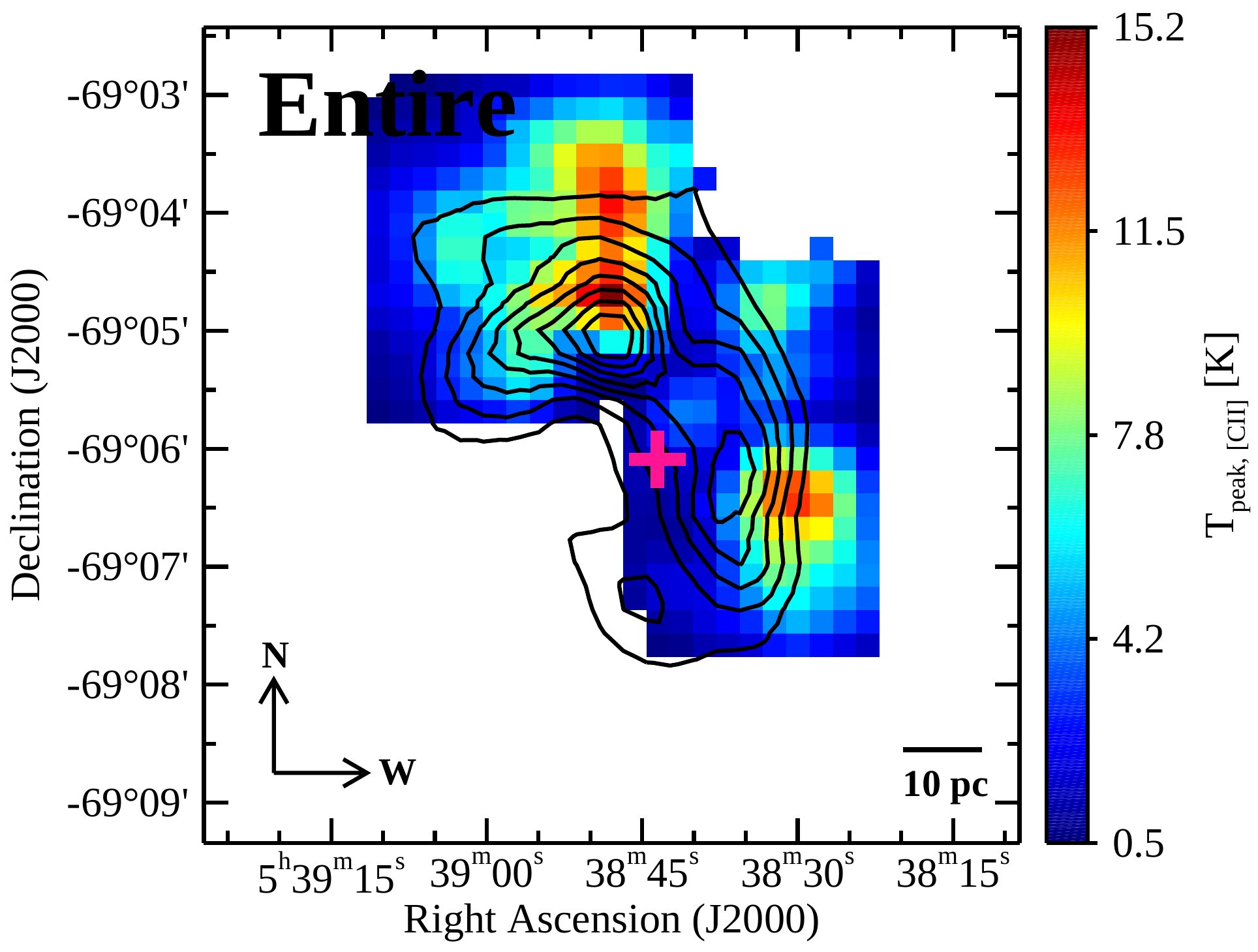} \hspace{0.1cm}
    \includegraphics[scale=0.25]{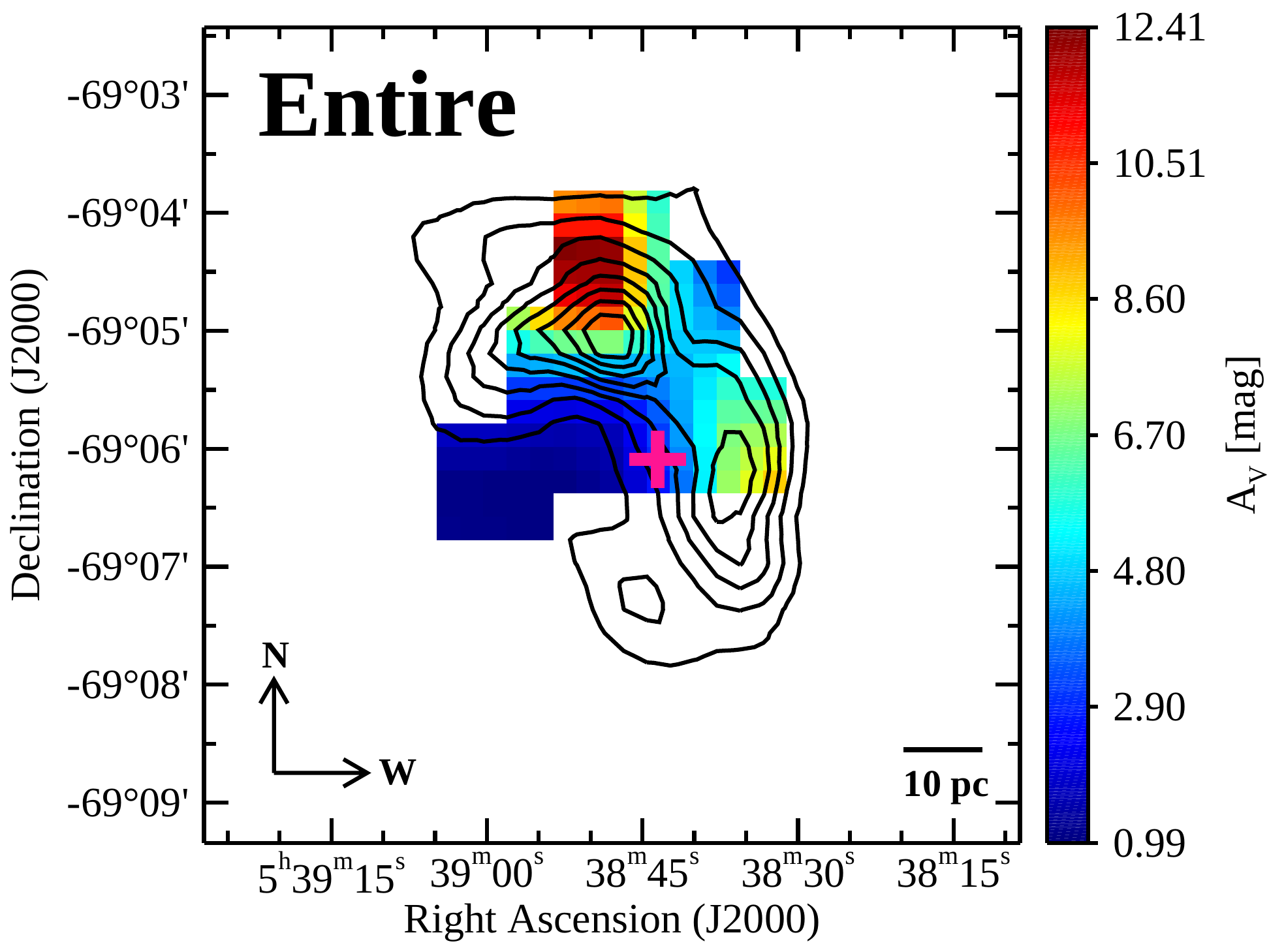}
	\caption{\label{f:cii_peak_av} {Spatial distributions of the \CII peak brightness temperature (left; 30$''$ scales) 
	and $V$-band dust extinction (right; 42$''$ scales).
	The entire velocity range was considered for the \CII peak temperature, 
	while the dust extinction was scaled by (30/42)$^{2}$ to match the angular resolution of the \HI data. 
	For both images, 1.4~GHz continuum emission is overlaid as the contours with levels ranging from 20\% to 90\% of the peak value ({448}~K) in steps of 10\%, 
	and the location of R136 is indicated as the red cross.} }
\end{figure*}

{Before applying the methodology of \citet{goldsmith2018} to 30 Dor, 
we examined if \CII emission is indeed optically thin. 
According to \citet{goldsmith2012}, the medium is in the ``effectively optically thin (EOT)'' regime   
if the main-beam brightness temperature is lower than $\sim$1/3 of the gas kinetic temperature (or $\lesssim$~20--30~K for typical interstellar clouds). 
In this case, the main-beam brightness temperature is linearly proportional to the C$^{+}$ column density regardless of the line opacity. 
The measured peak main-beam brightness temperature of \CII emission in 30 Dor reaches up to $\sim$15~K only (Figure~\ref{f:cii_peak_av}), 
suggesting that the [C~\textsc{ii}]-emitting gas is likely in the EOT regime. }

{Next, we estimated the molecular hydrogen column densities for B1, B3, and B4 
based on the $V$-band dust extinction data from \citet{lee2019}. 
To do so, we derived the total hydrogen column density for the foreground components (including B1, B2, B3, and B4; ``Entire'') as follows: }

{\begin{equation} \label{eq:N_H_Av_based} 
\begin{split}
N(\textrm{H})_{\textrm{f,Entire}}~(\textrm{cm}^{-2}) & = N(\textrm{H}~\textrm{\textsc{i}})_{\textrm{f;Entire}} + 2N(\textrm{H}_{2})_{\textrm{f,Entire}} \\ 
                                                     & = (5.56 \times 10^{21}) q A_{V} \\ 
\end{split}
\end{equation} }

\noindent 
{Here we adopted the same $q = 0.5$ as we did for the determination of \HI properties (Section~\ref{s:ts_derivation}) 
and the gas-to-dust ratio of $N(\textrm{H})/A_{V} = 5.56 \times 10^{21}$~cm$^{-2}$~mag$^{-1}$ for the LMC from \citet{pineda2017}. 
In addition, we scaled the 42$''$-scale $A_{V}$ from \citet{lee2019} by (30/42)$^{2}$ to match the angular resolution of our \HI data, 
assuming that dust grains uniformly fill the beams. 
To divide the total hydrogen column density into the individual velocity bands, 
we calculated the fractional contribution of each velocity band to the total atomic hydrogen column density 
$R_{i} = N(\textrm{H~\textsc{i}})_{\textrm{f,i}} / N(\textrm{H~\textsc{i}})_{\textrm{f,Entire}}$ ($i$~=~B1, B3, and B4) 
and applied it to the total hydrogen column density as follows: }

{\begin{equation} \label{eq:N_H2_Av_based} 
2N(\textrm{H}_{2})_{\textrm{f,i}} = R_{\textrm{i}} \left(N(\textrm{H})_{\textrm{f,Entire}} - N(\textrm{H~\textsc{i}})_{\textrm{f,Entire}}\right). 
\end{equation} }

\noindent 
{We also derived the C$^{+}$ column densities for B1, B3, and B4
by applying the carbon fractional abundance of $[\textrm{C/H}] = 7.94 \times 10^{-5}$ for the LMC from \citet{dufour1982} 
to the total hydrogen column density: }

{\begin{equation} \label{eq:N_C+}
\begin{split}
N(\textrm{C}^{+})_{\textrm{f,i}} & = 7.94 \times 10^{-5} N(\textrm{H})_{\textrm{f,i}} \\ 
				 & = 7.94 \times 10^{-5} \left(N(\textrm{H~\textsc{i}})_{\textrm{f,i}} + 2N(\textrm{H}_{2})_{\textrm{f,i}}\right).
\end{split}
\end{equation} }

\noindent 
{With this C$^{+}$ column density, 
the ratio of the C$^{+}$ column density to the integrated intensity for B1, B3, and B4 can be estimated as follows: }

{\begin{equation} \label{eq:X}
\begin{split}
	X_{\textrm{f,i}} & = \frac{N(\textrm{C}^{+})_{\textrm{f,i}}}{q W(\textrm{[C~\textsc{ii}]})_{\textrm{i}}} \\
	                 & =  \frac{N(\textrm{C}^{+})_{\textrm{f,i}}}{0.5 W(\textrm{[C~\textsc{ii}]})_{\textrm{i}}}. \\
\end{split}
\end{equation} }

\begin{figure*} 
	\centering 
	\includegraphics[scale=0.17]{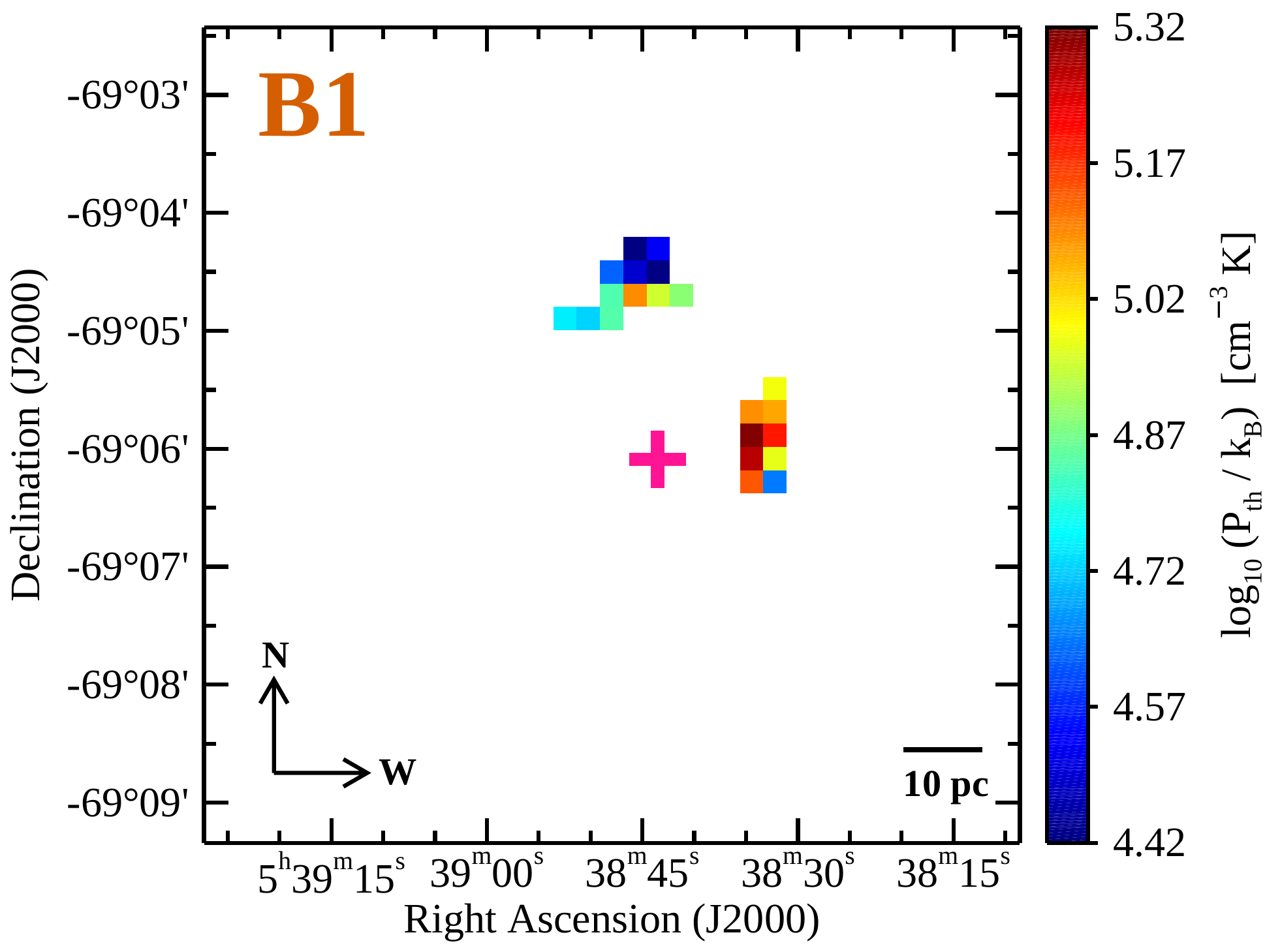} \hspace{0.1cm}
	\includegraphics[scale=0.17]{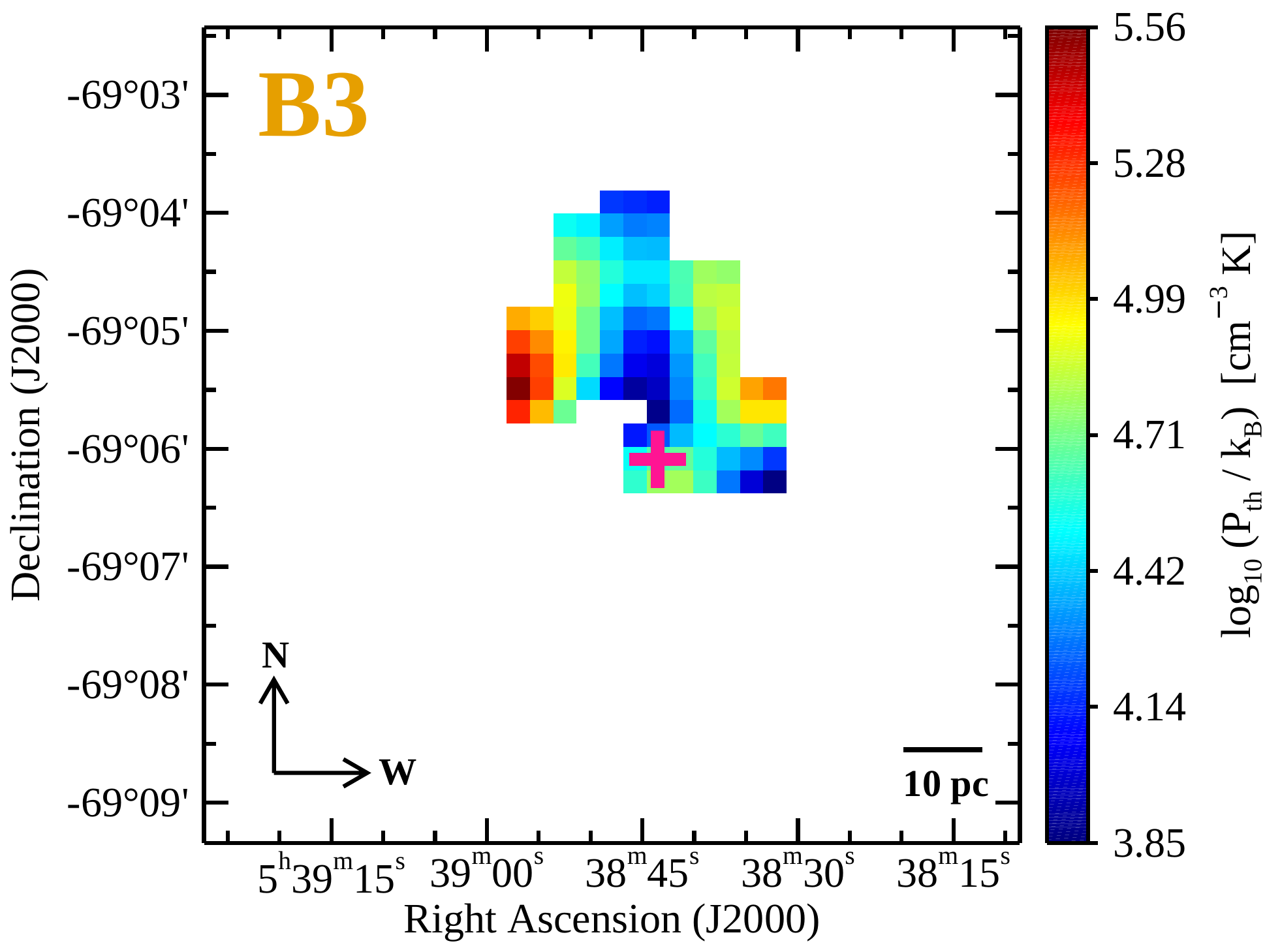} \hspace{0.1cm}
	\includegraphics[scale=0.17]{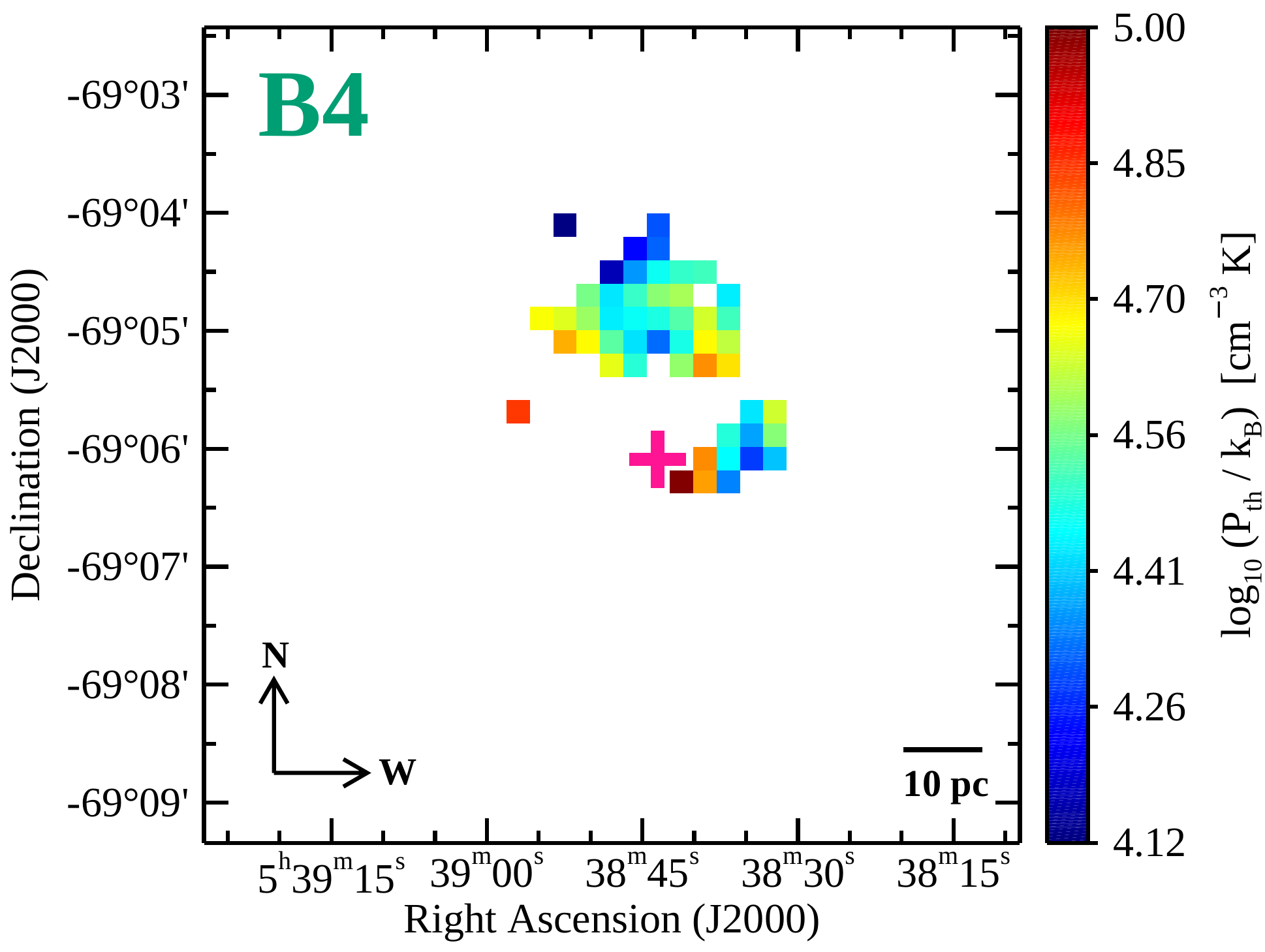}
	\caption{\label{f:pth_per_band_using_Av} {Spatial distribution of the thermal pressure for each velocity band (B1, B3, and B4). 
	All three maps have an angular resolution of 30$''$ with a pixel size of 12$''$, 
	and the location of R136 is indicated as the red cross.} }
\end{figure*}

\begin{figure*}
    \centering
    \includegraphics[scale=0.3]{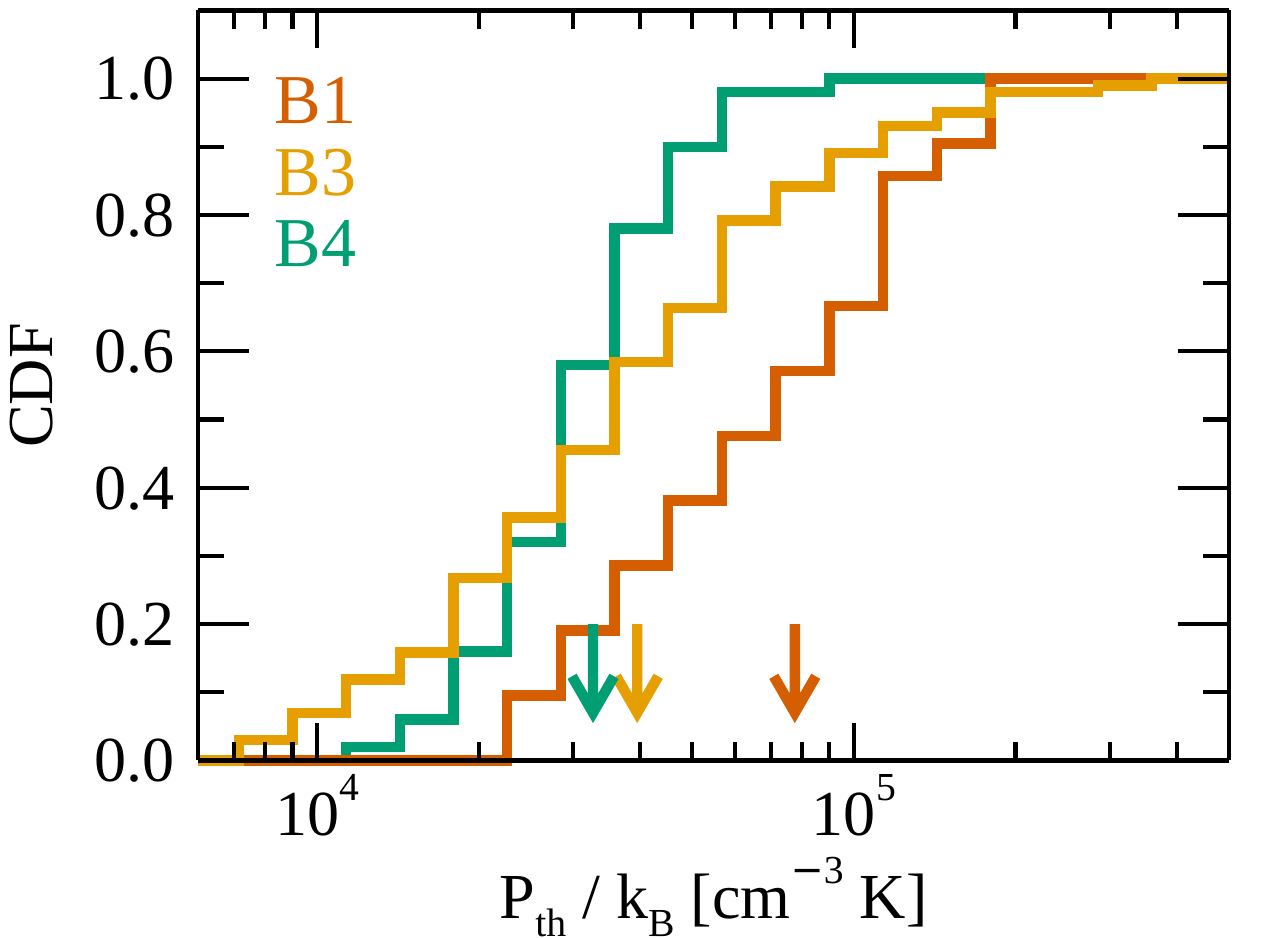}
	\caption{\label{f:cdfs_using_Av} {CDFs of the thermal pressures. The median of each distribution is marked as the arrow.} }
\end{figure*}

{To calculate the thermal pressure of the neutral ISM, we assumed that atomic and molecular hydrogens  
have the same number densities ($f_{\textrm{n}}$(H$^{0}$)~=~$f_{\textrm{n}}$(H$_{2}$)~=~0.5) 
and kinetic temperatures ($T_{\textrm{k}}$~=~<$T_{\textrm{s}}$>) 
and applied the corresponding Equations~\ref{eq:R_ul_mix} and \ref{eq:X} to Equation~\ref{eq:total_n}. 
For our calculation, we selected the pixels where $N$(H~\textsc{i})$_{\textrm{f,i}}$, $N$(H)$_{\textrm{f,i}}$, and $W$([C~\textsc{ii}])$_{\textrm{f,i}}$ 
($i$~=~B1, B3, and B4) are higher than their 3$\sigma$ uncertainties and 
present the results in Figures~\ref{f:pth_per_band_using_Av} and \ref{f:cdfs_using_Av} and Table~\ref{t:derived_quantities}. }

{While keeping in mind the limitation of the simple assumptions in our calculation  
(e.g., $q = 0.5$, fixed gas-to-dust ratio, and same densities/temperatures for H$^{0}$ and H$_{2}$), 
we found that B1 has systematically higher thermal pressures than B3 and B4. 
For example, the median thermal pressure of 7.8~$\times$~10$^{4}$~cm$^{-3}$~K for B1 is 
a factor of two higher than that of (3.3--3.9)~$\times$~10$^{4}$~cm$^{-3}$~K for B3 and B4. 
The higher thermal pressures of B1 are consistent with the expectation for the observed weakness in \HI emission (Section~\ref{s:outflow_inflow}).
Finally, we note that our estimates of (0.7--36.7)~$\times$~10$^{4}$~cm$^{-3}$~K for B1, B3, and B4 are largely in agreement with \citet{pineda2017}, 
who found (5.0--12.6)~$\times$~10$^{4}$~cm$^{-3}$~K for the LOSs associated with 30 Dor 
based on the same methodology as ours except for few differences (e.g., different densities/temperatures for H$^{0}$ and H$_{2}$). }

\begin{deluxetable}{l c c c c c c c}
\centering
\tablecaption{\label{t:derived_quantities} {Results for B1, B3, and B4} }
    \tablewidth{0pt}
    \setlength{\tabcolsep}{15pt}
    \tabletypesize{\small}
    \tablehead{
	    \colhead{Band} & \colhead{\# of Pixels} & \colhead{<$T_{\textrm{s}}$>} &\colhead{$2N({\rm H_{2})}_{\rm f}$} & \colhead{$f(\textrm{H}_{2})_{\textrm{f}}$} & \colhead{$qW({\rm [C~\textsc{ii}]})$} & \colhead{$X_{\textrm{f}}$} & \colhead{$P_{\textrm{th}}/k_{\textrm{B}}$} \\
	    \colhead{} & \colhead{} & \colhead{(K)} & \colhead{(10$^{21}$ cm$^{-2}$)} & \colhead{} & \colhead{(K~km~s$^{-1}$)} & \colhead{($10^{16}$~cm$^{-2}$~(K~km~s$^{-1}$)$^{-1}$)} & \colhead{(10$^{4}$~cm$^{-3}$~K)} \\
	    \colhead{(1)} & \colhead{(2)} & \colhead{(3)} & \colhead{(4)} & \colhead{(5)} & \colhead{(6)} & \colhead{(7)} & \colhead{(8)}
    }
    \startdata
	B1 & 21 & 43.6--138.4 & 0.5--1.3 & 0.79--0.91 & 1.7--5.1 & 1.4--5.9 & 2.6--20.7 \\
	   &    & 80.6 & 0.9 & 0.84 & 2.8 & 3.1 & 7.8 \\
    B3 & 101 & 72.0--183.1 & 0.9--10.9 & 0.47--0.91 & 1.3--31.1 & 1.2--22.9 & 0.7--36.7 \\
      &    & 114.9 & 4.3 & 0.81 & 8.5 & 4.5 & 3.9 \\
	B4 & 50 & 200.5--558.7 & 0.5--4.3 & 0.53--0.91 & 1.4--5.2 & 2.4--20.4 & 1.3--9.9 \\
           &    & 310.7 & 2.1 & 0.83 & 2.7 & 7.8 & 3.3 \\
    \enddata
    {\tablecomments{\small 
	(1) Velocity band; 
	(2) Number of pixels that were considered for the derivation of the thermal pressure. 
	These pixels were selected to have $N$(H~\textsc{i})$_{\textrm{f}}$, $N$(H)$_{\textrm{f}}$, and $W$([C~\textsc{ii}])$_{\textrm{f}}$~>~3$\sigma$ uncertainties. 
	For B1, two pixels with negative thermal pressures were masked;
	(3) LOS average spin temperature;
	(4) Dust-based H$_{2}$ column density (Equation~\ref{eq:N_H2_Av_based}). 
	The derived H$_{2}$ column densities are all higher than 3$\sigma$ uncertainties; 
	(5) Molecular fraction, as defined by 2$N$(H$_{2}$)$_{\textrm{f}}$/$N$(H)$_{\textrm{f}}$; 
	(6) \CII integrated intensity; 
	(7) Ratio of the C$^{+}$ column density to the integrated intensity (Equation~\ref{eq:X}); 
	(8) Thermal pressure.  
	For each of the columns (3)--(8), the range and median value for the selected pixels are listed. 
    } }
\end{deluxetable} 

\clearpage

\bibliography{gpark}
\bibliographystyle{aasjournalv7}

\end{document}